\documentclass[twocolumn]{aastex631}

\usepackage{cancel}

\usepackage{amsmath}
\usepackage{graphicx}
\usepackage{float}
\usepackage{listings}
\usepackage{cancel}
\usepackage{url}
\usepackage{natbib}
\usepackage{wrapfig}
\usepackage{nicefrac}
\usepackage{placeins}
\usepackage[ruled,lined]{algorithm2e}\usepackage{algpseudocode}
\usepackage[english]{babel}
\usepackage{silence}
\defcitealias{Castor1975}{CAK}

\shorttitle{Line-Driving Radiation Force for AGN Outflows}

\begin{document}

\title{A Self-Consistent Treatment of the Line-Driving Radiation Force for Active Galactic Nuclei Outflows: New Prescriptions for Simulations}

\correspondingauthor{Aylecia S. Lattimer}

\author[0000-0002-2004-5084]{Aylecia S. Lattimer}
\affiliation{Department of Astrophysical and Planetary Sciences,
Laboratory for Atmospheric and Space Physics,
University of Colorado, Boulder, CO 80309, USA}

\author[0000-0002-3699-3134]{Steven R. Cranmer}
\affiliation{Department of Astrophysical and Planetary Sciences,
Laboratory for Atmospheric and Space Physics,
University of Colorado, Boulder, CO 80309, USA}

\begin{abstract}
Flows driven by photons have been studied for almost a century, and a quantitative description of the radiative forces on atoms and ions is important for understanding a wide variety of systems with outflows and accretion disks, such as active galactic nuclei. Quantifying the associated forces is crucial to determining how these outflows enable interactive mechanisms within these environments, such as AGN feedback. The total number of spectral lines in any given ion of the outflow material must be tabulated in order to give a complete characterization of this force. Here we provide calculations of the dimensionless line force multiplier for AGN environments. For a wide array of representative AGN sources, we explicitly calculate the photoionization balance at the proposed wind-launching region above the accretion disk, compute the strength of the line-driving force on the gas, and revisit and formalize the role of the commonly-used ionization parameter $\xi$ in ultimately determining the line-driving force.  We perform these computations and analyses for a variety of AGN central source properties, such as black hole mass, initial wind velocity, and number density. We find that, while useful, the ionization parameter provides an incomplete description of the overall ionization state of the outflow material. We use these findings to provide an updated method for calculating the strength of the radiative line-driving using both the X-ray spectral index $\Gamma_X$ and the ionization parameter.
\end{abstract}

 
\section{Introduction} \label{sec:intro}
Active Galactic Nuclei (AGN) are believed to be powered by accretion onto a supermassive black hole (SMBH). Quantifying and understanding the associated mass and energy flows are crucial to understanding the observed properties of quasars, blazars, Seyfert, and radio galaxies, as well as how these objects evolve and interact with their surroundings. These powerful outflows are suspected of being capable of influencing star formation rates far beyond the small-scale size of the AGN itself. This long-range ``feedback'' is likely responsible for producing several known correlations between the SMBH mass and a range of large-scale properties of the host galaxy and intergalactic medium \citep{Magorrian1998,Silk+Rees1998,Cavaliere+2002,Ostriker+2010,Fabian2012,King+Pounds2015,Hopkins+2016,Harrison+2018}.

\begin{figure*}
    \centering
    \includegraphics[width=0.8\textwidth]{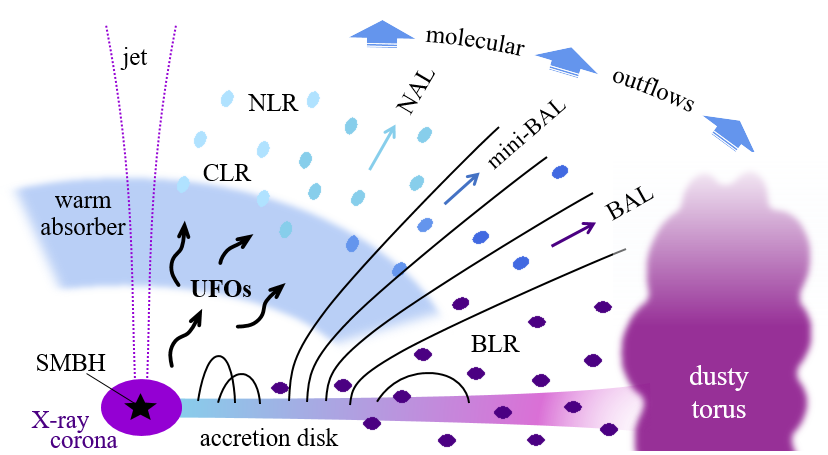}
    \caption{Idealized example illustration of AGN disk geometry showing the variety of possible observed outflow features and regions. See text for acronym definitions.}
    \label{fig: AGN_example}
\end{figure*}

As seen in Figure \ref{fig: AGN_example}, various outflow components have been proposed to be present in many AGN \citep[see also][]{Crenshaw2003,Beckmann+Shrader2012,Laha2021}. The large number of observed outflow characteristics has resulted in a proliferation of taxonomic AGN types. Along with the development of these AGN categories came many attempts to build unified models that explain the wind types not as products of fundamental structural differences, but as a result of, e.g., viewing angle or evolutionary stage \citep{Antonucci1993,Elvis2000,Netzer2015,Padovani+2017}.
For example, features associated with the broad-line region (BLR) are believed to sit at lower latitudes than the larger-scale distribution of features associated with the narrow-line region (NLR). Energetic outflows in those regions are observed in UV lines as broad absorption line (BAL) or narrow absorption line (NAL) clouds, with features of an intermediate nature called ``mini-BAL.'' X-ray spectra indicate the presence of a dynamic coronal line region (CLR), some outflow regions that are highly ionized and escaping at nearly relativistic speeds (e.g., ultra-fast outflows; hereafter UFOs), and some with lower velocities and a range of ionization states (i.e., the elusive ``warm absorber''). The exhibited overlap in the measured properties of these features suggest that it is possible some could arise from the same highly structured and stratified structures being viewed with different spectral diagnostics.

Multiple explanations for AGN wind acceleration have been proposed. For example, thermal forces, magnetocentrifugal, and other magnetohydrodynamic (MHD) effects have all shown promise in being able to explain various key observations. 
It is likely that some combinations of effects may be responsible for accelerating some features. 
Although the physical processes responsible for accelerating AGN winds have not been conclusively identified, another possible mechanism driving for these outflows is radiative driving \citep{Giustini+Proga2019}. This is often referred to as radiation pressure, where the force of radiation acts on the spectral lines in the material of the wind or outflow. The number of spectral lines in any given ion have a dominant effect on this force, which can act on the plasma as a whole, or on subsets of ions \citep[][hereafter CAK]{Castor1974,Castor1975}. 
The subsequent absorption and re-emission of photons in a given spectral line results in a net radial transfer of momentum, giving rise to a line-driven wind. 

We now know that photon-driven outflows may play a role in various other astrophysical environments, such as massive OB stars, as well as other accreting objects such as protostars and cataclysmic variables. This type of driving is also often thought to occur in the BAL region of AGN, where the blueshift of the spectral lines suggest that outflow velocities can reach up to $0.2c$. Line-driven disk winds are a promising hydrodynamical scenario for these outflows, due in part to the large number of strong resonance lines present in the spectra of many broad absorption line quasars \citep[BALQSOs;][]{Murray+1995, Proga+2000, Proga2007,Risaliti&Elvis2010,Higginbottom2014,Nomura+2016,Zhu+2022}.

However, in order for line-driving to launch and maintain an outflow from the accretion disk, there must exist a sufficient number of bound atoms to produce those lines. Thus, the gas must be relatively lowly ionized. This presents a challenge to the efficiency of line-driving for these winds, since the material is question is in close proximity to the central source and can easily become over-ionized \citep{Higginbottom2014}. A proposed solution to this issue is a so-called ``failed wind'' near the interior of the accretion disk, which acts to shield the outer material from becoming over-ionized \citep[see, e.g.,][]{Proga+2000,Proga&Kallman2004, QueraBofarull2020}.

The main parameter used to characterize these outflows is the line-force multiplier (discussed in detail in Section \ref{subsec: M(t)}), which describes both the overall strength of the line-driving as well as how it varies in different acceleration regimes of an outflow. 
This work aims to produce improved models of these line-driving forces by building upon previous force multiplier calculations for AGN and other high-energy systems containing photoionized outflows \citep{Stevens&Kallman1990,Stevens1991,Arav+Li1994,Dannen+2019ApJ,Chelouche+Netzer2003,Everett2005,Chartas+2009,Saez+Chartas2011}. 
In this paper, we adapt the process outlined for massive stellar sources in \citet{Lattimer2021} to calculate the force multiplier for winds launched from AGN disks, using 
an even larger database of atomic line-strength parameters. This paper also computes the line-driving force for a large grid of background AGN parameters (i.e., black-hole mass, accretion rate, and the radial location, speed, and acceleration of the outflow), whereas many earlier models limit these calculations to a smaller number of representative cases. 
Additionally, we explore the importance of the previously mentioned
failed wind in launching and maintaining a line-driven disk wind.

In Section \ref{sec: set-up} we describe the assumed geometry of the system and the initial parameters of the models. Section \ref{sec:Atomic Data} describes the physics taken into account in the model and the calculation of the thermal equilibrium, ionization balance, and luminosity along the radial line of sight (LOS) from the central source. In Section \ref{sec:line strengths} we discuss the calculation of the weighted strengths of the spectral lines. We end with a discussion of our results and our conclusions in Sections \ref{sec:Results} and \ref{sec:Discussion} respectively.

\section{Model Geometry and Set-Up}\label{sec: set-up}
\subsection{SED of the Central AGN}
The purpose of this paper is to describe the variety of possible forms the line-driving force multiplier $M(t)$ may take in realistic AGN  environments. To do this, we stop short of attempting to produce self-consistent dynamical models of the AGN outflows themselves. We include only the properties of the gas required to compute the {\em ionization states} of the elements producing spectral lines that affect $M(t)$. Thus, the densities, velocities, and temperatures that we specify at various distances from the AGN central source are meant to be  representative examples of the conditions likely to be found in these regions, not complete solutions to fluid or MHD conservation equations.

\begin{figure}
    \centering
    \includegraphics[width=\linewidth]{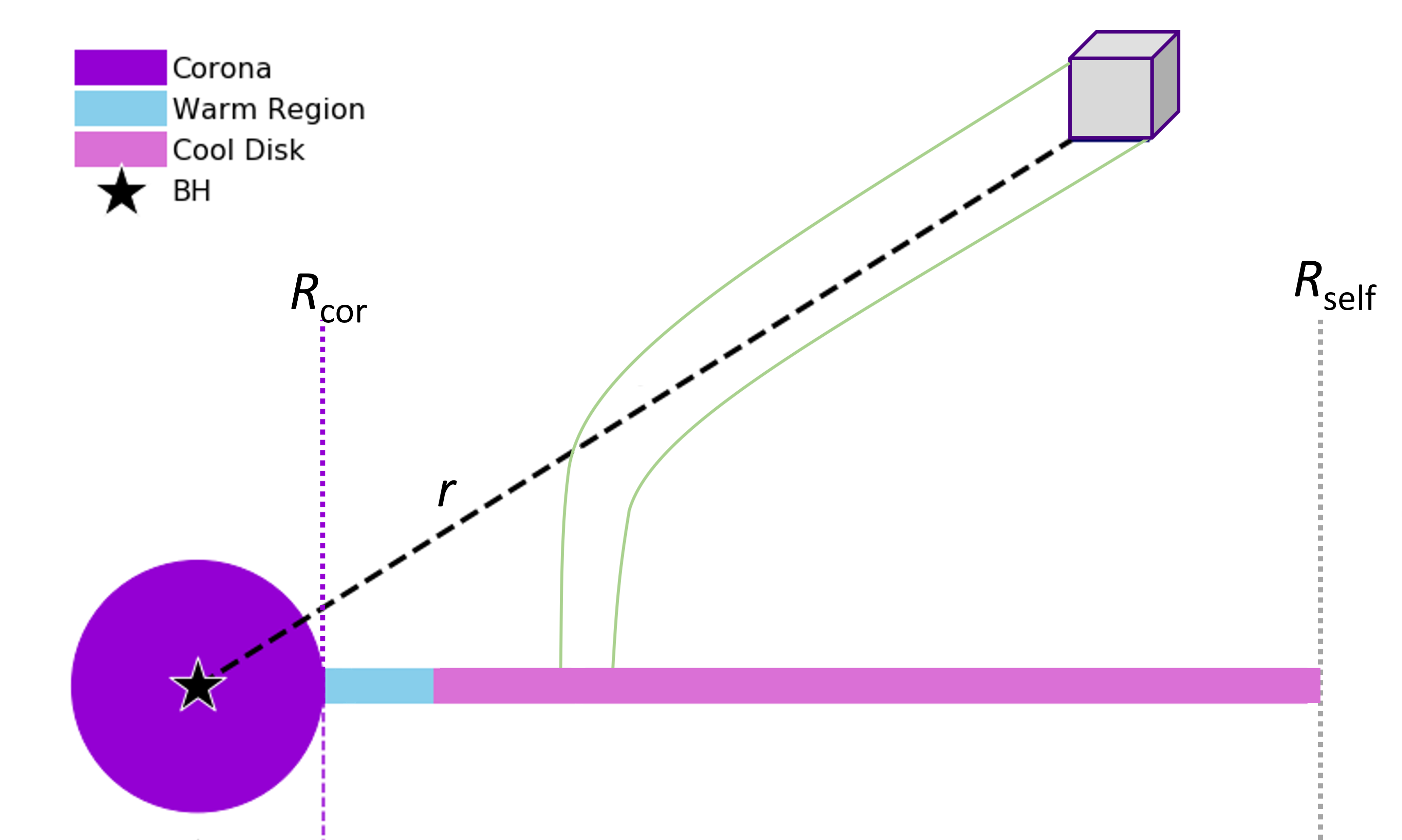}
    \caption{Modeled radial line-of-sight geometry at an arbitrary angle $\theta$ above the disk. Black dashed line shows the modeled line of sight and gray box represents the gas parcel. The radius of the hot corona and the edge of the disk are denoted by $R_{\rm cor}$ and $R_{\rm self}$ respectively. A representative version of the flux tube, shown in green, is included for completeness. Distances and sizes are not to scale.}
    \label{fig: geo}
\end{figure}

For our models, we consider an AGN central source embedded in an accretion disk. The initial SED of the central source is computed using a modified version of the python module QSOSED \citep{QSOSED}. This module recreates the \textit{qsosed} model of XSPEC \citep{XSPEC}, following the method outlined in \cite{Kubota+2018}. These model SEDs are made up of three characteristic regions: the outer cool disk, the warm Comptonising inner disk, and the hot inner Comptonising region (i.e. the X-ray corona). For the warm Comptonising region, QSOSED follows the passive disk scenario as discussed in \cite{Petrucci+2018,Kubota+2018}. The general geometry followed by both QSOSED and our models is shown in Figure \ref{fig: geo}.

The two main free parameters used for computing the SED are the mass of the black hole $M_{\rm BH}$ and the black hole mass accretion rate in Eddington units $\dot{m}$ \citep{Beckmann+Shrader2012}, given by
\begin{equation}
    \dot{m} = \frac{\dot{M}_{\textrm{acc}}}{\dot{M}_{\textrm{Edd}}}.
\end{equation}
Additionally, for all models we assume a black hole dimensionless spin absolute value $a_\ast$ of zero, meaning that all black holes in our models are non-rotating. This also determines that the radius $r_{\rm isco}$ of the innermost stable circular orbit (ISCO) will be constant across all models when given in units of gravitational radii $R_g$, defined as
\begin{equation}
    R_g = \frac{GM_{\textrm{BH}}}{c^2}.
\end{equation}

For a given set of AGN properties, we compute the line force multiplier $M(t)$ along a radial ray of points that starts at the radius of the corona $R_{\textrm{cor}}$ and extends to some outermost distance $r_{\rm out}$. This ray allows us to compute the line force at many possible locations where there may be a line-driven wind. This is also used to compute the absorption of the SED due to material interior to each radial distance $r$ along it. For the remainder of this work, we define $r_{\rm out}$ as 3 kpc for each model. This was chosen to ensure that the outermost radial step of the model exceeds the largest distances reported for observed AGN outflow features \citep[see, e.g.][]{Laha2021}. For completeness, we additionally define the outer edge of the disk to be the self-gravity radius $R_{\rm self}$ as given in  \cite{Laor+Netzer1989}. 
Figure \ref{fig: geo} shows a generalized diagram of the modeled geometry. We note that although Figure \ref{fig: geo} includes the curvature of the flux-tube for the sake of completeness, this curvature is not taken into account by the mass conservation equation used in this work.

\begin{figure*}[t]
    \centering
    \includegraphics[width=\linewidth]{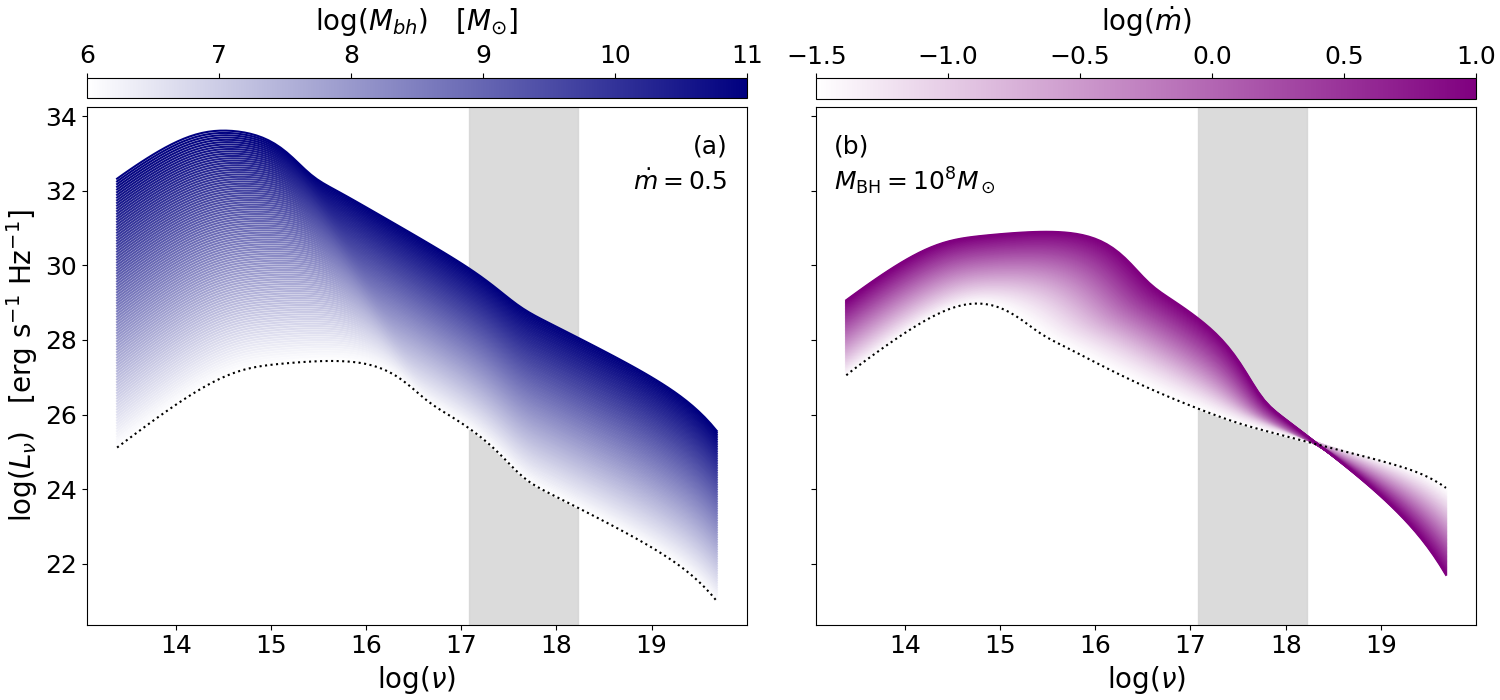}
    \caption{Luminosity $L_\nu$ as calculated by QSOSED at the first radial step $R_{\rm cor}$ for (a) varying black hole mass $M_{\rm BH}$ and (b) varying accretion rate in Eddington units $\dot{m}$. The curves corresponding to the lowest value of $M_{\rm BH}$ and $\dot{m}$ respectively are marked with a dotted black line in each panel. The portion of the X-ray region (0.5--7 keV) used to define X-ray spectral index $\Gamma_X$ is shaded in gray.}
    \label{fig: qsosed_example}
\end{figure*}

The luminosity of the central source at the first radial step ($R_{\rm cor}$), is calculated from the SED flux $F_\nu$ that is produced by QSOSED:
\begin{equation} \label{eq:Lnu_init}
    L_{\nu,0} = 4\pi R_{\textrm{cor}}^2F_\nu
\end{equation}
Likewise, the initial bolometric luminosity
\begin{equation}
    L_{\textrm{bol}} = \int L_\nu \,d\nu
\end{equation}
is also defined by QSOSED at $R_{\textrm{cor}}$. Figure \ref{fig: qsosed_example} shows the calculated luminosity as produced by QSOSED at $R_{\rm cor}$ for a range of black hole masses and accretion rates. 
We see a linear correlation between the luminosity $L_\nu$ and $M_{\rm BH}$. Thus, as seen in \cite{Kubota+2018}, we can set the initial SED using solely the black hole mass $M_{\rm BH}$ and the accretion rate $\dot{m}$. We also see that the SED at X-ray frequencies depends strongly on $\dot{m}$, flattening significantly with decreasing accretion rate. Similarly, \cite{Kubota+2018} found that the hot Comptonization region must truncate at the corona ($R_{\rm cor}$) for low accretion rates ($\dot{m}\lesssim 0.2$). This is due to lower limits on the X-ray spectral index $\Gamma_X$ \citep{Haardt+Maraschi1991, Malzac+2005,Stern+1995}, which describes the slope of the X-ray portion of the SED (see Section \ref{sec:Discussion} for further discussion of $\Gamma_X$). Thus, we assume for all models that the disk does not extend inward beyond the radius of the hot corona such that the disk begins at $R_{\rm cor}$ and extends to $r_{\rm out}$. This subsequently allows us to use $R_{\rm cor}$ as the first radial point along the primary LOS.

\begin{figure}[t!]
    \centering
    \includegraphics[width=\linewidth]{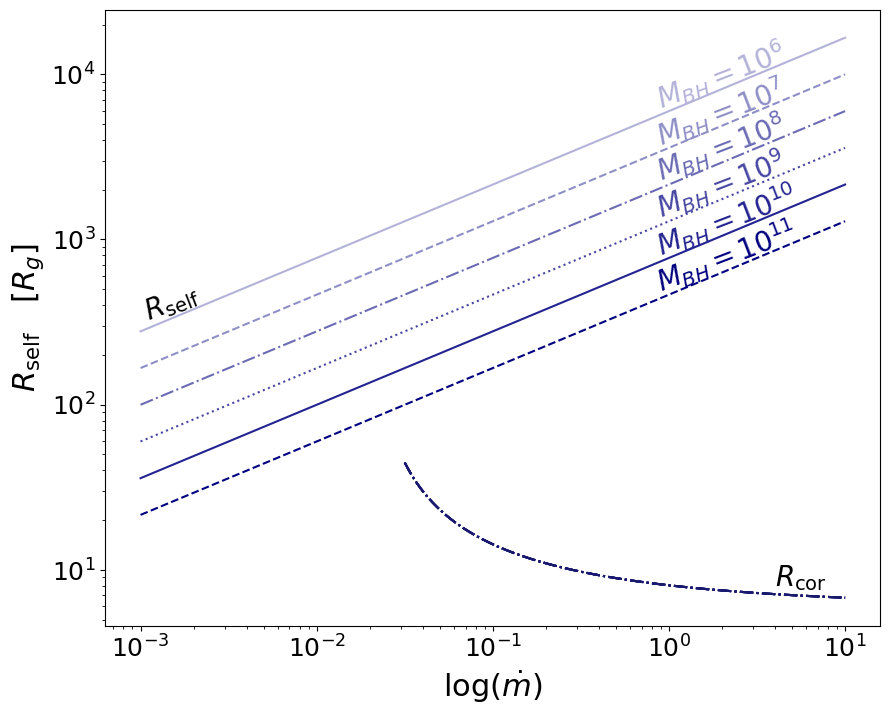}
    \caption{Dependence of the self-gravity radius $R_{\rm self}$ and the corona radius $R_{\rm cor}$ on the mass accretion rate onto the black hole in Eddington units, $\dot{m}.$ $R_{\rm cor}$ shows little dependence upon the mass of the central black hole ($M_{\rm BH}$).}
    \label{fig: mdot_dep}
\end{figure}

Figure \ref{fig: mdot_dep} shows the dependence of both $R_{\rm self}$ and $R_{\rm cor}$ on the mass accretion rate onto the black hole $\dot{m}$. From Equation (18) of \cite{Laor+Netzer1989}, we expect that the self-gravity of the disk will be lower for large $M_{BH}$, due to the strong gravity of the central source. Conversely, for large values of $\dot{m}$, the larger amount of gas in the disk should increase the self-gravity, pushing $R_{\rm self}$ outward. This is confirmed in Figure \ref{fig: mdot_dep}, where we see that $R_{\rm cor}$ starts to exceed $R_{\rm self}$ for Eddington accretion rates below a few times $10^{-2}$, putting a lower limit on the values of $\dot{m}$ that we use in these models. 

\subsection{Gas Outflow Properties \& Monte Carlo Process} \label{sec: initial conditions}
To create our grid of models, we construct ranges of reasonable values for our initial parameters. The first of these are the black hole mass and the accretion rate as described above, which set the properties of the source SED. Additional input parameters include launch radius of the wind $r_L$, number density exponent $\beta$, a shielding factor $S$, the wind fraction $f_{\rm wind}$, and the outflow velocity $u_{\rm wind}$. Grids of possible input values for each initial parameter are given in Table \ref{Table: param_grids}, and we describe each of these parameters in more detail below.

We must also define the radial grid that represents the line of sight between the parcel of gas and the central AGN. Beginning at $R_{\rm cor}$, each model steps radially through the line of sight of the gas parcel to the final observing radius $r_{\rm out}$ from the central source. We found that 500 points along this line of sight were sufficient to achieve convergence in our final calculations of the force multiplier. Taking into account the truncation of the disk at $R_{\rm cor}$ discussed above, and the fact that the calculation of the initial SED includes the SED of the corona itself, we therefore use the corona radius as the first radial step along the LOS. This precludes the ability to begin the radial grid inside the corona, at, for example, $r_{\rm isco}$. 
The rapid attenuation of the SED with increasing radius requires that the distance between each radial step be smaller at distances close to $R_{\rm cor}$.
Because the radial grid must by necessity span several orders of magnitude, we space our steps by first creating a condensed grid, defined by bounds $x_1$ and $x_2$, given by
\begin{align}
    x_1 & = (\log R_{\textrm{cor}})^{1/2} \\
    x_2 & = (\log r_{\text{out}})^{1/2}
\end{align}
We then space steps evenly in this condensed grid. To construct the final set of radial points to be used in our models, we convert these linearly spaced steps to fit the original bounds of our radial range (i.e. $R_{\rm cor}$ and $r_{\rm out}$). For example, for a given value $x_1<x<x_2$ in the condensed grid, the radius in the final radial grid along the primary LOS would be given by $\log(r)=x^2$.
This method preferentially clusters the radial points of the final grid near $R_{\rm cor}$, allowing us to reach consistent results with far fewer radial steps. For this work, we determined the number of radial points necessary to reach converged results by calculating results for a fixed representative subset of models using 100, 250, 500, 1000, and 1500 radial points. Using the method described above, we reached consistent results, where results from a given number of radial points matched those from the highest number of points, using only 500 radial steps. This drastically reduces the computation time required for a single model.
Therefore, for each model we used a grid of 500 radial steps, beginning at $R_{\rm cor}$ and ending at $r_{\rm out}$.

Motivated by the assumption that AGN outflows are launched at the surface of the accretion disk and often become reoriented in the radial direction \citep[see, e.g.][]{Proga&Kallman2004,QWIND3}, we define a so-called launch radius $r_L$ that falls somewhere between 100 and 1,000 $R_g$ for each model. We can then specify the hydrogen number density and outflow speed differently inside and outside this radius using
\begin{align}
    \label{eq: nH}
    n_H & = 
    \begin{cases}
        Sn_L & \text{, } R_{\textrm{cor}}<r<r_L \\
        n_L(r_L/r)^\beta & \text{, } r \geq r_L 
    \end{cases}\\
    \label{eq: u_r}
    u_r & =
    \begin{cases}
        0 & \text{, } R_{\textrm{cor}}<r<r_L \\
        u_{\textrm{wind}} & \text{, } r \geq r_L 
    \end{cases}
\end{align}
Here, $S$ is a dimensionless shielding enhancement factor introduced to vary the number density before the launch radius, in order to probe the importance of a proposed inner failed wind \citep[see, e.g.][]{Murray+1995,Chelouche+Netzer2003}. For this work, $\log(S)$ is sampled uniformly between $0$ and $2$. 

\begin{figure}
    \centering
    \includegraphics[width=\linewidth]{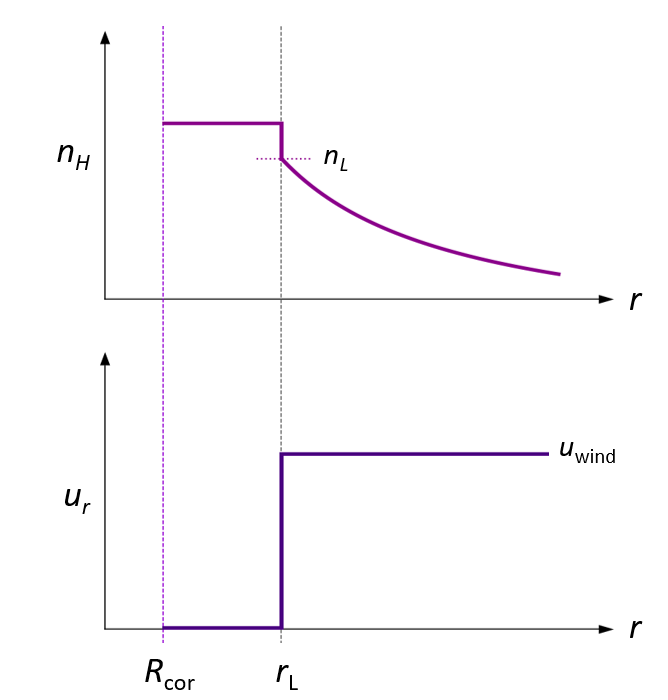}
    \caption{Representative versions of the hydrogen number density $n_H$ (Equation (\ref{eq: nH})) and radial flow speed $u_r$ (Equation (\ref{eq: u_r})). The radius of the hot corona and launch radius of the wind are denoted by $R_{\rm cor}$ and $r_L$ respectively. The number density at the launch radius, $n_L$, is also marked.}
    \label{fig: nh_ur}
\end{figure}

Note that at radii less than the launch radius, we assume a more-or-less static atmosphere with a large enough scale height to safely assume a constant density.  At radii greater than the launch radius, our radial grid intercepts a presumed outflow flux-tube that has a representative radial speed $u_{\rm wind}$ and a density that decreases with increasing distance.
The exponent $\beta$ is varied between values of 1.01 and 4. For a time-steady spherically symmetric outflow with a constant outflow speed, mass-flux conservation would give $\beta = 2$. Larger values would be consistent with a radially increasing velocity, so $\beta$ can be considered as a parameterization of the local degree of acceleration in the modeled outflow. 
Conversely, smaller values tend toward the commonly-used assumption of a constant representative density at different distances \citep[see, e.g.][]{Dannen+2019ApJ}.  However, we do not use values $\beta \leq 1$ as these values would lead to an infinitely large column density (see Equation (\ref{eq: N_H})). Figure \ref{fig: nh_ur} shows representative versions of $n_H$ and $u_r$ with increasing radius.


The number density at the launch radius $n_L$ is given by mass-flux conservation:
\begin{equation}\label{eq: nL}
    n_L = \frac{\dot{M}_{\textrm{wind}}}{4\pi r^2_Lm_Hu_{\textrm{wind}}}
\end{equation}
where $m_H$ is the mass of a hydrogen atom, and  $\dot{M}_{\textrm{wind}}$ is a spherically-symmetric approximation of the wind's mass-loss rate. This is given as some fraction $f_{\textrm{wind}}$ of the black hole's mass accretion rate $\dot{M}_{\textrm{acc}}$ \citep[see also][]{Daly2021}:
\begin{equation}
    \dot{M}_{\textrm{wind}} = f_{\textrm{wind}} \dot{M}_{\textrm{acc}}.
\end{equation}
We constrain this $f_{\textrm{wind}}$ ``efficiency factor'' to fall within the range of $10^{-2}$ and $10$ \citep[see, e.g.][]{King2010,King+Pounds2015,Mou+2017,Yi+2017,Leighly+2018}.

In Equation (\ref{eq: nL}), $u_{\rm wind}$ is the outflow velocity, which we set at a constant value for all points along the line of sight, beginning at the launch radius, as in Equation (\ref{eq: u_r}). The wind velocity for each model is chosen from a range of 100 to 70,000 km s$^{-1}$. This range ensures that our models encompass the full range of possible outflow velocities. The lower limit of $100$ km s$^{-1}$ is consistent with low-velocity NAL outflows, while the upper limit of $70,000$ km s$^{-1}$ represents high velocity BALs and mini-BALs \citep{Giustini+Proga2019}. Because we dictate that the launch radius of the wind fall between 100 and 1,000 $R_g$, we do not consider velocities that are typical of UFOs ($\sim0.4c$), as these are often launched below this range, at $\sim10$ $R_g$.


From the hydrogen number density $n_H$ (Equation (\ref{eq: nH})), the hydrogen column density $N_H$ at each radial step can also be found:
\begin{equation} \label{eq: N_H}
    N_{\rm H} \, = \, \int_r^{\infty} n_{H}(r')  dr'   
\end{equation}
Note that in Equation (\ref{eq: N_H}) we adopt the convention of calculating the column density as the intervening column density between the parcel of gas at its radially dependent position and an observer at $r=\infty$. See Section \ref{subsec: Lum & Flux} for an in-depth description of how the SED absorption is computed along the ray.

Additionally, we calculate the effective hydrogen-ionizing luminosity, $L_X$, of the current SED, given by
\begin{equation} \label{eq: L_X}
\begin{aligned}
    L_X & = {} \int_{\nu_{\rm min}}^{\nu_{\rm max}} L_\nu \,d\nu \\
    & = \int_{\nu_{\rm min}}^{\nu_{\rm max}}  16\pi^2r^2J_\nu d\nu.
\end{aligned}
\end{equation}
Here, $J_\nu$ is the mean intensity for a given frequency. It is useful to define $L_X$ in terms of $J_\nu$, as we eventually use the mean intensity to account for the full radiation field seen by the parcel of gas at a given radius (see Section \ref{subsec: Lum & Flux}). The frequency limits $\nu_{min}$ and $\nu_{max}$ are defined as the frequencies corresponding to 1 and 1,000 Rydbergs, respectively \citep{Fukumura+2022}. It is worth noting that $L_X$ is sometimes given as the integral over all energies above 1 Rydberg (13.6 eV). However, in this work we find the difference between this and the chosen limits of 1 to 1,000 Rydbergs to have a negligible difference on the values of $L_X$. 

From $L_X$ we can calculate the ionization parameter $\xi$ at each radial step, which is defined as 
\begin{equation} \label{eq: xi}
    \xi = \frac{L_X}{n_Hr^2}
\end{equation}
as originally given by \cite{Tarter+1969}. The ionization parameter is often used to describe the full ionization balance of the wind. 
However, it may fail to take into account the full ionization and recombination processes that may be present \citep[see, e.g ][]{Shields1974,Stevens1991,Osterbrock+Ferland2006,Kritcka+2022}.

\begin{deluxetable}{lccr} \label{Table: param_grids}
    \tablecolumns{4}
    \tablewidth{\linewidth}
    \tablecaption{The ranges of values and the associated units used to select the initial input parameters for each model run. Since these ranges by necessity cover many orders of magnitude for the majority of these parameters, the random sampling of each is done in logspace, with the exception of the density exponent $\beta$ and the launch radius, $r_L$.}
    \tablehead{
    \colhead{Parameter} & \colhead{\hspace{.5cm}Units\hspace{.5cm}}\hspace{.5cm} & \colhead{\hspace{.5cm}Lower\hspace{.5cm}}\hspace{.5cm} & \colhead{\hspace{.5cm}Upper\hspace{.5cm}}}
    \startdata
    $\log(M_{BH})$        & $M_\odot$   & $6$        & $11$      \\
    $\log(f_{\rm wind})$  &             & $-2$       & $1$       \\
    $\log(S)$             &             & $0$        & $2$       \\
    $\beta$               &             & $1.01$     & $4$       \\
    $r_L$                 & $R_g$       & $100$      & $1000$       \\
    $\log(u_{\rm wind})$  & $\rm km~s^{-1}$        & $2$        & $\sim 4.85$  \\
    $\log(\dot{m})$       & $\dot{M}/\dot{M}_{\rm Edd}$ & $-1.5$     & $1$    \\
    $\log(\gamma)$        &             & $-10$      & $-1$      \\
    \enddata
\end{deluxetable}

\section{Atomic \& Ion Microphysics}\label{sec:Atomic Data}

\subsection{Thermal Equilibrium} \label{subsec:Therm. Eq.}
In order to compute the ionization balance of each relevant element, we require values of the local temperature $T$ at each point along the radial ray.  To compute $T$, we assume that the parcel is in time-steady thermal equilibrium. \cite{Dyda+2017} found that this equilibrium consists essentially of radiative heating/cooling and adiabatic cooling. This can be specified as
\begin{equation} \label{eq:therm_eq}
    Q_{\textrm{rad}}(T) + Q_{\textrm{ad}}(T) = 0. 
\end{equation}
The radiative term is specified by an analytic fit that contains terms for bremsstrahlung, Compton, X-ray, and UV-line processes. We define
\begin{equation} \label{eq:Q_rad}
    Q_{\textrm{rad}}(T) = n^2_H(\Gamma-\Lambda).
\end{equation}
The quantity $\Lambda - \Gamma$ is given by Equation 8 of \cite{Dyda+2017} as a function of $T$ and the X-ray temperature $T_X$, which we compute from the radially dependent SED. The X-ray temperature $T_X$ can be calculated from the mean photon energy:
\begin{equation} \label{eq: TX}
    \langle h\nu\rangle = k_BT_X = \frac{\int_{\nu_{\rm min}}^{\nu_{\rm max}} L_\nu h\nu \textrm{d}\nu}{\int_{\nu_{\rm min}}^{\nu_{\rm max}} L_\nu\textrm{d}\nu}.
\end{equation}
The integration limits $\nu_{\rm min}$ and $\nu_{\rm max}$ used here are the frequencies that correspond to the traditionally accepted energy bounds on what are considered X-rays, 
at $0.1$ and $100~\rm keV$ respectively. These are chosen to represent the full range of wavelengths that fall in the X-ray portion of the spectrum when calculating the X-ray temperature. We note here that these limits are used solely in the calculation of $T_X$. The entire frequency range of the source SED is used when calculating the other quantities (e.g., $\xi$) that are used in determining the temperature used at each radial step.
\cite{Dyda+2017} used a fixed X-ray temperature of $T_X = 1.2\times 10^8$ K. 

\begin{figure}[b!]
    \centering
    \includegraphics[width=1.15\linewidth]{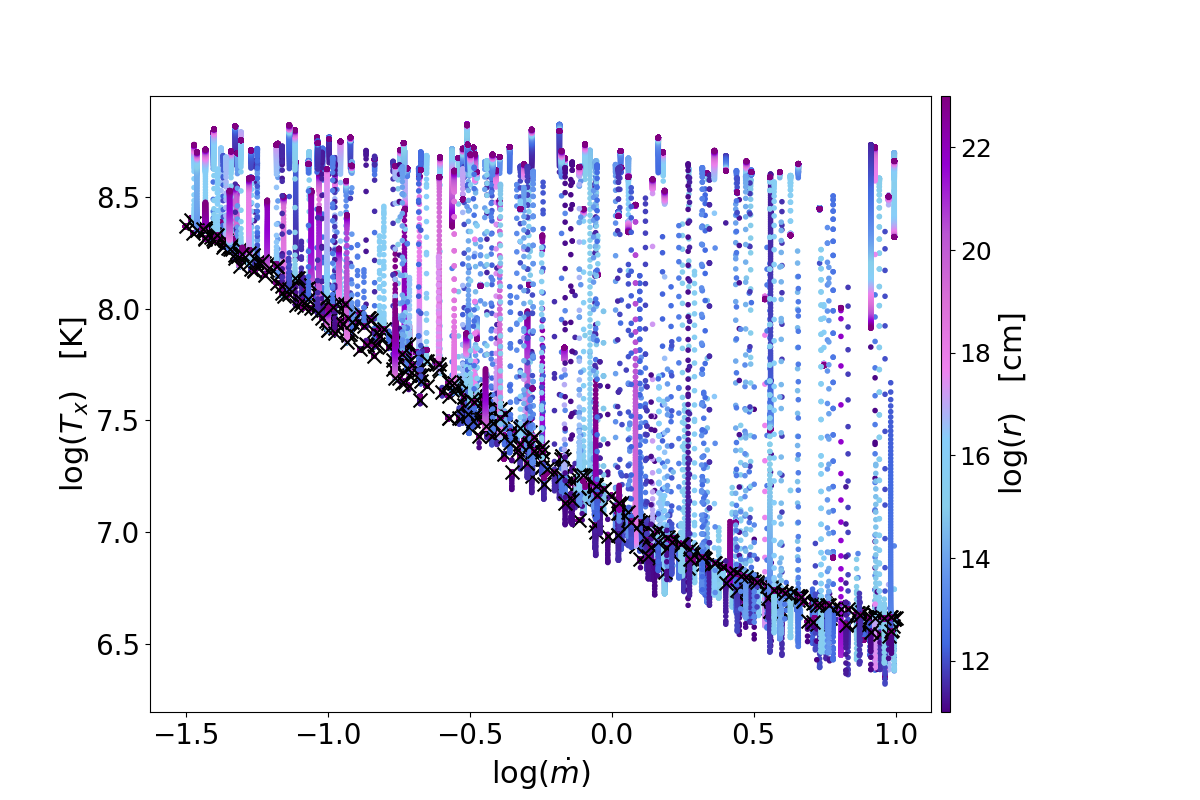}
    \caption{Calculated X-ray temperature $T_X$, plotted against the accretion rate $\dot{m}$ as it varies along the primary LOS. Black crosses indicate the basal value of $T_X$ at the first radial step $R_{\rm cor}$.}
    \label{fig: TX}
\end{figure}

We see from Figure \ref{fig: TX},  which shows the calculated values of $T_X$ along each step of the radial grid for our set of $\sim 550$ models, that many of our models at some point along the primary LOS fall near this value. 
The number and criteria of what we consider complete models are discussed in further detail later in this section. We see a clear anti-correlation between $\dot{m}$ and $T_X$ at the coronal base radius $R_{\rm cor}$.  This is a consequence of the steepening in the SED at X-ray frequencies seen in Figure \ref{fig: qsosed_example}(b), with steeper spectra (i.e., correspondingly lower values of $T_X$) occurring for larger accretion rates. We also see that as radial distance increases, $T_X$ frequently increases. The changes in SED shape that are responsible for this variation occur due to frequency-dependent absorption, and are discussed further below in Section \ref{subsec: Lum & Flux}.

Rearranging Equation (8) of \cite{Dyda+2017}, we can define $\Gamma - \Lambda$ as
\begin{equation} \label{eq: Lambda-Gamma}
\begin{aligned}
\Gamma - \Lambda & =\left[\mathcal{A}_c \xi(T_X - 4T)\right] \\
& +\left[\mathcal{A}_x \frac{\xi^{1/4}}{T^{1/2}}(1-\frac{T}{T_X})\right] \\
& -\left[\mathcal{A}_b \sqrt{T}\right] \\
& -\left[\mathcal{A}_l \frac{e^{-T_l/T}}{\xi\sqrt{T}}+10^{-24}\right], \\
\end{aligned}
\end{equation}
As in \cite{Dyda+2017}, the terms of Equation (\ref{eq: Lambda-Gamma}) 
represent the heating from Compton processes ($\mathcal{A}_c$) and X-rays ($\mathcal{A}_x$) and cooling from bremsstrahlung ($\mathcal{A}_b$) and spectral line emission ($\mathcal{A}_l$). Taking all quantities to be in cgs units, we set $\mathcal{A}_c = 8.9\times 10^{-36}$, $\mathcal{A}_x = 1.5\times 10^{-21}$, $\mathcal{A}_b = 1.3\times10^{-26}$, $\mathcal{A}_l = 1.7\times 10^{-18}$, and the line temperature $T_l = 1.3\times 10^5$ K \citep[see][]{Blondin1994, Dyda+2017}.

The adiabatic cooling term, which is present only when there is a nonzero outflow speed, was specified by \cite{Cranmer+2007} for flux-tube streamlines with cross-sectional area $A(r)$ as
\begin{equation} 
    Q_{ad}(T) = -\frac{3u}{2}\frac{\partial P}{\partial r} - \frac{5P}{2A}\frac{\partial}{\partial r}(uA)
\end{equation}
where $u$ is the outflow speed along the flux tube and $P$ is the gas pressure. If we assume power-law radial dependencies for these quantities, we can express everything in terms of the local quantities and not their derivatives. Thus, we assume $T\propto r^{-\delta}$ and $\rho \propto r^{-\beta}$. Combining this with mass-flux conservation along the expanding tube (i.e., $\rho uA =$  constant), we get
\begin{equation} \label{eq:Q_ad}
    Q_{ad}(T) = \frac{uP}{r}\left(\frac{3\delta}{2}-\beta\right).
\end{equation}
We set $\delta = 0$ for our model outflows. Setting $\delta = 0$ assumes the temperature varies more slowly with distance than the density, and it also ensures that $Q_{ad}$ is always negative, making it a true cooling term. Finally, for an ionized, hydrogen-dominated plasma we use the ideal gas law to specify $P \approx 2n_Hk_BT$.  

Combining Equations (\ref{eq:Q_rad}), (\ref{eq: Lambda-Gamma}), and (\ref{eq:Q_ad}), solving for $T$ becomes a root-finding exercise, which for this work was done using the \texttt{optimize} module available in SciPy \citep{scipy}. We also implement a temperature floor $T_{\rm floor}$, which we set at $T_{\rm floor} = 1000$ K. In the cases where the algorithm fails to find a root, the model is terminated and marked as a failed run. 
In order to achieve an acceptable level of statistical significance, we initiated 576 models. After accounting for these failed runs, likely resulting from randomly occurring unphysical combinations of initial parameters, we were left with a total of 556 successful modeled outflows. All analyses in the remainder of this work use this total number of model runs.

\begin{figure*}
    \centering
    \includegraphics[width=\textwidth]{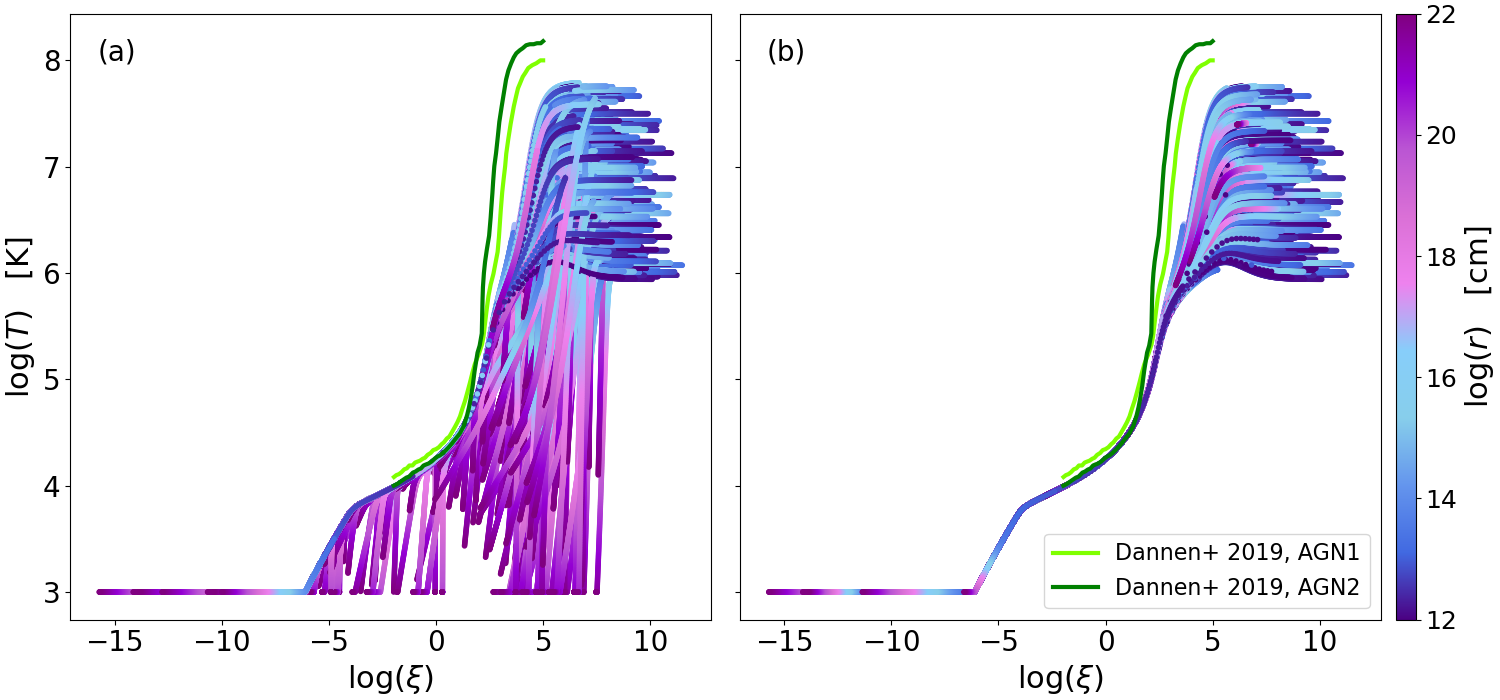}
    \caption{The calculated temperature $T$, varying with ionization parameter $\xi$ and compared to \cite{Dannen+2019ApJ}. Panel (a) shows $T$ as calculated from the full form of Equation (\ref{eq:therm_eq}). Panel (b) gives $T$ as calculated neglecting the adiabatic cooling term in Equation (\ref{eq:therm_eq}).}
    \label{fig: temp}
\end{figure*}

Figure \ref{fig: temp} shows the calculated temperature for each completed model as a function of the ionization parameter $\xi$. For all successful runs, we hereafter assume that the temperature $T$ found by the thermal equilibrium calculation described above is equal to the electron temperature of the material, such that $T_e=T$. Note that if all of our models had used identical values of $T_X$ and ignored adiabatic cooling, the results in Figure \ref{fig: temp} would all fall along a single curve, similar to the results of \cite{Dannen+2019ApJ} that are also displayed. Further quantitative comparison to previous work is given detailed in Appendix \ref{quantitative comparisons appendix}.

However, our models show additional variability that depends on different combinations of the input parameters. Specifically, the variability at small radial distances (corresponding mostly to $\xi \gtrsim 10^6$) is mostly due to variations in $T_X$ at the coronal base, which in turn are driven by different values of $\dot{m}$ (see Figure \ref{fig: TX}). The variability among our models at larger radii ($r \gtrsim 10^{18}$~cm, or $\xi \lesssim 10^5$) is mostly due to different models exhibiting differing relative amounts of adiabatic cooling. 
At inner radii, corresponding to high ionization regions beginning at $\sim\xi \gtrsim 10^3$, our results differ from the two \cite{Dannen+2019ApJ} curves due to the adiabatic cooling included in our thermal equilibrium calculation (see also Section \ref{subsec: Line-Driving Forces}). 
We note also that models with $\xi \lesssim 10^{-6}$ tend to correspond to plasmas with equilibrium temperatures at or below the imposed floor of 1000~K. In subsequent sections we often neglect these models as not corresponding to ``realistic'' AGN conditions.

\subsection{Ionization/Recombination Balance} \label{subsec:Ion. Bal.}
After obtaining the local temperature $T$, the ionization/recombination equilibrium of the gas parcel at the current radius can be computed. There are five processes that need to be accounted for in this calculation: collisional ionization ($CI_i$), autoionization ($AI_i$), photoionization ($PI_i$), radiative recombination ($RR_{i+1}$), and dielectric recombination ($DI_{i+1}$). We note that for this work we do not include three-body (collisional) recombination ($3BR_{i+1}$), as this effect generally does not represent a significant contribution to the ionization balance except in the high-density limit, which the AGN environments considered here traditionally do not experience \citep[see, e.g.,][]{Rees+1989}. The five rates considered are given in units of inverse seconds (s$^{-1}$). For a time steady case, we can write the ionization balance as
\begin{equation} \label{eq:ion_balance_tot}
    \frac{n_{\rm ion+1}}{n_{\rm ion}}= \frac{CI_{\rm ion} + AI_{\rm ion} + PI_{\rm ion}}{RR_{\rm ion+1} + DR_{\rm ion+1}}.
\end{equation}
Here the subscript $\rm ion$ indicates processes that ionize out of a given ionization state $\rm ion$ and into $\rm ion+1$, and the subscript $\rm oin+1$ in the denominator terms indicated processes that recombine from state $\rm ion+1$ down to $\rm ion$. Hence the numerator accounts for ionization processes, and the denominator accounts for recombination.

The collisional ionization ($CI_i$) rate is given by
\begin{equation} \label{eq:CI}
    CI_{\rm ion} = n_eq_{c,\rm ion}(T)
\end{equation}
where $n_e$ is the electron number density and $q_{c,i}(T)$ is a temperature dependent collision rate coefficient, with units of [cm$^3$ s$^{-1}$]. Fitting formulas for $q_{c,\rm ion}(T)$, for all elements up to Ni ($Z = 28$), were provided by \cite{Voronov1997} and are computed by a routine distributed by Dima Verner of the University of Kentucky. We estimate the coefficients for $Z =29$ and $Z=30$ according to the method developed in \cite{Cranmer2000}. 

We can write the autoionization ($AI_i$) rate as 
\begin{equation}
    AI_{\rm ion} = n_e q_{a,\rm ion}(T).
\end{equation}
Data for the rate coefficient fits to $q_{a,i}(T)$ were provided in \cite{Landini+Fossi1990} for 13 of the most abundant elements. Where available, we use these fits to calculate the autoionization contribution to the overall ionization balance. Where the rate coefficients were not available, i.e. for less abundant elements, the autoionization contribution was set to zero. 

The photoionization rate ($PI_{\rm ion}$) can be described by
\begin{equation} \label{eq:PI}
    PI_{\rm ion} = \int \frac{4\pi J_\nu}{h\nu}\sigma_\nu \textrm{d}\nu
\end{equation}
where $\sigma_\nu$ is the photoionization cross-section and $J_\nu$ is the mean intensity, as discussed further in Section \ref{subsec: Lum & Flux}. 

The rate for radiative recombination can be written as
\begin{equation}
    RR_{\rm ion+1} = n_e\alpha_{r,\rm ion}(T).
\end{equation}
Note that here the $\alpha$ rate coefficient is labeled with the stage that is being recombined into. For elements up to Zn ($Z = 30$), these rates were tabulated by \cite{Verner+Ferland1996} 

We describe dielectric recombination as
\begin{equation}
    DR_{\rm ion+1} = n_e\alpha_{d,\rm ion}(T).
\end{equation}
We calculate the dielectric recombination rates for elements up to $Z = 28$ from \cite{Mazzotta+1998}. However, we neglect dielectric recombination for elements with $Z>28$. 


Once the overall ionization balance is calculated from Equation (\ref{eq:ion_balance_tot}) for each ion of each element considered, we can then compute the number density $n_{\rm ion}$ for each ion. This is done following the iterative process described in Section 2.1 of \cite{Lattimer2021}. At each radial step the initial estimate of $n_e$ was given by $n_e = 0.8n_H$, where $n_H$ is the hydrogen number density of the wind (Equation (\ref{eq: nH})). The initial estimate was then refined using an undercorrection technique until convergence to a final value. This was done at the end of each iteration over the ionization balance by tabulating a new estimate of $n_e$ from the calculated ionization balance, which was then multiplied by the previous estimate. The square root of this product was then used as the estimate of $n_e$ for the next iteration. We found that 35 iterations were sufficient to reach a stable and self-consistent value. 
At each radial step we also calculate the fraction $n_{\rm ion}/n_{\rm el}$, as well as the mean ionization number $\mathcal{I}$, which we define using the ion number $i$:
\begin{equation} \label{eq: mean ion num}
   \mathcal{I} = \frac{\sum_{i}[(n_{\textrm{ion}}/n_{\textrm{el}})i]-1}{Z}.
\end{equation}
This provides a uniform method of showing the relative distance between an element being completely neutral (${\cal I} = 0$) and being fully ionized (${\cal I} = 1$). 
We also provide ionic fractions $n_{\rm ion}/n_{\rm el}$ for selected elements in Appendix \ref{quantitative comparisons appendix}, as well as a comparison of these values to those obtained from other photoionization codes.

\begin{figure*}[htb!]
    \centering
    \includegraphics[width=\textwidth]{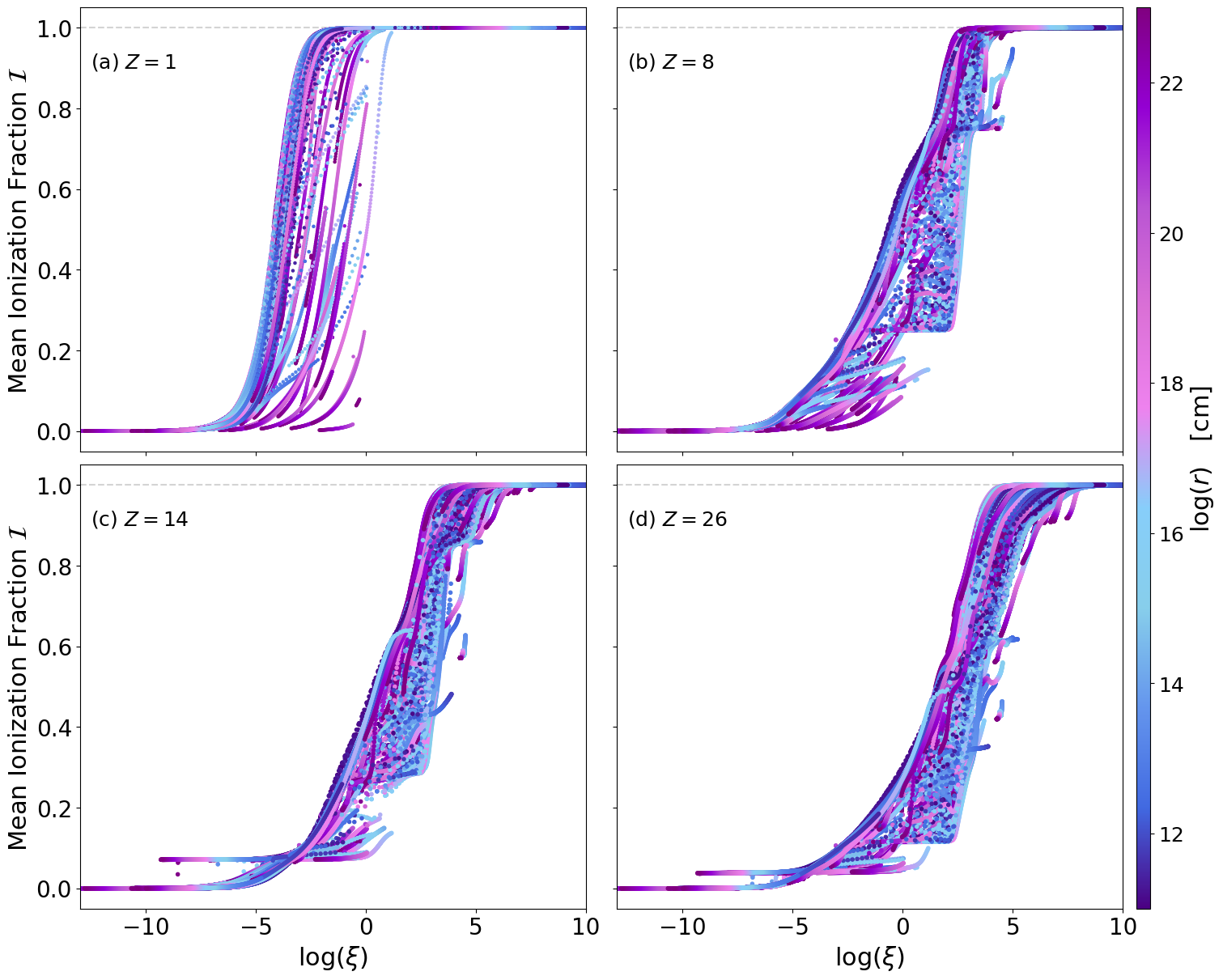}
    \caption{Mean ionization number $\mathcal{I}$, given by Equation (\ref{eq: mean ion num}) for (a) $Z=1$ (b) $Z=8$ (c) $Z=14$ and (d) $Z=26$ with respect to ionization parameter $\xi$. We see that high ionization parameter generally corresponds to radial points closer to the central source, i.e. high $\xi \rightarrow$ low $r$.}
    \label{fig: mean_ion_frac2}
\end{figure*}

Figure \ref{fig: mean_ion_frac2} shows the mean ionization number $\mathcal{I}$, varying with the ionization parameter $\xi$ for each model. Traditionally, $\xi$ has been used to describe the total ionization state of the outflow material. However, we see from Figure \ref{fig: mean_ion_frac2} that multiple values of $\mathcal{I}$ are possible for the same value of the ionization parameter, suggesting that $\xi$ does not provide an adequately detailed and thus unique description of the outflow's ionization balance \citep[see also, e.g.][]{Kallman&McCray1982,Stevens1991,Devereux&Heaton2013,Dannen+2019ApJ,Kallman+2021}. We discuss the implications of this further in Section \ref{sec:Discussion}.  

\subsection{Luminosity \& Flux Calculation}\label{subsec: Lum & Flux}

At each radial step along the line of sight, the local temperature of the gas $T$ (Section \ref{subsec:Therm. Eq.}), the ionization/recombination balance (Section \ref{subsec:Ion. Bal.}), and the luminosity $L_\nu$ of the source are recalculated to account for the attenuation of the SED due to both the increased distance from the central source and absorption from the intervening material. When
calculating the large-scale absorption of the SED, we take into account continuum and bound-free sources of opacity. We also adopt the convention of assuming that the line opacity is involved only in the calculation of the line-driving force \citepalias[see, e.g.,][]{Castor1975}. The frequency-dependent opacity of the material is then given by
\begin{equation} \label{eq: opacity}
    \chi_\nu = n_e\sigma_T + \sum_{\textrm{ion}=1}^{Z} n_{\textrm{ion}}\sigma_\nu
\end{equation}
where $\sigma_T$ is the Thomson scattering cross section. 
The cross-section for photoelectric absorption $\sigma_{\nu}$ is given here as the
photoionization cross section, and is calculated for each ion according to Equation (1) of \cite{Verner+Yakovlev1995}. The necessary analytic fitting coefficients are taken from \cite{Verner1994}. $\sigma_\nu$ is often equal to zero below a threshold ionization frequency. These threshold frequencies for each ion are also taken from \cite{Verner1994,Verner+Yakovlev1995}. The number density of each ion $n_{ion}$ is taken from the calculated photoionzation balance at each radial step along the line of sight (see Section \ref{subsec:Ion. Bal.}). 

The above sources of opacity give us the optical depth $\Delta\tau$ of the material between two radial steps $\Delta r$:
\begin{equation} \label{eq: optical depth}
    \Delta \tau = \chi_\nu\Delta r
\end{equation}

The attenuated luminosity to be used in the next radial step's calculations can be found using the optical depth $\Delta\tau$ and the current SED $L_\nu$. The current radial step is denoted by the subscript $n$, and the next step along the radial grid is denoted by the subscript $n+1$. Putting together Equations (\ref{eq: opacity}), (\ref{eq: optical depth}), and the luminosity at the current $L_{\nu,n}$ step we have
\begin{equation} \label{eq: attenuated L_nu}
    L_{\nu,n+1} = L_{\nu,n} e^{-\Delta\tau} 
\end{equation}

The incident luminosity (described above) is only one component of the local radiation field at an arbitrary distance $r$ from the central AGN source. Below, we describe how we model the entire radiation field as an effective flux $F_{\rm eff}$ that arises from two distinct sources of radiation. 

We consider two lines of sight: the primary LOS from the central source, and the secondary LOS that is a proxy for photons that undergo multiple scattering. The primary LOS is calculated according to Equation (\ref{eq: attenuated L_nu}) and a standard hydrogen density model (see Section \ref{sec: initial conditions}, Equation (\ref{eq: nH})). Along the primary LOS, the AGN’s SED undergoes attenuation due to both Thomson scattering and photoelectric absorption.

In addition to the incident luminosity from the primary LOS, ionizing photons that contribute to the total flux due to scattering and reprocessing must also be accounted for in order to calculate the total effective flux $F_{\textrm{eff}}$ at the observer \citep{Higginbottom2014}. This would ideally be done with a fully self-consistent hydrodynamic and radiative transfer model; however, for simplicity we do this by implementing a luminosity ``floor'' at each radial step. At the first radial step ($n=0$), the total flux is approximated as simply the flux due to the luminosity of the source, since at this point we assume there has been no attenuation of the SED. At subsequent radial steps, we implement an approximate luminosity floor $L_{floor}$ on the attenuation of the SED. 

We use the secondary LOS to calculate $L_{floor}$, which differs from the primary LOS in two ways: 1) we assume no shielding ($S=1$), and 2) it is only affected by Thomson scattering. The lack of shielding is consistent with our interpretation of this path as circumnavigating the failed-wind region due to scattering and reprocessing. The assumed lack of photoelectric absorption allows us to set the attenuated secondary LOS SED as a simple ``gray'' fraction of the unattenuated SED. 

We have assumed that the secondary LOS is a proxy for a path that involves multiple scattering events. Thus, only a fraction of the luminosity $L_{floor}$ will reach the observer’s location in the form of a diffuse radiation field. The probability for multiple scattering is lower than that for single scattering, but computing this accurately is beyond the scope of this paper \citep[see, e.g.][]{Higginbottom2014}. Thus, we account for it using an arbitrary fraction $\gamma$, which is also randomly selected for each model from a range of reasonable values (see Table \ref{Table: param_grids}), such that the value of $L_{floor,0}$ is given by
\begin{equation}
    L_{floor,0} = \gamma L_{\nu,0},
\end{equation}
where $L_{\nu,0}$ is taken from Equation (\ref{eq:Lnu_init}).

At each additional step along the radial grid, we follow a process similar to that used for the primary LOS to calculate the value of $L_{floor}$, which is given by
\begin{equation}
    L_{floor,n+1} = L_{\nu,n} e^{-\Delta rn_e\sigma_T},
\end{equation}
where $n_e$ is the electron number density as calculated from the ionization balance in Section \ref{subsec:Ion. Bal.}, $\sigma_T$ is the Thomson scattering cross-section, and $\Delta r$ is the size of the radial step. We can subsequently define the floor flux $F_{floor}$ at the next radial step as
\begin{equation}
    F_{floor} = \frac{L_{floor,n+1}}{4\pi r_{n+1}^2}
\end{equation}

\begin{figure}
    \centering
    \includegraphics[width=1.15\linewidth]{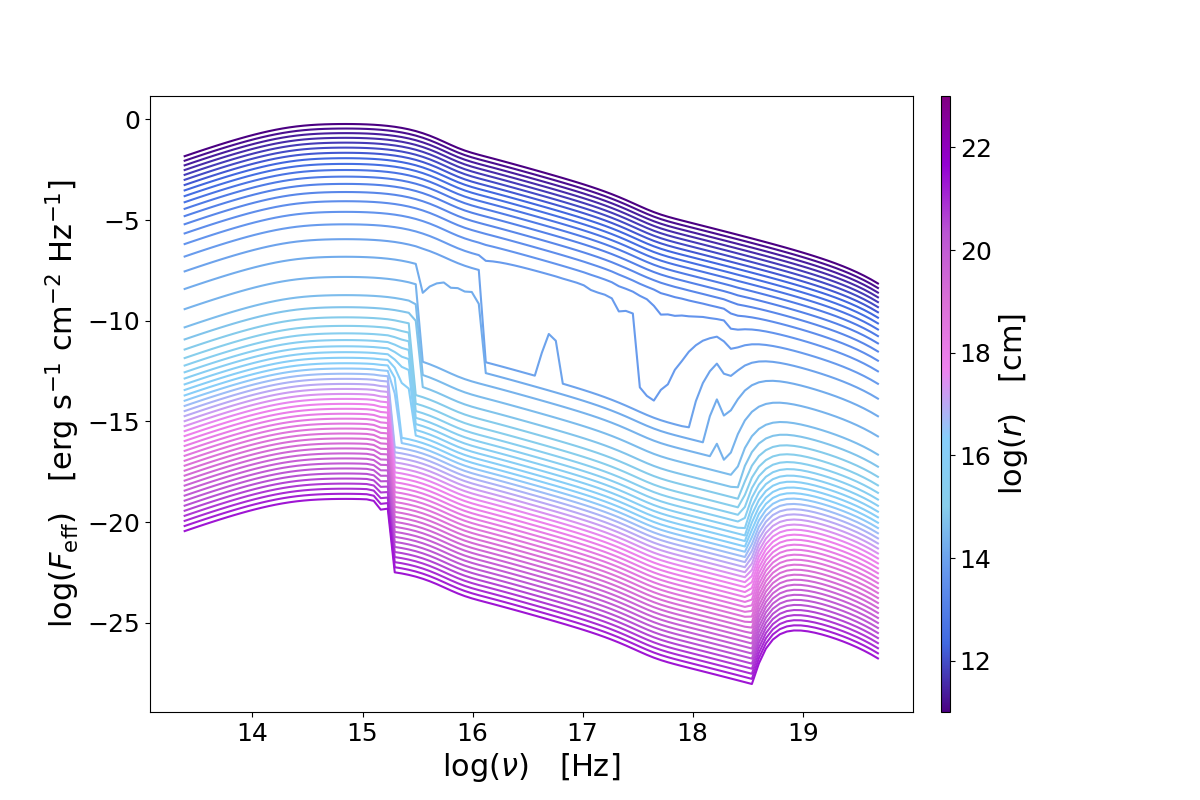}
    \caption{Final effective flux $F_{\rm eff}$, as calculated from Equation (\ref{eq: F_eff}), for a single example model. The initial input parameters for this example model are given by $\log(M_{\rm BH}) \approx 9.4 M_\odot$, $\log(f_{\rm wind}) \approx -0.5$, $S \approx 2.1$, $\beta \approx 1.3$, $r_L \approx 903 R_g$, $u_{\rm wind} \approx 320$ km s$^{-1}$, $\log(\dot{m}) \approx -0.1$,  and $\log(\gamma) \approx -5.6$.}
    \label{fig: F_eff_final}
\end{figure}

Finally, we compute the total effective flux $F_{\rm eff}$ that reaches the observer at radial distance $r_{n+1}$ from the central AGN. The flux $F_{LOS}$ from the primary radial LOS (i.e., the maximally attenuated SED) is similarly calculated from Equation (\ref{eq: attenuated L_nu}) as
\begin{equation}
    F_{LOS} = \frac{L_{\nu,n+1}}{4\pi r_{n+1}^2}.
\end{equation}
We include the impact of the secondary LOS only when it exceeds the most attenuated parts of the primary SED, such that the total flux at the observing point for a given frequency $\nu$ at the next radial step $r_{n+1}$ is given by 
\begin{equation}\label{eq: F_eff}
    F_{\textrm{eff},n+1} =
    \begin{cases}
        F_{LOS}, & F_{LOS}>F_{floor} \\
        F_{floor}, & F_{LOS}<F_{floor}
    \end{cases}    
\end{equation}
Figure \ref{fig: F_eff_final} shows the evolution with radii of the effective flux $F_{\rm eff}$ for an example model. For this model, the floor imposed on the flux by the secondary LOS takes effect at a radius of $\sim 10^{14}$ cm. 

This newly calculated flux $F_{\textrm{eff},n+1}$ is then used to determine the mean intensity at each wavelength $J_{\nu}$ for the next radial step:
\begin{equation} \label{eq: Jnu}
    J_\nu = \frac{F_{\textrm{eff}}}{4\pi}.
\end{equation}

$J_\nu$ is then used to recalculate the local gas temperature, X-ray temperature $T_X$ (Equation (\ref{eq: TX})), hydrogen-ionizing luminosity $L_X$ (Equation (\ref{eq: L_X})), and ionization parameter $\xi$ (Equation (\ref{eq: xi})) for the next radial step. We then repeat the calculations detailed in Sections \ref{subsec:Therm. Eq.} and \ref{subsec:Ion. Bal.} to find the ionization balance of the new radial step, and so on. This process is continued until the final observational radius $r_{\rm out}$ has been reached.

\section{Putting the Lines in Line-Driving}\label{sec:line strengths}
Next, we describe the computation of the radiative-driving line-force multiplier $M(t)$ for all 500 radial grid zones in each of the successful Monte Carlo model outflows. Thus, we aim to provide realistic calculations of $M(t)$ for a comprehensively broad range of AGN-relevant physical environments.

\subsection{Spectral Line Database Selection}\label{sec:Databases}

It has been shown historically that the inclusion of as many spectral lines as possible is critical in order to achieve a an accurate characterization of the line-driving acceleration \citep[see, e.g.,][]{Gormaz-Matamala2019}. For example, when CAK performed their analysis of the radiative driving using 900 C~III multiplets, they predicted mass loss rates greater than 100 times those initially found by \cite{Lucy&Solomon1970}, who used 12 doublets of C, N, Si, and S.

Since a complete tabulation of the line force requires the inclusion of as many lines as possible, the atomic line data used in this work was retrieved from a variety of sources. As in \cite{Lattimer2021}, we used multiple atomic databases to fill in any gaps in the available data wherever possible. Databases used include the National Institute of Standards and Technology (NIST) \citep{NIST_ASD}, version 10.0 of the CHIANTI database \citep{CHIANTI_1,CHIANTI_v10}, the database of lines used by the radiative transfer code CMFGEN \footnote{\url{https://kookaburra.phyast.pitt.edu/hillier/web/CMFGEN.htm}} \citep{Hillier1990,H&M1998,H&L2001}, the Opacity Project's TOPbase \citep{TOPbase_1,TOPbase_2}, and the XSTAR database \citep{XSTAR3}. 
For the AGN central sources considered here, the more highly ionized states of each element become increasingly important. Therefore we include, where available, all ionization states for all elements up to Zn ($Z=30$). The CHIANTI and CMFGEN databases additionally include both theoretical and observed spectral line data. For the sake of completeness, our line list is comprised of both theoretical (where applicable) and observed transitions for the remainder of this work. The final line list contains 5,671,602 lines, an improvement of approximately 25\% over the list used in \cite{Lattimer2021}. An overview of the database selection process is provided in Appendix \ref{database appendix}, and specific line counts and the database used for each ion are detailed in Table \ref{Table: database by ion}. The compiled line list is also available for public use online\footnote{also available at \url{https://github.com/ASLattimer/FLOWR}} \citep{Zenodo_FLOWR_2024}.

\subsection{Weighted Line Strengths}
As in \cite{Lattimer2021}, we follow \citet{Gayley1995} in characterizing the distribution of spectral line strengths as a set of dimensionless ratios $q_i$. This represents the ratio of the radiative force due to a single spectral line to the radiative force due to Thomson scattering. We also calculate the dimensionless weighting factors $\widetilde{W}_i$, which weight the line strength ($q_i$) according to the illumination of the atom from the local SED. The product $q_i \widetilde{W}_i$ thus gives the full ratio of radiative acceleration due to a specific line $i$ to the acceleration on free electrons. The dimensionless line strength parameter $q_i$ can be expressed by
\begin{equation} \label{eq:q_i}
    q_i \equiv \frac{3}{8}\frac{\lambda_0}{r_e}f_{ij}\frac{n_i}{n_e}\left(1-e^{-h\nu_0/k_{\rm B}T}\right)
\end{equation}
and the dimensionless weighting factor $\widetilde{W}_i$ by 
\begin{equation} \label{eq:W_i}
    \widetilde{W}_i = \frac{\nu_0 L_\nu(\nu_0)}{L_{\textrm{bol}}} = \frac{\nu_0 L_\nu(\nu_0)}{\int L_\nu \,d\nu}.
\end{equation}
In Equation \ref{eq:q_i}, $n_i$ is the number density of ions of a given stage that are also in the lower bound level of a given line.
In Equation (\ref{eq:W_i}), $L_\nu(\nu_0)$ denotes the luminosity $L_\nu$ in $\textrm{erg}\textrm{ s}^{-1}\textrm{Hz}^{-1}$ at the specific rest-frame frequency $\nu_0$ of the line being considered.

As in \cite{Gayley1995}, we also define the sum of the line strengths $\overline{Q}$ as
\begin{equation} \label{eq: Qbar}
    \overline{Q} = \sum_i q_i \widetilde{W}_i.
\end{equation}

For simplicity, we assume Boltzmann excitation equilibrium for atomic level populations. We then use the total number of particles in a given ionization state to express the number of particles in the lower transition level $n_i$ as
\begin{equation}\label{eq:ni/nion}
    \frac{n_i}{n_{\rm ion}} = \frac{g_i e^{-(E_i - E_0)/k_{\rm B}T} }{U_{\rm ion}(T)},
\end{equation}
where the partition function $U_{\rm ion}(T)$ is found using a version of the \cite{Cardona2010} method described in Section 3.1 of \cite{Lattimer2021}. 
The use of Boltzmann excitation equilibrium is not rigorously valid
for an AGN environment, and in future work this assumption will be revisited in order to explore how significantly atomic level populations could be affected by non-LTE processes.

We are able to determine $n_{ion}/n_e$ for each ion using the full ionization balance, which is calculated as described in Section \ref{subsec:Ion. Bal.}. To do this, we follow the procedure outlined in Section 2.2 of \cite{Lattimer2021}. We can then rewrite Equation (\ref{eq:q_i}) in terms of known quantities: 
\begin{equation} \label{eq: qi_fin}
        q_i = \frac{3}{8}\frac{\lambda_0}{r_e}g_if_{ij}
        \frac{ e^{-(E_i - E_0)/k_{\rm B}T} }{U_{\rm ion}(T)}\frac{n_{ion}}{n_e}
        \left(1-e^{-h\nu_0/k_{\rm B}T}\right).
\end{equation}

Using Equations (\ref{eq:W_i}) and (\ref{eq: qi_fin}) respectively, we can calculate the weighting function $\widetilde{W}_i$ and the dimensionless line strength $q_i$  for any line in our list from the calculated ionization balance and flux at any given radial point along the primary LOS. 

\begin{figure*}
    \centering
    \includegraphics[width=1\linewidth]{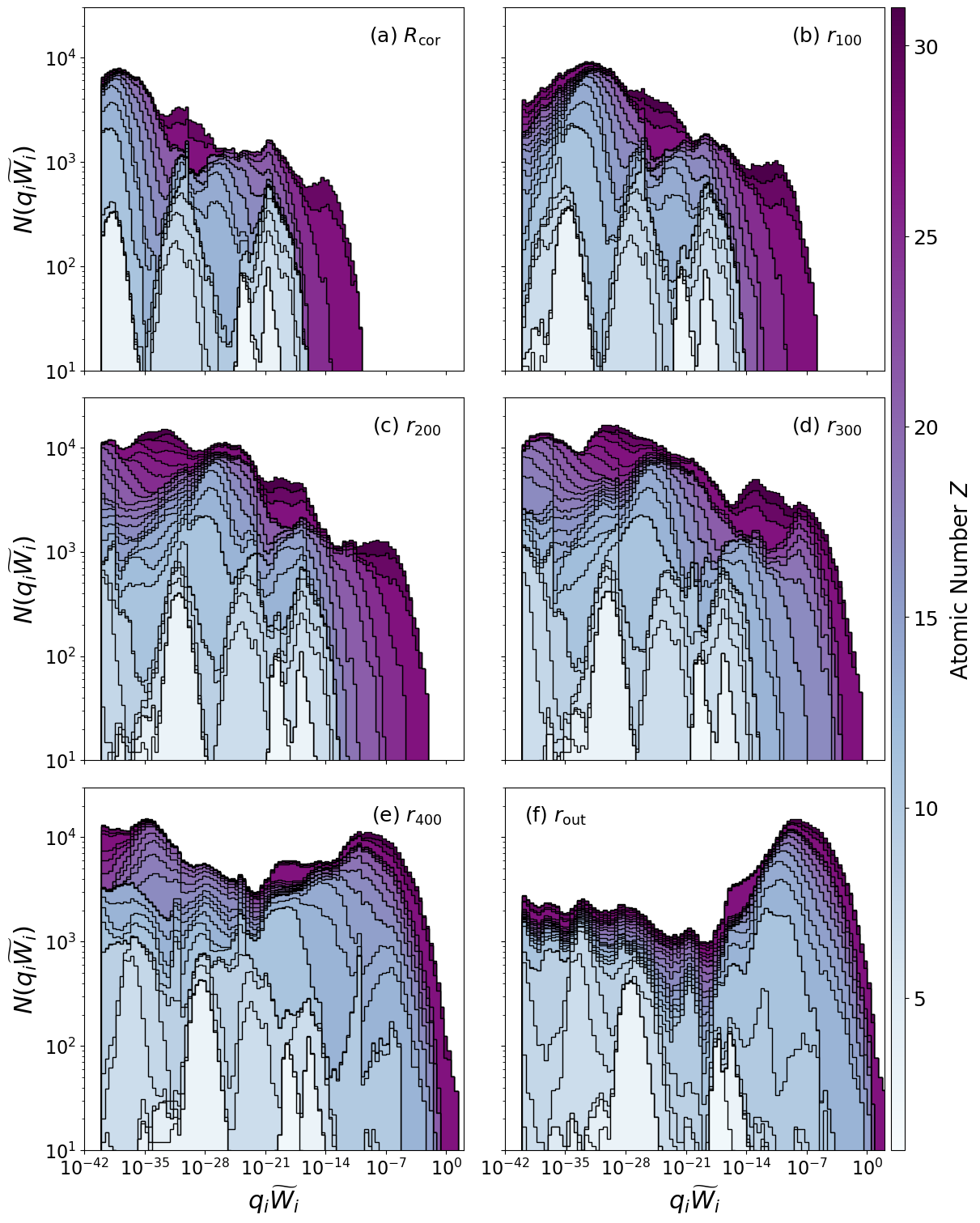}
    \caption{Histogram of the contributions by element to the total distribution of weighted line strengths $q_i\widetilde{W}_i$ for an example model run, calculated at increasing radial steps spanning from $R_{\rm cor}$ to $r_{\rm out}$, with subscripts in panels (b), (c), (d), and (e) denoting the index of the radial step. The initial input parameters for this example model are given by: $\log(M_{\rm BH})\approx10.6 M_\odot$, $\log(f_{\rm wind}) \approx -1.3$, $S \approx 1.2$, $\beta \approx 1.7$, $r_L \approx 657R_g$, $u_{\rm wind} \approx 25,134$ km s$^{-1}$, $\log(\dot{m}) \approx -0.001$, and $\log(\gamma) \approx -1.9$.}
    \label{fig: qW hist}
\end{figure*}

Figure \ref{fig: qW hist} shows the distribution of the weighted line strengths $q_i\widetilde{W}_i$ for an example model at various radial distances spanning from $R_{\rm cor}$ to $r_{\rm out}$. 
For many models, such as the one shown here, the high ionization of the wind material at the inner radii close to $R_{\rm cor}$ leads to a prevalence of weak lines, as even the strongest lines (i.e. those occurring at the highest values of $q_i\widetilde{W}_i$) are pushed to extremely low values of $q_i\widetilde{W}_i$ $(\sim 10^{-6})$. However, as we move to larger radii, we see that the strongest lines shift to occur at much higher strengths ($q_i \widetilde{W}_i \gtrsim 10^{1}$). This is due to the decrease in the overall ionization of the wind material via the absorption of the ionizing X-rays by the material at inner radii. This suggests that the shielding by the inner, highly ionized region does indeed play an important role in allowing the outer regions of the wind to become less ionized and reach higher line strengths, thus allowing radiative pressure to become a viable driving mechanism.

\subsection{The Line-Force Multiplier} \label{subsec: M(t)}

We define the line acceleration as the radiative acceleration due to electron scattering multiplied by the line force multiplier $M(t)$. Here, $t$ is a dimensionless optical depth parameter that that does not depend on the line strength, given by
\begin{equation}\label{eq: CAK_t}
    t = \kappa_e \rho v_{\rm th} \left|\frac{dv}{dr}\right|^{-1}
\end{equation}
for expanding atmospheres \citepalias{Castor1975}.
In this case, $t$ is less than the electron scattering optical depth, while for static atmospheres $t$ is equivalent to this depth. 

For calculating the force multiplier, we assume that the outflow is a) supersonic, b) contains no overlapping lines, and c) that we can apply the Sobolev approximation due to large velocity gradients in the outflow. 
This approximation holds in rapidly accelerating outflows where $|dv/dr|$ is much larger than $v_{\rm th}/H$, the thermal speed divided by the outflow's radial scale height \citep{Sobolev1957,Sobolev1960}.

For an expanding wind, the full form of $M(t)$ depends on various radiative transfer effects, such as line ``self-shadowing'', which arises due to differences in the Doppler-shifted local reference frames \citep{Gayley1995}. 
Using our compiled line list, we therefore write the full calculation of the force multiplier $M(t)$ as 
\begin{equation}\label{eq: line list M(t)}
  M(t) \, = \, \eta \, \sum_{i} q_i \widetilde{W}_i
  \left( \frac{1 - e^{-\tau_i}}{\tau_i} \right)
\end{equation}
where the geometric finite-disk factor $\eta$ is the same ratio $F$ given by Equation (21) of \cite{Gayley1995}, defined as the ratio of the true line force to that derived in the limit of purely radial photons. Because this effect can be modeled independently of the atomic physics, we use a nominal value of $\eta = 1$ for the remainder of  this work. 

Following the standard convention, we use the proton thermal velocity $v_{\rm th}^2 = 2k_{\rm B}T / m_p$ to calculate the dimensionless optical depth $\tau_i$, which depends on the CAK $t$ parameter:
\begin{equation}\label{eq: tau_1}
    \tau_i = \frac{c}{v_{\rm th}}q_i t.
\end{equation}

\citetalias{Castor1975} proposed a power-law form of $M(t)$, expressed as 
\begin{equation}\label{eq:complete CAK M(t)}
  M(t) \, = \, \eta \, k t ^{-\alpha}
\end{equation}
where $k$ and $\alpha$ are the fit constants of the power-law. Given the computationally intensive manner of calculating the full form of $M(t)$ from compiled line lists, this form has since been widely used. 

Although $t$ can be calculated at any point in a radiatively-driven outflow, for this work we evaluate $M(t)$ over a grid of 100 evenly spaced values of $t$ ranging from $t=10^{-25}$ to $t=10^{15}$. 
In the limit of $t \gg 1$, the corresponding regions in an AGN environment are not likely to be relevant to the line-driven outflows considered here because they are either deep in the optically thick corona or in the vicinity of the ISCO. We therefore include values of $t>1$ in our calculations only for the purposes of evaluating any trends in the form of the force multiplier, as $t=1$ roughly corresponds to the ``photosphere'' of the source, and thus these high values represent a realistically unobservable region of the accretion disk.

\section{Results}\label{sec:Results}
\subsection{Plasma Properties}

In order to capture as complete a picture of the outflow as possible for each model, we calculate the properties of the wind at each radial step, rather than only at $r_{\rm out}$. The evolution of the ionization parameter (Equation (\ref{eq: xi})) as the model steps radially outward in the wind is particularly important.
In general, ionization decreases with increasing radius from the central source; at small radii very close to the black hole, the wind material will be highly ionized due to the high incident luminosity at these radii. 

\begin{figure}
    \centering
    \includegraphics[width=1.15\linewidth]{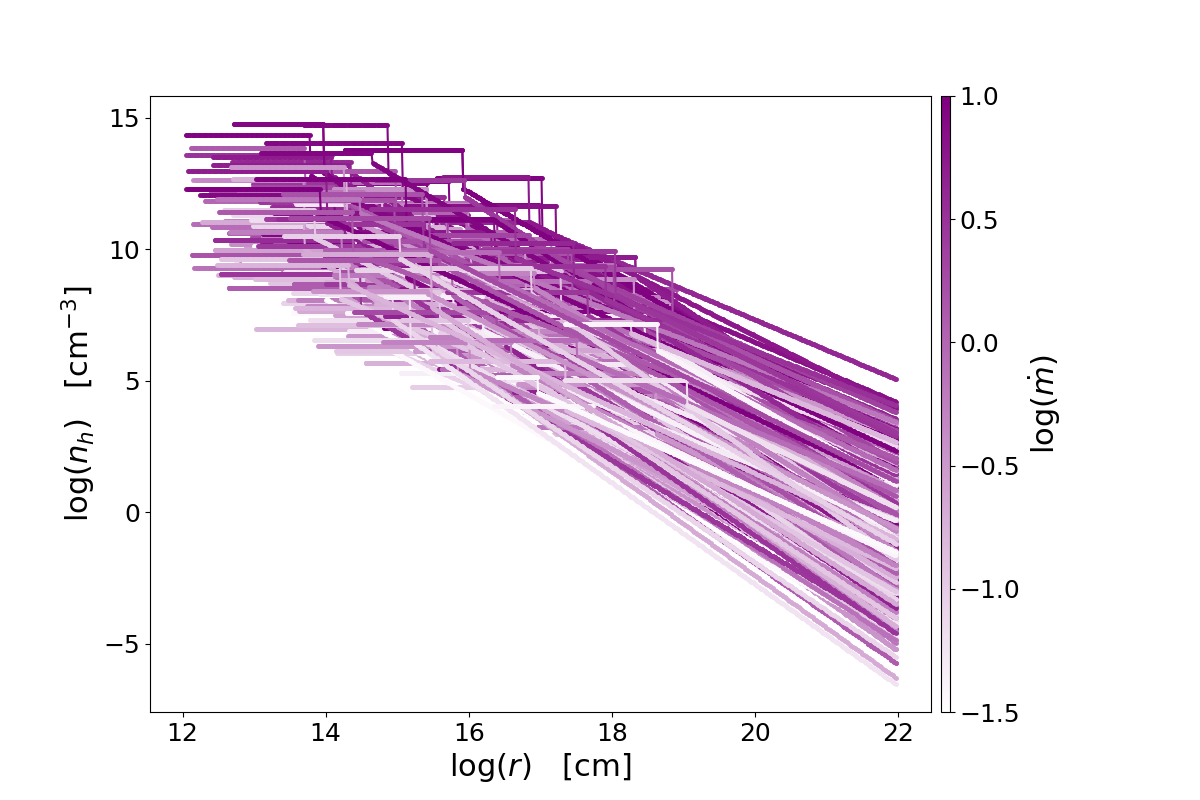}
    \caption{The hydrogen number density $n_H$ varying with radius for a representative subset of models, colored according to the input accretion rate $\dot{m}$.}
    \label{fig: n_h}
\end{figure}

In ballistic wind codes such as QWIND \citep{QWIND3}, the dense inner region close to the SMBH is often incapable of successfully launching an outflow, due to the subsequent over-ionization of the wind material. However, this resulting failed wind region is thought to play a crucial role in shielding the material at larger radii from becoming similarly over-ionized. In our models, we simulated this effect by introducing a shielding parameter $S$, as discussed in Section \ref{sec: initial conditions}, to vary the number density of this inner region prior to the launch radius. Figure \ref{fig: n_h} shows how the hydrogen number density $n_H$ varies along the radial line of sight for each model, according to the input parameters $S$ and $\beta$. The step-function drops in $n_H$ that are seen here clearly indicate the point at which the shielded failed-wind region of each model transitions to the corresponding outflowing region.

We use our grid of $t$ to calculate the force multiplier at every radial step along the primary LOS that exhibits a value of $N_H$ that falls within the range of reasonable values for the wide variety of observed AGN outflow features. To ensure that any edge behavior is included, we define this range as $17<\log(N_H)<25$, which is slightly wider than the range given in \cite{Laha2021}. Approximately 71.8\% of our modeled points fell within this region, with $\sim 28\%$ corresponding to what we consider unobservable values.

\begin{figure}
    \centering
    \includegraphics[width=1.15\linewidth]{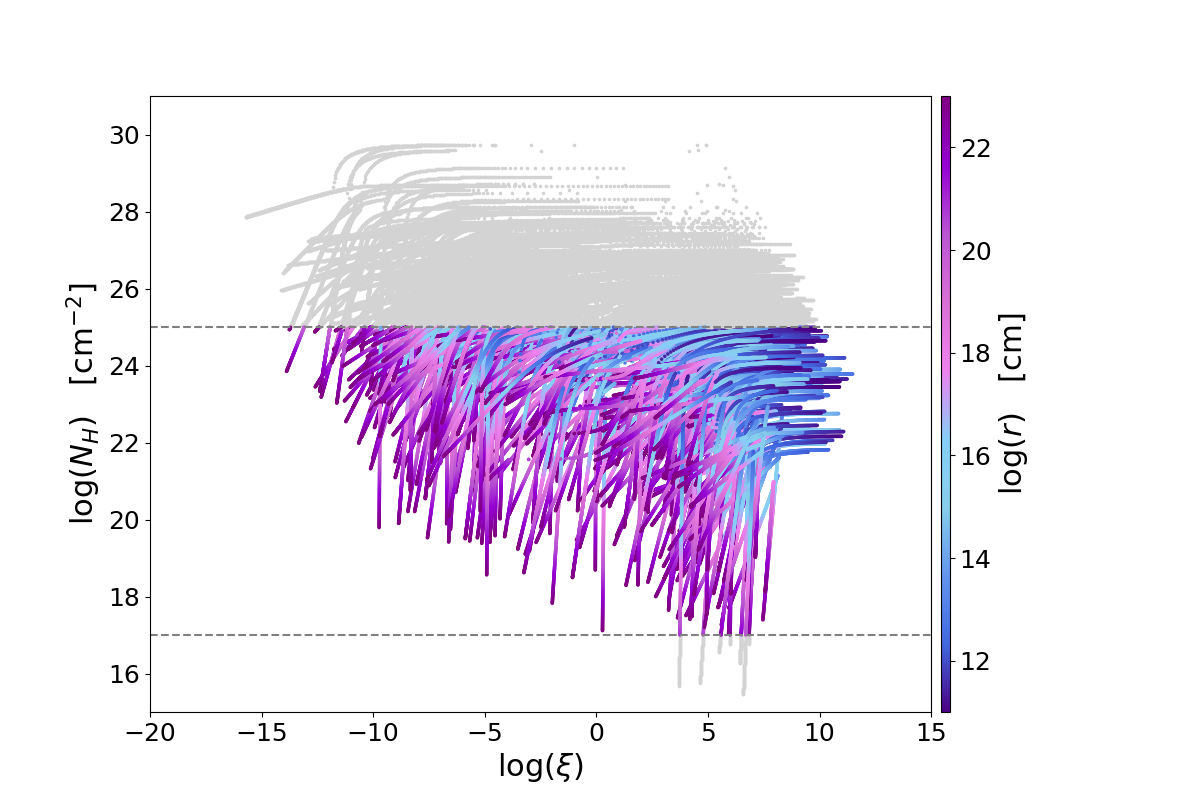}
    \caption{Values of the hydrogen column density $N_H$ along the primary LOS for each model run 
    as calculated from Equation (\ref{eq: N_H}). 
    Dashed lines show the limits of what we consider a reasonable values of $N_H$. Colored points indicate radii for which the force multiplier $M(t)$ was calculated, i.e. those that we consider to fall within the observable range.}
    \label{fig: cuts in N_H}
\end{figure}

\begin{figure*}
    \centering
    \includegraphics[width=\textwidth]{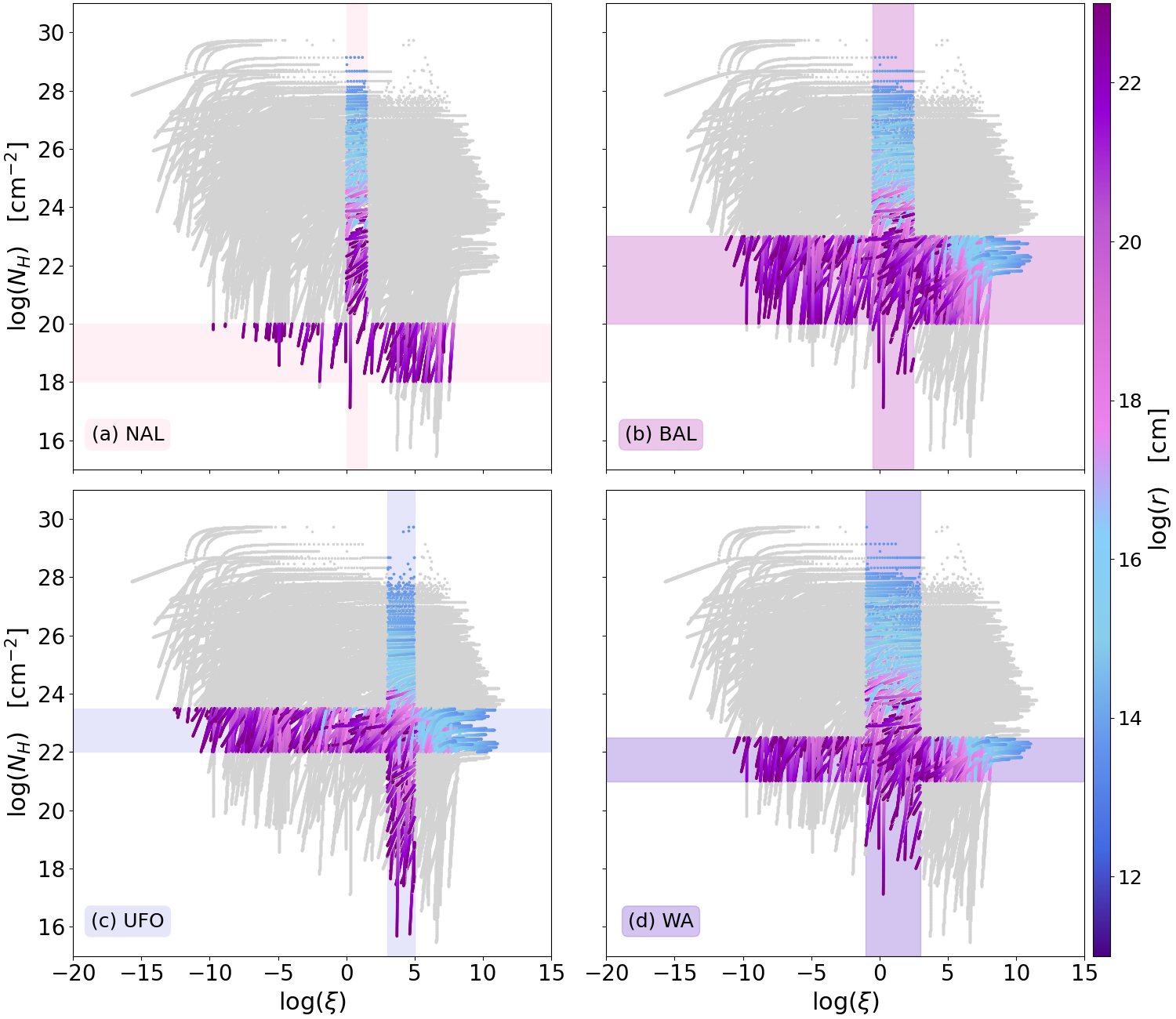}
    \caption{Colored points indicate the model-calculated values of $N_H$ that fall within ranges of $N_H$ and $\xi$ (shaded) corresponding to (a) NAL, (b) BAL, (c) UFO, and (d) Warm Absorber AGN observations. The complete set of points of $N_H$ as calculated from Equation (\ref{eq: N_H}) are plotted in grey for each panel.}
    \label{fig: AGN N_H regions}
\end{figure*}

Figure \ref{fig: cuts in N_H} shows the calculated hydrogen column density in terms of radius and $\xi$ 
for all models, with the points neglected as falling outside the observable range shaded in grey. 
Figure \ref{fig: AGN N_H regions} additionally highlights the ranges of $N_H$ and $\xi$ for various types of AGN, taken from \cite{Laha2021}, and the fraction of our modeled points that fall within those regions. The NAL, BAL, UFO, and Warm Absorber $N_H$ regions encompass approximately $4\%$, $27\%$, $23\%$, and $15\%$ respectively of our modeled points, and the observable ranges of $\xi$ account for approximately $5\%$, $11\%$, $15\%$, and $15\%$ respectively.
In total, $\sim 40\%$ of our modeled points fall within the $N_H$ regions and $\sim 30\%$ fall within $\xi$ ranges that correspond to at least one of these four AGN types. 

In Figures \ref{fig: cuts in N_H}-\ref{fig: AGN N_H regions}, the regions shaded in grey, which we designate as unphysical, are the result of our random selection of initial parameters. While each initial parameter is individually selected from ranges of reasonable values, it is likely that not all model runs resulted in physically feasible combinations; some unphysical combinations of parameters will inevitably arise. This is also discussed in Section \ref{subsec:Therm. Eq.} in regards to the thermal equilibrium calculation. We therefore use the ranges of calculated column density discussed above to reduce our total set of models (556 completed runs out of 576 initiated models) to only those that fall within observed ranges of column density after all the models have been completed.

\subsection{Line-Driving Forces} \label{subsec: Line-Driving Forces}
\begin{figure}
    \centering
    \includegraphics[width=1.15\linewidth]{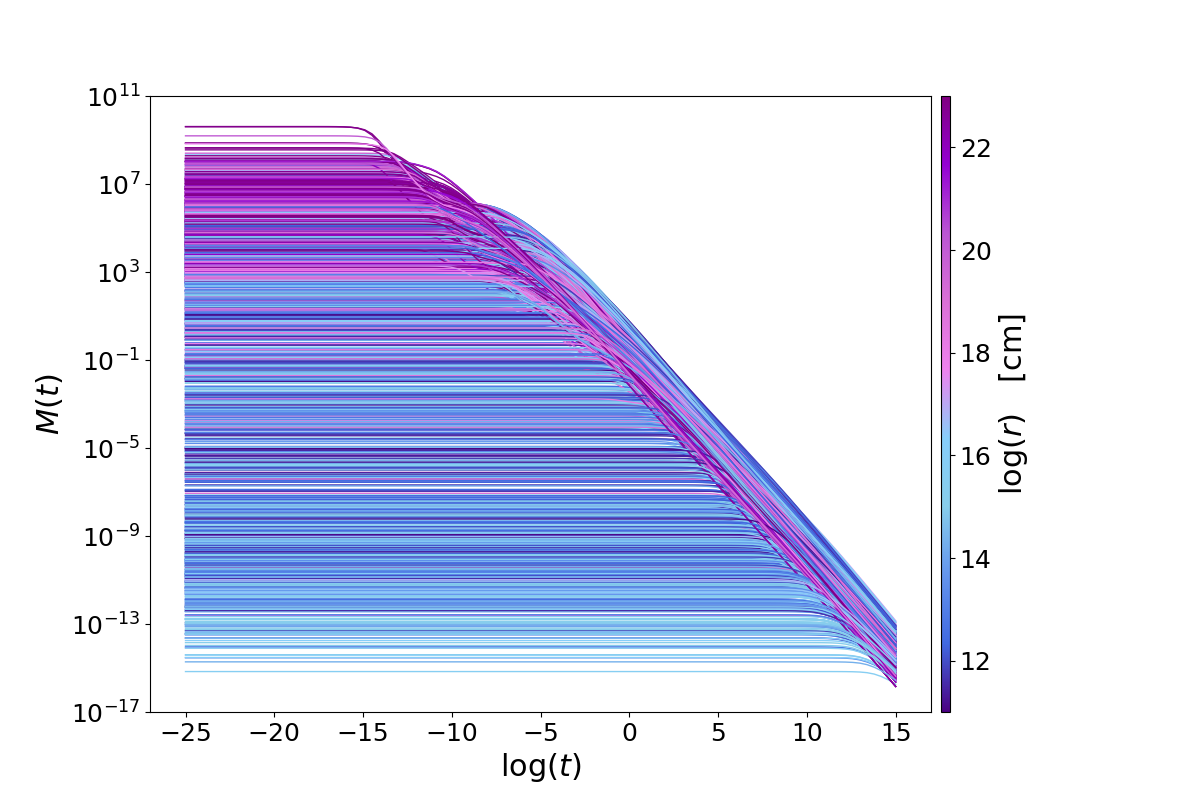}
    \caption{Calculated values of $M(t)$ for a representative subset of models and radii, colored according to radius.}
    \label{fig: Mt}
\end{figure}

Accounting for the calculation of the force multiplier at each radial step, each model produces up to 500 $M(t)$ curves. After discarding the radii that do not fall within the reasonable range of $N_H$ (78,373 points), we are left with 199,627 valid radial steps, and thus the same number of $M(t)$ curves, with each curve having $100$ optical depth points. Figure \ref{fig: Mt} shows a randomly selected subset of these curves.
As in \cite{Lattimer2021}, we find that $M(t)$ takes the form of a saturated power-law, with a generic power-law component at generally higher values of $t$ that then flattens into a constant value as $t\rightarrow0$. This is described by
\begin{equation} \label{eq: SatPow}
    M(t) = \frac{Ck}{(k^s+C^st^{\alpha s})^{1/s}}
\end{equation}
where $C,k,\alpha,$ and $s$ are fit constants that set the shape of the curve. Here, $C$ is a constant that sets the magnitude of the flat, saturated portion as $t\rightarrow 0$. In our $M(t)$ curves, we easily see that $C=\overline{Q}$. The parameter $\alpha$ sets the slope of the power-law region at high values of $t$, and the parameters $k$ and $s$ set the position and sharpness of the transition from the high- to low-$t$ regions respectively. 

We note from Figure \ref{fig: Mt} that this turnover to a constant value of $M(t)=\overline{Q}$ occurs at varying values of $t$ for different models and radii; this is sometimes as low as $t \approx 10^{-12}$. At the smallest radii, $M(t)$ is essentially constant and the turnover to the final saturation value can occur at optical depths as high as $t \approx 10^{15}$. In general, the value of $t$ that marks the transition from the power-law to constant regimes decreases as radius increases.

\begin{figure}[b!]
    \centering
    \includegraphics[width=\linewidth]{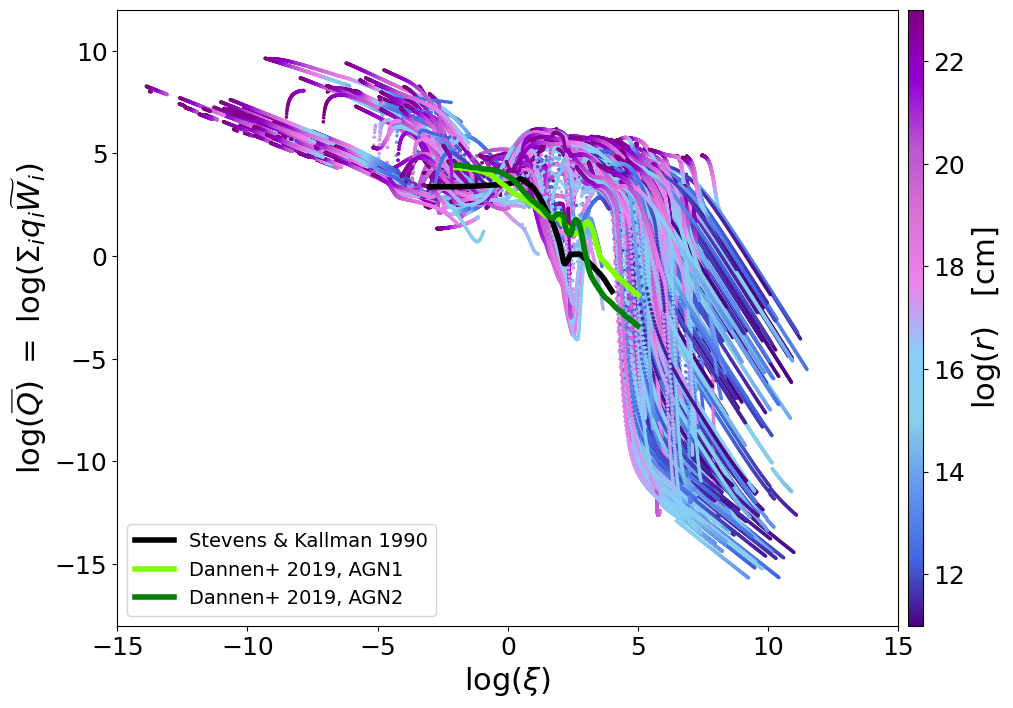}
    \caption{The evolution of $\overline{Q}$ with radius, plotted against the ionization parameter $\xi$ for each model. Solid black line indicates values taken from \cite{Stevens&Kallman1990}, and solid green lines correspond to the values found in \cite{Dannen+2019ApJ} for the two AGN considered}.
    \label{fig: qbar(xi)}
\end{figure}

\begin{figure}
    \centering
    \includegraphics[width=1.15\linewidth]{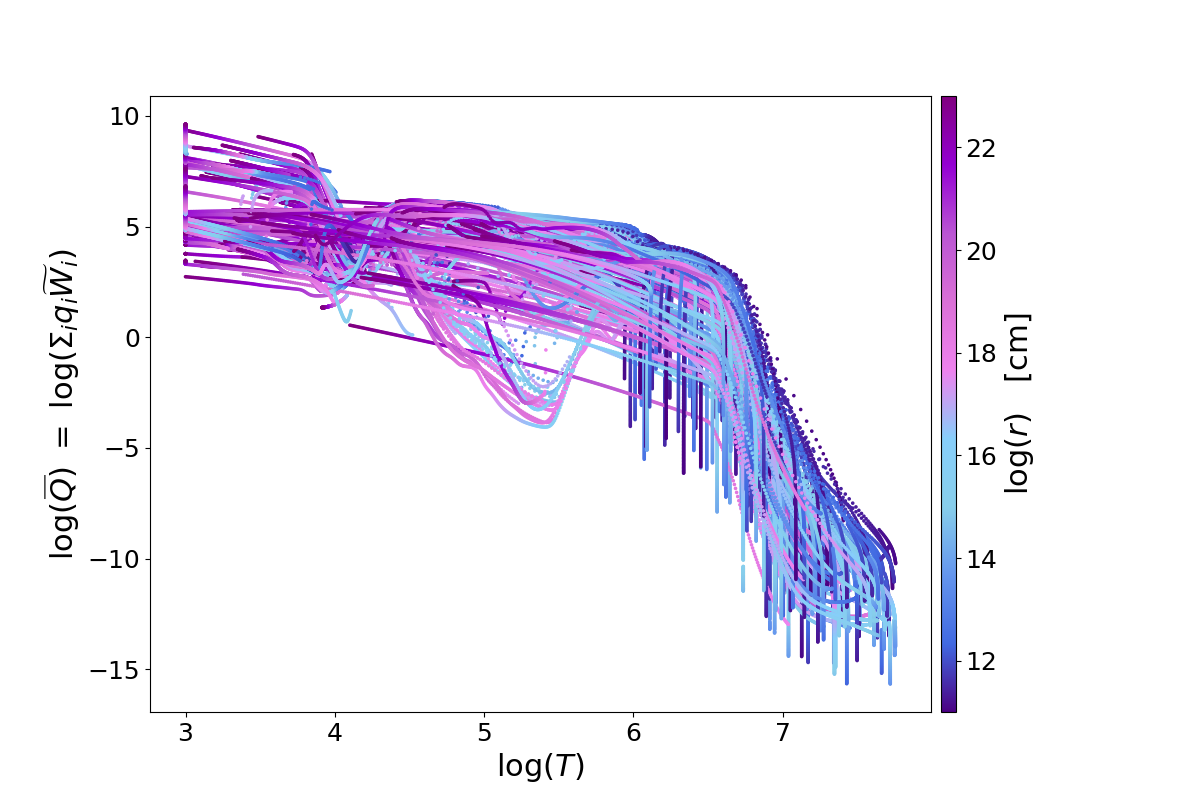}
    \caption{The evolution of $\overline{Q}$ with radius, plotted against the calculated temperature $T$ for each model.}.
    \label{fig: qbar(T)}
\end{figure}

Likewise, the saturation value $\overline{Q}$ varies across models and radii. 
Figure \ref{fig: qbar(xi)} shows $\overline{Q}$ (Equation (\ref{eq: Qbar})) with respect to the ionization parameter $\xi$, and Figure \ref{fig: qbar(T)} shows the evolution of $\overline{Q}$ as the temperature $T$ varies along the radial LOS.  
As anticipated, the values of $\overline{Q}$ follow a general upward trend with increasing radius; i.e, $\overline{Q}$ (and thus the strength of the radiative outflow) increases as overall ionization decreases. In addition to this general trend, we find that all models exhibit a steep decrease in $\overline{Q}$ as $\xi \rightarrow \infty$, starting at $\xi\approx 10^5$.

We also compare our values of $\overline{Q}$ to those taken from \cite{Stevens&Kallman1990} and \cite{Dannen+2019ApJ}. Our models generally agree with these values within the applicable range of $\xi$, with the notable exception of $10^{2.5} \lesssim \xi \lesssim 10^{5}$. The disagreement in this region is likely due to our inclusion of adiabatic cooling in the thermal equilibrium calculations, which was not considered in \cite{Stevens&Kallman1990}. \cite{Dannen+2019ApJ} computed the cooling rate as a function of $\xi$ using \texttt{XSTAR} \citep{XSTAR3,Dyda+2017}. These differences are likely responsible for the discrepancies shown in Figure \ref{fig: qbar(xi)}. Our increased cooling estimates consequently lower the calculated temperatures along the radial LOS (see Figure \ref{fig: temp}), which in turn decreases the overall ionization of the material. As $M(t)$, and thus $\overline{Q}$, is directly dependent on the ionization of the material, this results in an increased final value of $\overline{Q}$. 

We see from Figures \ref{fig: Mt}, \ref{fig: qbar(xi)}, and \ref{fig: qbar(T)} that there is great diversity in the possible $M(t)$ curves. It is therefore useful to normalize the values of $M(t)$ by each curve's corresponding value $\overline{Q}$. 
Additionally, we designate the optical depth where the $M(t)$ curve transitions from the saturated $\overline{Q}$ portion to the power-law portion as $t^*$. We define this turnover point as the value of the optical depth $t$ at which the slope of the $M(t)$ curve $\alpha$ differs significantly from 0. Here, we use a conservative condition of $\alpha>0.25$, where $\alpha$ is given by
\begin{equation}\label{eq: alpha slope}
  \alpha =-\frac{d \log M}{d \log t}.
\end{equation}

\begin{figure*}
    \centering
    \includegraphics[width=\textwidth]{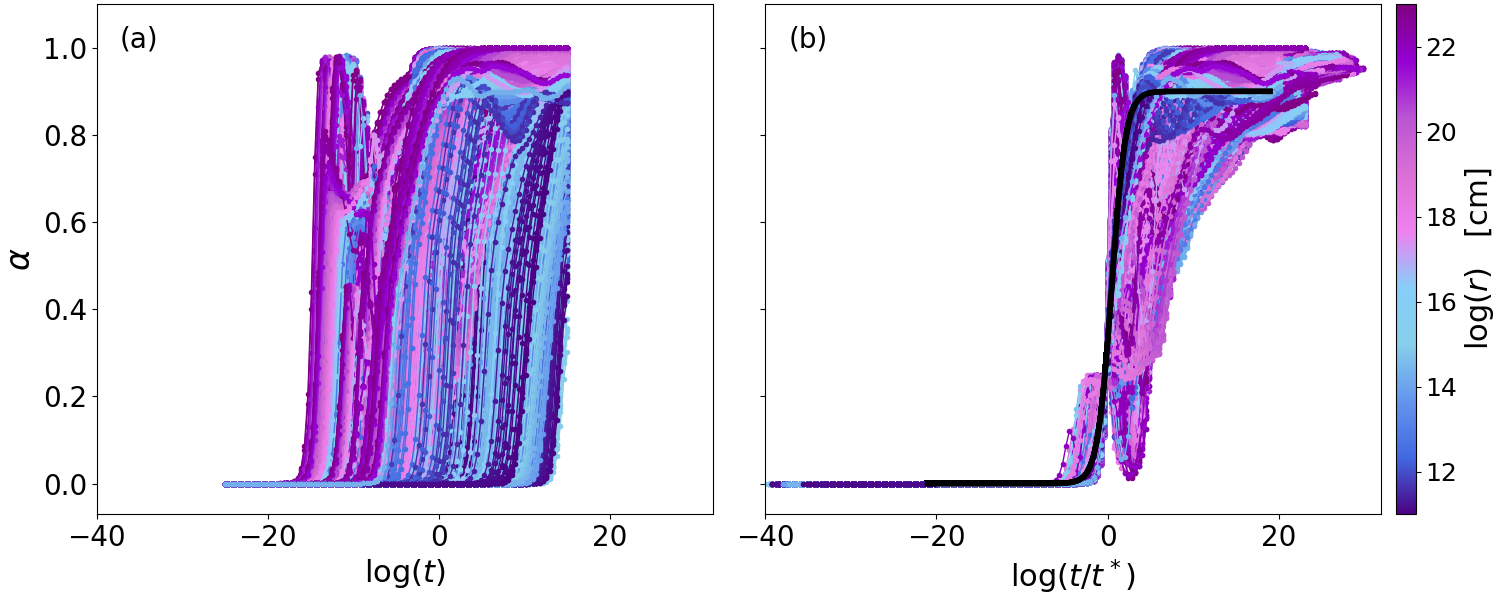}
    \caption{$M(t)$ slope $\alpha$ as defined in Equation (\ref{eq: alpha slope}), varying with (a) $\log(t)$ and (b) $\log(t/t^*)$. Solid black line in panel (b) corresponds to the calculated slope $\alpha$ for the best fit curve shown in Figure \ref{fig: Mnorm}. }
    \label{fig: alphas}
\end{figure*}

The variation of the slope $\alpha$ with both the original optical depth $t$ and the normalized optical depth $t/t^*$ is shown in Figure \ref{fig: alphas}. We see here that $\alpha$ increases rapidly between $0 \lesssim \log(t/t^*) \lesssim 10$. Many of the models achieve a final slope of $\alpha \approx 1$ at high $t/t^*$.
The asymptotic value of $\alpha\rightarrow1$ as $t\rightarrow\infty$ is understandable from Equation (\ref{eq: line list M(t)}) as the remnant behavior of the strongest lines in the limit of infinite optical depth. Note that the widely used CAK-type values of $\alpha \approx$~0.5--0.7 occur within the slow transition over many orders of magnitude of $t$ between these two limiting behaviors.

\begin{figure}[b!]
    \centering
    \includegraphics[width=1.1\linewidth]{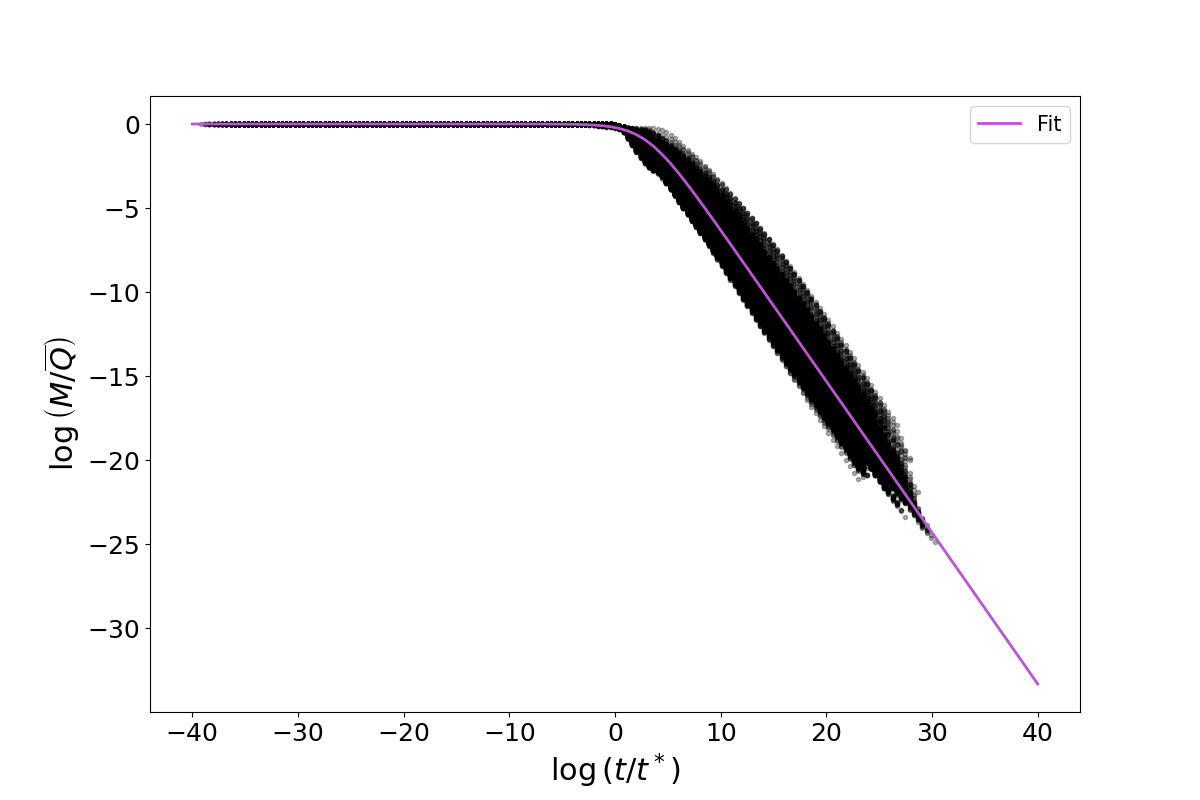}
    \caption{Normalized force multiplier $M(t/t^*)/\overline{Q}$ for the set of successful models, shown here in black. The best fit curve given by Equation (\ref{eq: Mnorm_fit}) and used in Algorithm \ref{alg1} is shown in purple (see Section \ref{subsec: normt}). The form of this fit is given by Equation (\ref{eq: SatPow}), with fit parameters $\alpha=0.9$, $s=0.3$, and $k=500$.}
    \label{fig: Mnorm}
\end{figure}

Figure \ref{fig: Mnorm} shows the same subset of force multiplier curves as in Figure \ref{fig: Mt}, normalized by the corresponding $\overline{Q}$ and $t^*$ values. Unsurprisingly, we find that all variation in the magnitude of $M(t)$ at $\log(t/t^*) \lesssim 0$ is dependent solely upon $\overline{Q}$, as all models normalize to a single curve at this point. Conversely, for $\log(t/t^*) \gtrsim 0$ the normalization of $M(t)$ and $t$ by $\overline{Q}$ and $t^*$ respectively does not remove all variation in the power-law segment of the curves.

\section{Discussion \& Conclusions}\label{sec:Discussion}
From Section \ref{sec:Results} we see that there are many factors that determine the final form of the force multiplier for any given outflow at a given radius. We have performed an explicit, radially dependent calculation of the ionization balance and the resulting strength of the line-driving force. While this type of explicit process may be ideal for characterizing the line-driving force, such calculations are not always computationally feasible. Therefore, we provide here a method for approximating the force multiplier using both our results and commonly used observational AGN properties.

\subsection{Determining $\overline{Q}$} \label{subsec: Qbar_est}
As $\overline{Q}$ sets the maximum value of the force multiplier for a given model and radius, its careful and accurate treatment is of the utmost importance. Although each initial input parameter (see Table \ref{Table: param_grids}) contributes to the final shape of the $M(t)$ curve and the value of $\overline{Q}$, we find that $\overline{Q}$ is most strongly dependent on the accretion rate $\dot{m}$ and the calculated X-ray temperature $T_X$. Recall from Figure \ref{fig: qsosed_example} and Equation (\ref{eq: TX}) that $T_X$ itself is strongly correlated with $\dot{m}$, as shown in Figure \ref{fig: TX}.  
However, $T_X$ depends not only on $\dot{m}$, but also on the degree of attenuation undergone by the SED between the coronal base at $R_{\rm cor}$ and the radial point being considered \citep{Stevens1991}. Therefore, when quantifying $\overline{Q}$ it is necessary to include the dependencies on both $T_X$ and $\dot{m}$.

\begin{figure*}
    \centering
    \includegraphics[width=\textwidth]{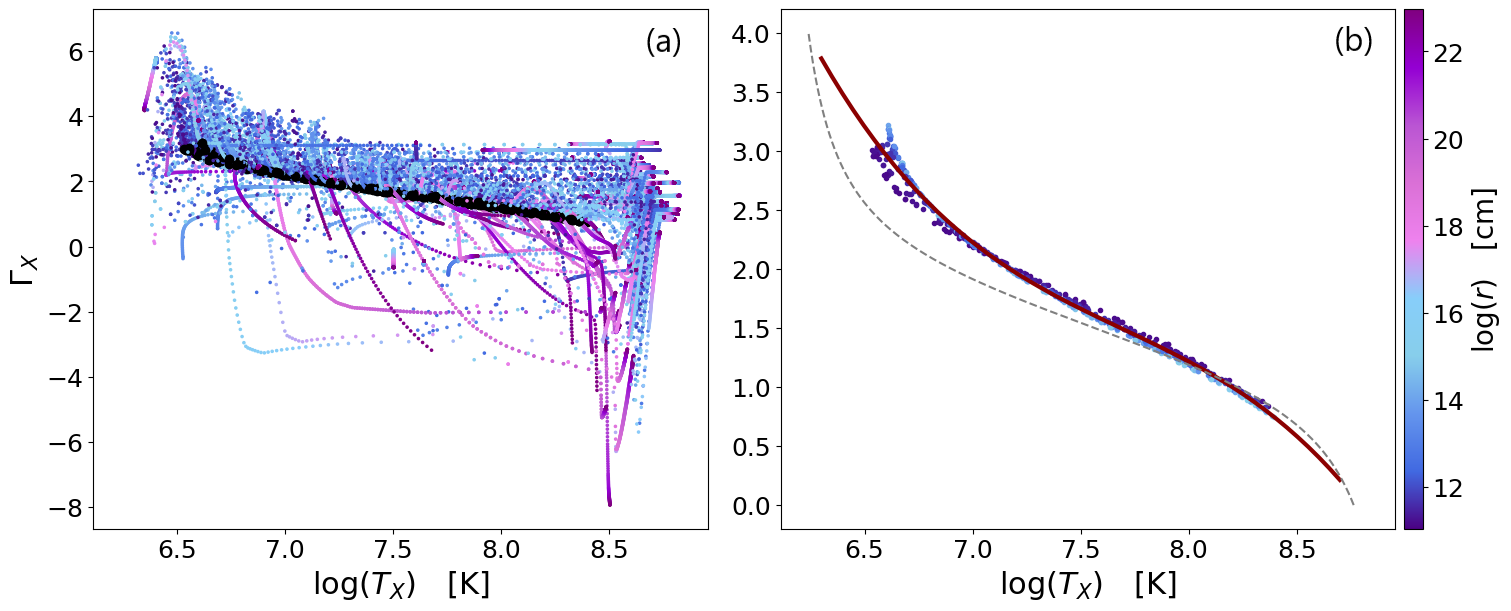}
    \caption{The evolution of the X-ray spectral index $\Gamma_x$ with X-ray temperature $T_X$ and radius $r$. Panel (a) shows $\Gamma_X$ for the entire set of modeled radial points, with the points corresponding to $R_{\rm cor}$ marked in black. Panel (b) shows $\Gamma_X$ at $R_{\rm cor}$. The analytic relationship corresponding to a pure power-law X-ray SED is marked with the grey dashed line, and the fit derived from our modeled points (Equation (\ref{eq: GammaX(TX)})) is denoted in red.}
    \label{fig: Gamma(TX)}
\end{figure*}

For the sake of accessibility and maintaining generality, it is useful to reparameterize $\overline{Q}$ in terms of its dependence on the ionization parameter $\xi$ and the X-ray spectral index $\Gamma_X$, which is more commonly used to characterize AGN spectra than X-ray temperature. This parameter describes the shape of the SED in the observed X-ray range, defined here as falling between 0.5 and 7 keV:
\begin{equation}
    L_\nu \propto \nu^{-\Gamma_X}.
\end{equation}
To find the value of $\Gamma_X$ for each of our models, we fit a simple power-law to the SED between this range at each radial step along the primary LOS. 
The relationship between $\Gamma_X$ and $T_X$ is illustrated in Figure \ref{fig: Gamma(TX)}. In log-space, the relationship between $\Gamma_X$ and the basal value of $T_X$ at $R_{\rm cor}$ is well defined by a simple cubic polynomial, given by
\begin{equation} \label{eq: GammaX(TX)}
    \Gamma_X = -0.4T_X^{3} + 9.25T_X^{2} - 72.04T_X + 190.69.
\end{equation}

We note that this formula differs slightly from the analytic solution that is found if the SED is assumed to follow a strict power-law across the entire X-ray frequency range, which is not the case for our attenuated SED models. This analytic solution is given by
\begin{equation}
  k_{\rm B} T_X \,\, = \,\, h
  \left( \frac{1-\Gamma_X}{2-\Gamma_X} \right)
  \left( \frac{\nu_2^{2-\Gamma_X} - \nu_1^{2-\Gamma_X}}
  {\nu_2^{1-\Gamma_X} - \nu_1^{1-\Gamma_X}} \right)
\end{equation}
where $\nu_1 = 0.1$~keV and $\nu_2 = 100$~keV, as in Equation (\ref{eq: TX}). The curve produced by this analytic solution is shown in panel (b) of Figure \ref{fig: Gamma(TX)}.

\begin{figure}[b]
    \centering
    \includegraphics[width=\linewidth]{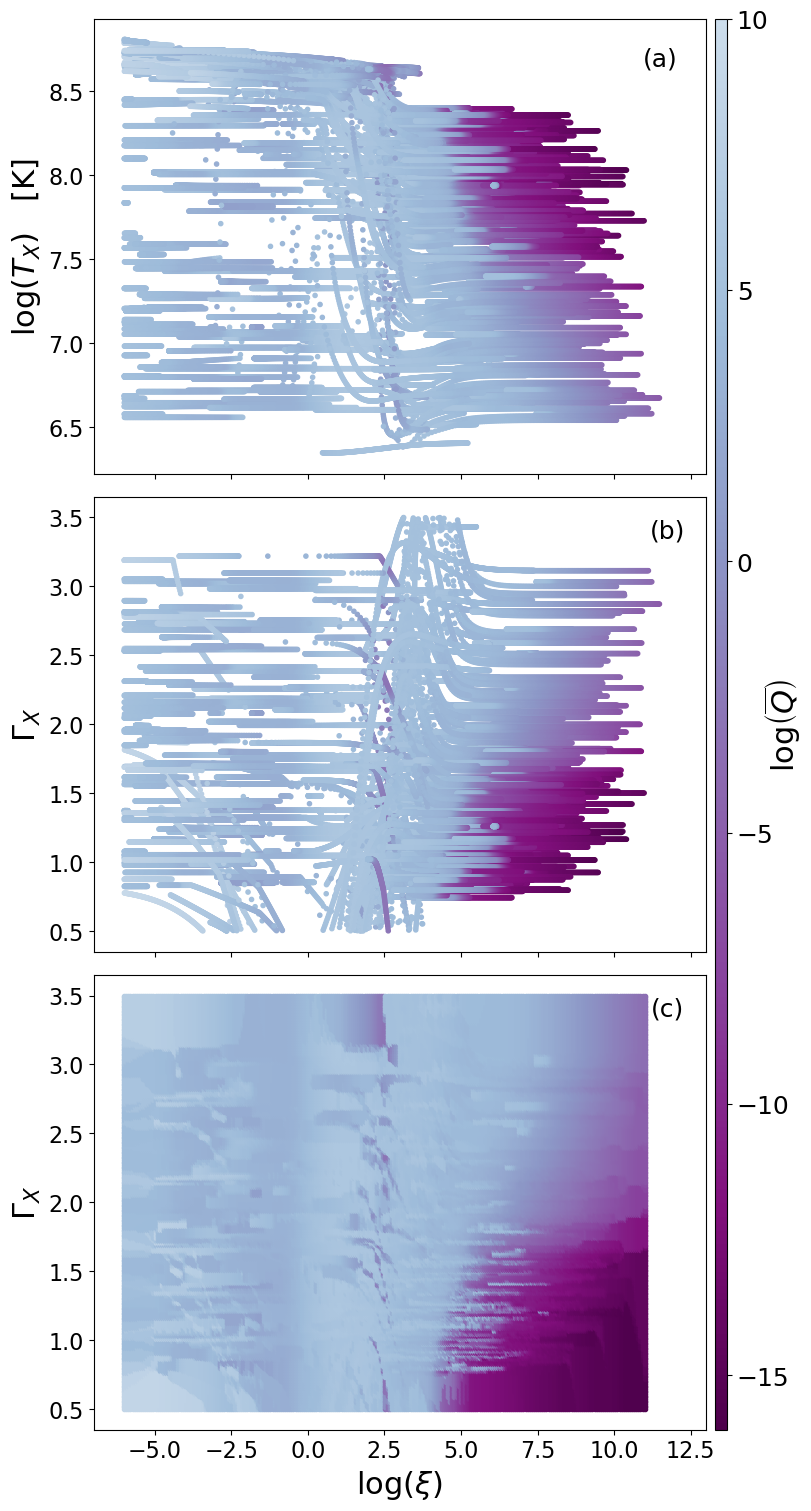}
    \caption{Dependence of $\overline{Q}$ on (a) ionization parameter $\xi$ and X-ray temperature $T_X$, and (b) $\xi$ and X-ray spectral index $\Gamma_X$. Panel (c) shows the dependence of $\overline{Q}_{KNN}$, as calculated via the KNN model described in Section \ref{subsec: Qbar_est}, for general grids of $\xi$ and $\Gamma_X$. }
    \label{fig: Qbar_Dependencies}
\end{figure}

For some models and radii, the specific combination of attenuation along the LOS produces an SED that does not follow a power-law in this range of frequency. These often result in anomalously high or low values of $\Gamma_X$, so for the remainder of this work we consider only SEDs with $0.5\leq \Gamma_X \leq 3.5$ \citep[see, e.g.,][]{Pounds+1990,Nandra+Pounds1994,Page+2005,Brightman+2013}. Approximately 94\% of the radial points in our set of successful models fall within this range.
Figure \ref{fig: Qbar_Dependencies} shows the dependence of $\overline{Q}$ on $\xi$, $T_X$, and $\Gamma_X$.
Note that the value of $\overline{Q}$ is determined more or less uniquely by its location in the two-dimensional plane defined by $\xi$ and $\Gamma_X$. However, there are combinations of parameters where there seem to exist multiple possible values of $\overline{Q}$, such as the vertical strip in the vicinity of $\log\xi\approx 2.5$. Unfortunately, the fitting formulae that we provide are not able to replicate the full complexity of the models in these regions.

We now turn our attention to characterizing $\overline{Q}$ in terms of $\xi$ and $\Gamma_X$. To do this, we use a k-nearest neighbors (KNN) algorithm \citep{Goldberger2004,vandermaaten2008}, which is trained and tested using our calculated values of these three parameters. As mentioned in Section \ref{sec:Atomic Data}, we neglect points where $\xi<10^{-6}$, and the corresponding values of $\overline{Q}$ and $\Gamma_X$, when training the KNN model. 

To construct this model, we used the KNN regression module for python available from \texttt{sci-kit learn}. In this method, the target data point is predicted by a local interpolation of the nearest points in the training set \citep{scikit-learn}. For constructing this model, each point in the local parameter-space made up by $\overline{Q},\xi,$ and $\Gamma_X$ contributed uniformly to the determination of a given query point. The initial training set was created by randomly splitting our model calculated values of $\overline{Q}$ and the corresponding $\Gamma_X$ and $\xi$ points. 
Approximately 80\% of the modeled points were allocated to the training set, with the remaining 20\% reserved to judge the efficacy of the model. After initial fitting and validation using the split sets of points, we reallocated the test points to the training set in order to provide as refined a fit as possible.

\begin{figure}[b]
    \centering
    \includegraphics[width=\linewidth]{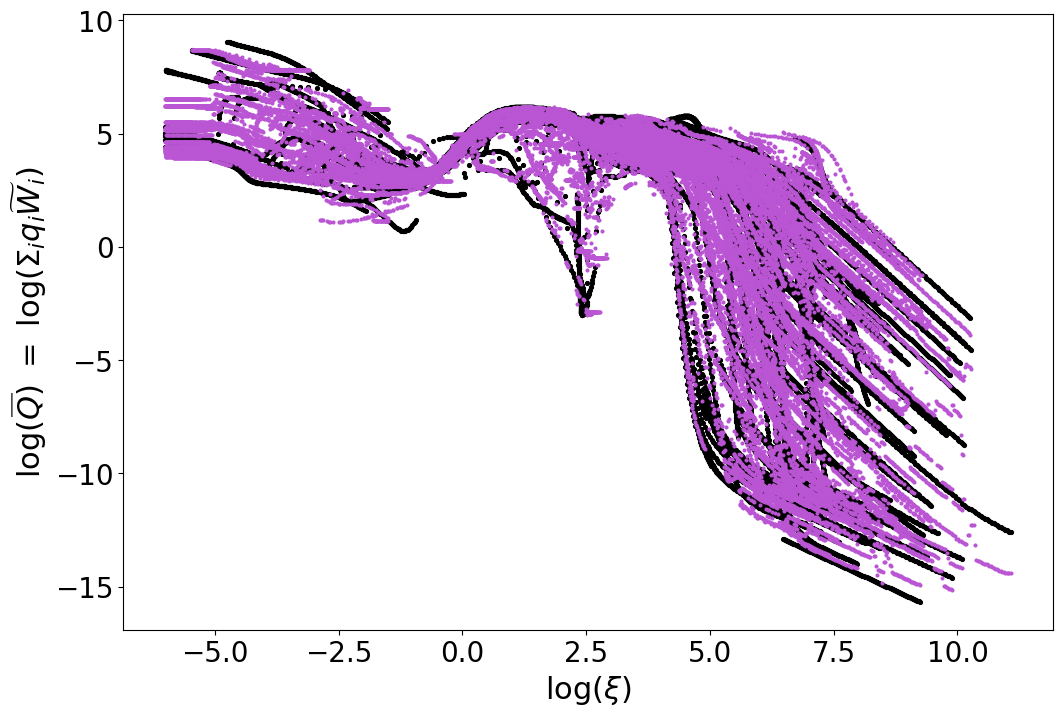}
    \caption{Comparison of the calculated and and KNN-predicted values of $\overline{Q}$ for the testing subset of points ($\sim 20\%$ of the total points). $\overline{Q}_{\rm calc}$ points are plotted in black, with $\overline{Q}_{\rm KNN}$ overlaid in purple.}
    \label{fig: KNN Qbar}
\end{figure}

\begin{figure}
    \centering
    \includegraphics[width=1.1\linewidth]{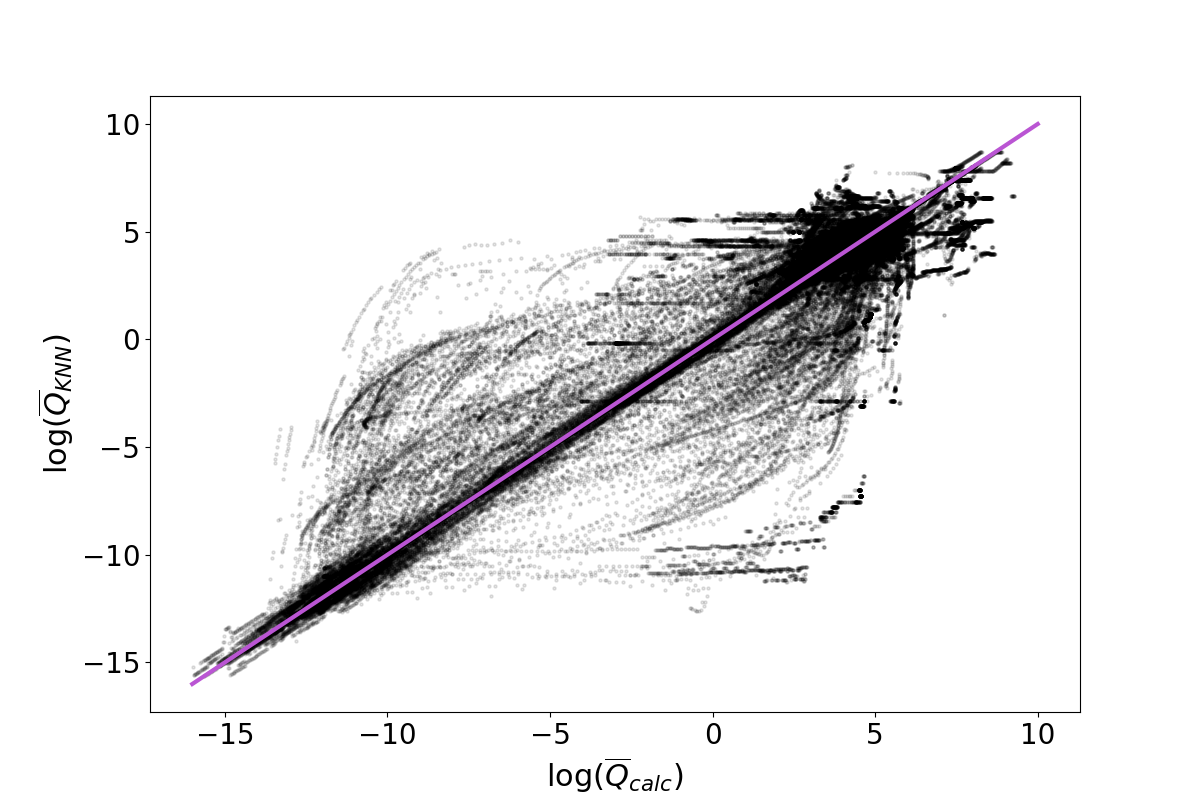}
    \caption{Correlation of $\overline{Q}_{\rm calc}$ and $\overline{Q}_{\rm KNN}$, plotted in black. Points are plotted with 50\% opacity, such that areas with high concentrations of points appear darker, and those with fewer points appear lighter. The line corresponding to $\overline{Q}_{\rm calc} = \overline{Q}_{\rm KNN}$ is marked in purple.}
    \label{fig: qbar calc vs knn }
\end{figure}

To validate the KNN model after training, we compare the KNN predictions $\overline{Q}_{KNN}$ with our calculated values $\overline{Q}$. Figure \ref{fig: Qbar_Dependencies} illustrates the dependence of $\overline{Q}$ and $\overline{Q}_{KNN}$ on $\xi$ and $\Gamma_X$. The final panel (c) of Figure \ref{fig: Qbar_Dependencies} gives $\overline{Q}_{KNN}$ as calculated from generic grids of $\xi$ and $\Gamma_X$, where $-6<\log(\xi)<11$ and $0.5<\Gamma_X<3.5$, with 250 points in each grid.
Figure \ref{fig: KNN Qbar} gives $\overline{Q}$ calculated from the full LOS ionization balance model, overlaid by the KNN-predicted values. From Figure \ref{fig: KNN Qbar}, we see the accuracy with which the KNN model is able to reproduce the values $\overline{Q}$ of the testing set when given the same input parameters.
To further quantify the validity of these results, we also plot $\overline{Q}$ against $\overline{Q}_{\rm KNN}$ (Figure \ref{fig: qbar calc vs knn }). The KNN model results in an $r$-value of $r=0.95$; thus, we conclude that it provides an acceptable prediction of the value of $\overline{Q}$ for a given input $\xi$ and $\Gamma_X$.

\subsection{Normalized Optical Depth $t^*$} \label{subsec: normt}
\begin{figure}
    \centering
    \includegraphics[width=\linewidth]{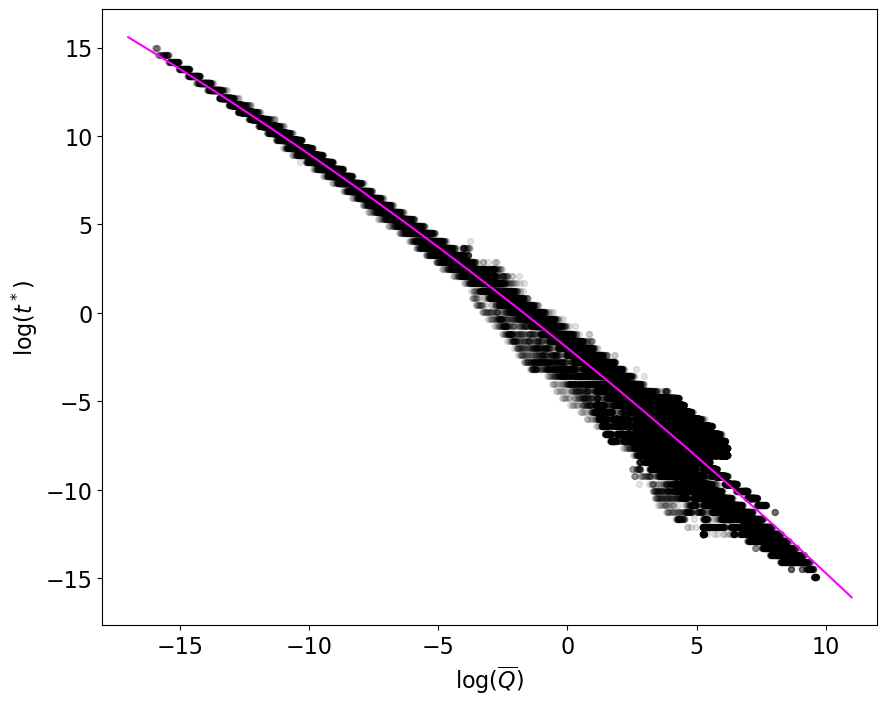}
    \caption{The normalized optical depth $t^*$, plotted against the saturation value $\overline{Q}$ for each model. The fit described by Equation (\ref{eq:t*(qbar)}) is shown in purple.}
    \label{fig: t*(Qbar)}
\end{figure}

It is also useful to calculate the normalized optical depth $t^*$ for a given $\overline{Q}$. We do this by fitting a second order polynomial to $t^*$ as a function of $\overline{Q}$, as shown in Figure \ref{fig: t*(Qbar)}. While there is some variation in $\overline{Q}$ for any given value of $t^*$, we find that the relationship between the two is well described by
\begin{equation} \label{eq:t*(qbar)}
    \log(t^*) = -0.01\log^{2}(\overline{Q}) - 1.18\log(\overline{Q}) - 1.97.
\end{equation}

By estimating the saturation value $\overline{Q}$ of $M(t)$ at low-$t$, and thus the turnover point $t^*$, we are able to deduce the normalized force multiplier curve. By fitting Equation (\ref{eq: SatPow}) to the normalized force multiplier curves of our modeled outflows, we determine the constants $\alpha$, $k$ and $s$. We find that the power-law slope of the high-$t$ region for our models is well described by $\alpha = 0.9$, and that the curves are overall reasonably characterized by setting the remaining parameters to $k=500$ and $s=0.3$ (shown in Figure \ref{fig: Mnorm}). The general shape of a given normalized $M(t)$ curve is therefore given by
\begin{equation} \label{eq: Mnorm_fit}
    M_{\textrm{norm}} = \frac{500}{(500^{0.3}+(t/t^*) ^{0.27})^{3.33}}.
\end{equation}

The normalized $M(t)$ curve in turn allows the actual force multiplier curve to be determined for a given input of $\xi$ and $\Gamma_X$. A self-contained function for this calculation, as outlined above and summarized in Algorithm \ref{alg1}, is available online for public use\footnote{also available at \url{https://github.com/ASLattimer/AGNwind}} \citep{Zenodo1_2024}. However, we stress that this is an estimation, and where possible the full radiative transfer effects should be considered.

\begin{algorithm}
\caption{Process for Estimating $M(t)$}\label{alg1}
\begin{algorithmic}
\State \textbf{Input:} $\xi$, $\Gamma_X$, $t$
\State Compute $\overline{Q}(\xi,\Gamma_X)$ from KNN model, as in Section \ref{subsec: Qbar_est}
\State Compute $t^*(\overline{Q})$ using Equation (\ref{eq:t*(qbar)}) from $\overline{Q}(\xi,\Gamma_X)$ 
\State Compute $t/t^*$ 
\State Compute $M(t)/\overline{Q}(\xi,\Gamma_X)$ from Equation (\ref{eq: Mnorm_fit}) \\
\hspace{20pt} $\rightarrow$ Compute $M(t)$ using $\overline{Q}$, $t^*$, and Equation (\ref{eq: Mnorm_fit}) 
\end{algorithmic}
\end{algorithm}

\subsection{Conclusions \& Future Work}
In this work we have provided new calculations of the radiative line-driving force in physical environments relevant to AGN outflows. We carried out a comprehensive and self-consistent treatment of the ionization balance in the outflow material and calculated the resulting force multiplier $M(t)$ for a variety of AGN and wind parameters.
Because these parameters varied over a large multi-dimensional space of reasonable values, we found a much broader variety in the magnitudes of the line-driving force multiplier than have previously been seen in the results of other work.
We found that the ionization parameter $\xi$, while often used to describe the ionization state of the material, is insufficient to completely describe the radiation force. Therefore, we provided an updated method of estimating the force multiplier using the results of the models that are presented in this paper. 

This result has important implications for calculations of outflow properties that currently depend on $\xi$, and may shed light on the nature of BAL outflows in general. It has been suggested by several BAL outflow studies that these outflows may exist at much larger radii than initially thought \citep[see, e.g.][]{deKool+2001,Moe+2009,Dunn+2010,Borguet+2013}.  For example, based on a sample of 42 outflows, \cite{Arav+2018} found that up to 50\% of quasar outflows are located at distances greater than $100$ pc. 
Given these results, it may naturally appear difficult to reconcile these kiloparsec scales with the much smaller launch radii (0.01-0.1 pc) typically exhibited by simulations of line-driven disk winds \citep{Matthews+2023,Borguet+2013}. This would pose questions such as whether BAL outflows should be considered ``disk winds'' in the traditional sense at all, or if these outflows are due to some combination of driving mechanisms \citep{Choi+2022,Matthews+2023}. However, the models presented here are applicable out to $R_{\rm out}=3000$ pc, and our results are consistent with the large spatial scales found by these previous studies. Our models suggest that line-driving remains a viable driving mechanism out to these large radii, with the inner shielding region of over-ionized material allowing these outer regions to reach high line strength and force multiplier values.

By estimating the saturation value $\overline{Q}$ of $M(t)$ at low-$t$, and thus the turnover point $t^*$, in terms of commonly used AGN parameters such as the ionization parameter $\xi$ and the X-ray spectral index $\Gamma_X$, we provide a new method of estimating the force multiplier $M(t)$ at any given point in the wind. $\Gamma_X$ is dependent on the initial conditions that at the base of the wind that set the initial SED. For example, the accretion rate $\dot{m}$ plays a prominent role in determining the initial SED, also sets $\Gamma_X$ at each radial point. The attenuation of the SED by the intervening gas also strongly affects the value of $\Gamma_X$. Because of these implicit dependencies, we choose to parameterize $\overline{Q}$ in terms of $\xi$ and $\Gamma_X$, rather than column density $N_H$ as suggested in \cite{Stevens1991}.

For future work, we plan to incorporate the force multiplier results and the model framework presented here into QWIND, a publicly available ballistic AGN-outflow code \citep{Risaliti&Elvis2010, QWIND3}. While these types of models neglect forces due to gas pressure gradients and MHD effects, they provide an efficient way to estimate the speeds, densities, and geometric streamline trajectories in the limiting case of supersonic radiation-driven flows. Currently, QWIND calculates the force multiplier as a function of both the optical depth parameter $t$ and the ionization parameter $\xi$. By replacing the current force multiplier with the $M(t)$ values calculated using the explicit ionization balance, we plan to explore to what extent the successful launching and subsequent strength of the outflow depends on the individual components of $\xi$ that are explored in this work. 
In addition, a full dynamical model would allow us to replace our parameterized expressions for adiabatic cooling, which may have overestimated the magnitude of this effect, with a more self-consistent treatment of the cooling processes.

It will also be useful to refine the assumed atomic level excitation balance used for calculating the ionization balance of the outflow. The ionization and recombination rates described in Section \ref{subsec:Ion. Bal.} should be updated with the most recent results from laboratory and theoretical studies \citep[e.g.,][]{Bryans+2006,Badnell2007,Dere+2009,Kallman+2021}. 
Also, computing non-LTE level populations presents a consistent challenge in the calculation of the force multiplier, as doing so requires complete atomic data that is often not available for the requisite number of spectral lines. As discussed, the models presented here likewise assume a Boltzmann excitation equilibrium for atomic level populations. However, some work suggests that this may result in the erroneous inclusion of many spectral lines, which in turn could increase the calculated values of $M(t)$ \citep[see, e.g.][]{Dannen+2019ApJ}. 
Therefore, replacing the current level populations with a non-LTE excitation equilibrium is an important next step that will shed light on the magnitude of this effect and consequently any influence it may have on the calculated properties of the modeled outflows.

Additionally, we note that although we implement an imposed floor on the SED, as discussed in Section \ref{subsec: Lum & Flux}, our absorption model is most accurately representative of a monolithic slab of wind material. In future work, it will be important to explicitly consider the stochastic or fractal structures of wind, which would attenuate the source SED differently to how we have calculated it here \citep{Jin+2022,Locatelli+2022}. 

Lastly, in some cases thermal \citep{Proga+Kallman2002} or magnetic \citep{Miller+2006} driving appears to be a more promising wind-driving scenario for AGN disk winds than radiative driving. For example, it has been suggested that the outflow features observed from the low-mass X-ray binary GRO J1655-40, composed of a stellar mass black hole ($M_{BH} = 5.4 \pm 0.3 M_\odot$) and a companion ($M = 1.45 \pm 0.35 M_\odot$), could be explained by magnetic driving \citep{Miller+2006,Miller+2008}. However, recent results from radiation hydrodynamic simulations have suggested that a thermal-radiative model may better match the observations \citep{Tomaru2023}. Furthermore, magnetic driving may not accurately reproduce the overall ion populations for the outflow material around GRO J1655-40 without including additional free parameters, implying that MHD driving is a less efficient method for launching and maintaining this outflow \citep{Tomaru2023}. Possible interactions between magnetic and radiative driving will be an interesting avenue to explore in future work.

Our models here provide a careful treatment of the radiative transfer processes, however, they do not currently incorporate the effects of MHD or thermal driving. In order to develop more comprehensive simulations of disk outflows, including determining which driving mechanisms are dominant, future work should include a thorough treatment of the MHD and thermal effects, as well as full time-dependence of the photoionization \citep{Proga&Kallman2004,Davis+Tchekhovskoy2020,Nomura+2021,Waters+2021,Rogantini2022,Higginbottom+2024}.
As these complex interactions are often simplified in modeling AGN disk winds, the development of a model that treats these self-consistently will be of utmost importance in determining how these effects contribute to the strength and sustainability of these outflows.

\begin{acknowledgments}
This work was supported by the Future Investigators in NASA Earth and Space Science and Technology (FINESST) program, through NASA grant 80NSSC22K1748.  It was also supported by start-up funds from the Department of Astrophysical and Planetary Sciences at the University of Colorado-Boulder.
CHIANTI is a collaborative project involving George Mason University, the University of Michigan (USA), University of Cambridge (UK) and NASA Goddard Space Flight Center (USA). This research made use of NASA's Astrophysics Data System (ADS). The authors would additionally like to thank the anonymous reviewer for their helpful insights.
\end{acknowledgments}

\software{Python v3.7.6 \citep{python},
NumPy \citep{numpy1,numpy2},
SciPy \citep{scipy},
matplotlib \citep{matplotlib},
AstroPy \citep{astropy1,astropy2,astropy3},
scikit-learn \citep{scikit-learn}}

\appendix 
\restartappendixnumbering
\section{Database Selection and Specific Line Counts}\label{database appendix}

To compile the line list used in this work, we follow the general process outlined in \cite{Lattimer2021}.
In cases where more than one database listed atomic data for a given ion, the database with the largest number of available transitions was used for each ionization state of each element. For several ions, the atomic data necessary to calculate the line strengths and weights were not available from the databases used. For example, the available sources do not provide data for a majority of the ions of both Cu and Zn. After compiling the line list from the total available data, any duplicate transitions were discarded, such that each line is only represented once in the final line list. Additionally, we exclude any listed lines produced by transitions where $i>j$ and those with anomalously large $gf$ values, which we define as $gf>10^2$. The CHIANTI and CMFGEN databases used here both include theoretical transitions which have not been observed experimentally. For the sake of completeness, in this work we do not differentiate between these two subsets of lines and include both observed and theoretical transitions in our final line list. Table \ref{Table: database by ion} shows a breakdown of the final line counts $n$ used by ion, with the source database for each also listed.

\startlongtable
\begin{deluxetable*}{llllllllllll}
\tabletypesize{\footnotesize}
\tablecolumns{12}

\tablewidth{\textwidth}
\tablecaption{Database and number of lines ($n$) for each ion. A dash (-) indicates that no data were available. CMFGEN, NIST, CHIANTI, and TOPbase are abbreviated as CM, N, CH, and T respectively.}
\label{Table: database by ion}

\tablehead{
\colhead{Ion} & \colhead{$n$} & \colhead{Database} & 
\colhead{Ion} & \colhead{$n$} & \colhead{Database} & 
\colhead{Ion} & \colhead{$n$} & \colhead{Database} &
\colhead{Ion} & \colhead{$n$} & \colhead{Database}}

\startdata
H I       &   386    &   CM   &  P XII     &   3416   &   CH   &  Ti IV     &   924    &   CM   &  Fe XXVI   &   67     &   CH   \\      
He I      &   2818   &   CM   &  P XIII    &   4041   &   CH   &  Ti V      &   4      &   N    &  Co I      &   118    &   N    \\      
He II     &   388    &   CM   &  P XIV     &   139    &   CH   &  Ti VI     &   11     &   N    &  Co II     &   61282  &   CM   \\      
Li I      &   38     &   N    &  P XV      &   65     &   CH   &  Ti VII    &   7      &   N    &  Co III    &   639484 &   CM   \\      
Li II     &   80     &   N    &  S I       &   12893  &   CM   &  Ti VIII   &   15     &   N    &  Co IV     &   68478  &   CM   \\      
Li III    &   7      &   N    &  S II      &   8414   &   CM   &  Ti IX     &   14     &   N    &  Co V      &   74477  &   CM   \\      
Be I      &   94     &   N    &  S III     &   4470   &   CM   &  Ti X      &   42     &   N    &  Co VI     &   73499  &   CM   \\      
Be II     &   66     &   N    &  S IV      &   3713   &   CM   &  Ti XI     &   9216   &   CH   &  Co VII    &   66918  &   CM   \\      
Be III    &   35     &   N    &  S V       &   3454   &   CM   &  Ti XII    &   2685   &   CH   &  Co VIII   &   83911  &   CM   \\      
Be IV     &   5      &   N    &  S VI      &   1532   &   CM   &  Ti XIII   &   5705   &   CH   &  Co IX     &   12001  &   CM   \\      
B I       &   83     &   N    &  S VII     &   72     &   N    &  Ti XIV    &   3      &   CH   &  Co X      &   5      &   N    \\      
B II      &   108    &   N    &  S VIII    &   53     &   N    &  Ti XV     &   24     &   CH   &  Co XI     &   7      &   N    \\      
B III     &   45     &   N    &  S IX      &   45     &   N    &  Ti XVI    &   722    &   CH   &  Co XII    &   6      &   N    \\      
B IV      &   49     &   N    &  S X       &   51     &   N    &  Ti XVII   &   27416  &   CH   &  Co XIII   &   33     &   N    \\      
B V       &   15     &   N    &  S XI      &   28401  &   CH   &  Ti XVIII  &   49     &   CH   &  Co XIV    &   13     &   N    \\      
C I       &   9639   &   CM   &  S XII     &   5076   &   CH   &  Ti XIX    &   3448   &   CH   &  Co XV     &   15     &   N    \\      
C II      &   7520   &   CM   &  S XIII    &   3467   &   CH   &  Ti XX     &   22     &   CH   &  Co XVI    &   9438   &   CH   \\      
C III     &   3980   &   CM   &  S XIV     &   20651  &   CH   &  Ti XXI    &   4906   &   CH   &  Co XVII   &   2621   &   CH   \\      
C IV      &   1252   &   CM   &  S XV      &   5297   &   CH   &  Ti XXII   &   9      &   N    &  Co XVIII  &   5651   &   CH   \\      
C V       &   1781   &   CM   &  S XVI     &   66     &   CH   &  V I       &   1093   &   N    &  Co XIX    &   3      &   CH   \\      
C VI      &   1316   &   CM   &  Cl I      &   75     &   N    &  V II      &   1401   &   N    &  Co XX     &   24     &   CH   \\      
N I       &   854    &   CM   &  Cl II     &   52     &   N    &  V III     &   21     &   N    &  Co XXI    &   59     &   CH   \\       
N II      &   7175   &   CM   &  Cl III    &   50     &   CH   &  V IV      &   239    &   N    &  Co XXII   &   24835  &   CH   \\   
N III     &   5989   &   CM   &  Cl IV     &   8488   &   CM   &  V V       &   10     &   N    &  Co XXIII  &   5298   &   CH   \\   
N IV      &   5105   &   CM   &  Cl V      &   3281   &   CM   &  V VI      &   4      &   N    &  Co XXIV   &   3362   &   CH   \\   
N V       &   1251   &   CM   &  Cl VI     &   2285   &   CM   &  V VII     &   7      &   N    &  Co XXV    &   4016   &   CH   \\   
N VI      &   1847   &   CM   &  Cl VII    &   1407   &   CM   &  V VIII    &   7      &   N    &  Co XXVI   &   8      &   N    \\   
N VII     &   1316   &   CM   &  Cl VIII   &   5      &   N    &  V IX      &   7      &   N    &  Co XXVII  &   9      &   N    \\   
O I       &   2425   &   CM   &  Cl IX     &   2      &   N    &  V X       &   13     &   N    &  Ni I      &   188    &   N    \\   
O II      &   8439   &   CM   &  Cl X      &   24     &   CH   &  V XI      &   57     &   N    &  Ni II     &   50092  &   CM   \\   
O III     &   6231   &   CM   &  Cl XI     &   58     &   CH   &  V XII     &   39     &   N    &  Ni III    &   65996  &   CM   \\   
O IV      &   7148   &   CM   &  Cl XII    &   28584  &   CH   &  V XIII    &   169    &   N    &  Ni IV     &   72012  &   CM   \\   
O V       &   3157   &   CH   &  Cl XIII   &   5167   &   CH   &  V XIV     &   16     &   N    &  Ni V      &   74011  &   CM   \\   
O VI      &   1521   &   CM   &  Cl XIV    &   3522   &   CH   &  V XV      &   19     &   N    &  Ni VI     &   76976  &   CM   \\   
O VII     &   2070   &   CM   &  Cl XV     &   3520   &   CH   &  V XVI     &   35     &   N    &  Ni VII    &   72405  &   CM   \\   
O VIII    &   1530   &   CM   &  Cl XVI    &   136    &   CH   &  V XVII    &   44     &   N    &  Ni VIII   &   68566  &   CM   \\   
F I       &   119    &   N    &  Cl XVII   &   67     &   CH   &  V XVIII   &   57     &   N    &  Ni IX     &   75491  &   CM   \\   
F II      &   2199   &   CM   &  Ar I      &   3618   &   CM   &  V XIX     &   23     &   N    &  Ni X      &   -      &   -    \\   
F III     &   9200   &   CM   &  Ar II     &   72676  &   CM   &  V XX      &   46     &   N    &  Ni XI     &   24723  &   CH   \\   
F IV      &   15     &   N    &  Ar III    &   6841   &   CM   &  V XXI     &   31     &   N    &  Ni XII    &   33740  &   CH   \\   
F V       &   11     &   N    &  Ar IV     &   11086  &   CM   &  V XXII    &   8      &   N    &  Ni XIII   &   191    &   CH   \\   
F VI      &   2162   &   T    &  Ar V      &   8250   &   CM   &  V XXIII   &   9      &   N    &  Ni XIV    &   2244   &   CH   \\   
F VII     &   1883   &   T    &  Ar VI     &   5      &   N    &  Cr I      &   45016  &   CM   &  Ni XV     &   23635  &   CH   \\   
F VIII    &   1626   &   T    &  Ar VII    &   35     &   CH   &  Cr II     &   65607  &   CM   &  Ni XVI    &   246    &   CH   \\   
F IX      &   985    &   T    &  Ar VIII   &   2732   &   CH   &  Cr III    &   -      &   -    &  Ni XVII   &   9464   &   CH   \\   
Ne I      &   2615   &   CM   &  Ar IX     &   5633   &   CH   &  Cr IV     &   66096  &   CM   &  Ni XVIII  &   2601   &   CH   \\   
Ne II     &   5757   &   CM   &  Ar X      &   4282   &   CH   &  Cr V      &   42633  &   CM   &  Ni XIX    &   5680   &   CH   \\   
Ne III    &   2159   &   CM   &  Ar XI     &   795    &   CH   &  Cr VI     &   4256   &   CM   &  Ni XX     &   4400   &   CH   \\   
Ne IV     &   8700   &   CM   &  Ar XII    &   677    &   CH   &  Cr VII    &   46     &   CH   &  Ni XXI    &   214    &   CH   \\   
Ne V      &   11418  &   CM   &  Ar XIII   &   28511  &   CH   &  Cr VIII   &   131    &   CH   &  Ni XXII   &   59     &   CH   \\   
Ne VI     &   4235   &   CM   &  Ar XIV    &   5196   &   CH   &  Cr IX     &   236    &   CH   &  Ni XXIII  &   24321  &   CH   \\   
Ne VII    &   4217   &   CM   &  Ar XV     &   3510   &   CH   &  Cr X      &   16     &   N    &  Ni XXIV   &   5281   &   CH   \\   
Ne VIII   &   24287  &   CH   &  Ar XVI    &   19341  &   CH   &  Cr XI     &   20     &   N    &  Ni XXV    &   3359   &   CH   \\   
Ne IX     &   193    &   CH   &  Ar XVII   &   5236   &   CH   &  Cr XII    &   72     &   N    &  Ni XXVI   &   14748  &   CH   \\   
Ne X      &   163    &   CH   &  Ar XVIII  &   97     &   CH   &  Cr XIII   &   9118   &   CH   &  Ni XXVII  &   4504   &   CH   \\   
Na I      &   1332   &   CM   &  K I       &   1111   &   CM   &  Cr XIV    &   2137   &   CH   &  Ni XXVIII &   67     &   CH   \\   
Na II     &   4747   &   CH   &  K II      &   34868  &   CM   &  Cr XV     &   5697   &   CH   &  Cu I      &   37     &   N    \\   
Na III    &   4342   &   CH   &  K III     &   220    &   CM   &  Cr XVI    &   3      &   CH   &  Cu II     &   544    &   N    \\   
Na IV     &   3713   &   CM   &  K IV      &   17189  &   CM   &  Cr XVII   &   23     &   CH   &  Cu III    &  -       &   -    \\   
Na V      &   8589   &   CM   &  K V       &   7196   &   CM   &  Cr XVIII  &   59     &   CH   &  Cu IV     &  -       &   -    \\   
Na VI     &   8766   &   CM   &  K VI      &   14608  &   CM   &  Cr XIX    &   26437  &   CH   &  Cu V      &  -       &   -    \\   
Na VII    &   4383   &   CM   &  K VII     &   69     &   N    &  Cr XX     &   5274   &   CH   &  Cu VI     &  -       &   -    \\   
Na VIII   &   3755   &   CM   &  K VIII    &   93     &   N    &  Cr XXI    &   3419   &   CH   &  Cu VII    &  -       &   -    \\   
Na IX     &   3879   &   CM   &  K IX      &   2743   &   CH   &  Cr XXII   &   2711   &   CH   &  Cu VIII   &  -       &   -    \\   
Na X      &   272    &   CH   &  K X       &   5600   &   CH   &  Cr XXIII  &   4757   &   CH   &  Cu IX     &  -       &   -    \\   
Na XI     &   65     &   CH   &  K XI      &   3      &   CH   &  Cr XXIV   &   9      &   N    &  Cu X      &  -       &   -    \\   
Mg I      &   2261   &   CM   &  K XII     &   24     &   CH   &  Mn I      &   164    &   N    &  Cu XI     &  -       &   -    \\   
Mg II     &   2594   &   CH   &  K XIII    &   59     &   CH   &  Mn II     &   46743  &   CM   &  Cu XII    &  -       &   -    \\   
Mg III    &   4617   &   CH   &  K XIV     &   28416  &   CH   &  Mn III    &   69673  &   CM   &  Cu XIII   &  -       &   -    \\   
Mg IV     &   5571   &   CM   &  K XV      &   5280   &   CH   &  Mn IV     &   71585  &   CM   &  Cu XIV    &  -       &   -    \\   
Mg V      &   5585   &   CM   &  K XVI     &   3486   &   CH   &  Mn V      &   76071  &   CM   &  Cu XV     &  -       &   -    \\   
Mg VI     &   11265  &   CM   &  K XVII    &   3940   &   CH   &  Mn VI     &   68569  &   CM   &  Cu XVI    &  -       &   -    \\   
Mg VII    &   10365  &   CM   &  K XVIII   &   143    &   CH   &  Mn VII    &   8188   &   CM   &  Cu XVII   &  -       &   -    \\   
Mg VIII   &   5212   &   CM   &  K XIX     &   67     &   CH   &  Mn VIII   &   47     &   CH   &  Cu XVIII  &  -       &   -    \\   
Mg IX     &   4426   &   CM   &  Ca III    &   520    &   N    &  Mn IX     &   137    &   CH   &  Cu XIX    &  -       &   -    \\   
Mg X      &   23012  &   CH   &  Ca IV     &   2      &   N    &  Mn X      &   236    &   CH   &  Cu XX     &  -       &   -    \\   
Mg XI     &   5512   &   CH   &  Ca V      &   9      &   CH   &  Mn XI     &   29     &   N    &  Cu XXI    &  -       &   -    \\   
Mg XII    &   66     &   CH   &  Ca VI     &   10     &   CH   &  Mn XII    &   19     &   N    &  Cu XXII   &  -       &   -    \\   
Al I      &   4312   &   CM   &  Ca VII    &   86     &   CH   &  Mn XIII   &   68     &   N    &  Cu XXIII  &  -       &   -    \\   
Al II     &   2528   &   CM   &  Ca VIII   &   200    &   CH   &  Mn XIV    &   9191   &   CH   &  Cu XXIV   &  -       &   -    \\   
Al III    &   2617   &   CH   &  Ca IX     &   9128   &   CH   &  Mn XV     &   2725   &   CH   &  Cu XXV    &  -       &   -    \\   
Al IV     &   5179   &   CH   &  Ca X      &   2750   &   CH   &  Mn XVI    &   5695   &   CH   &  Cu XXVI   &  -       &   -    \\   
Al V      &   6292   &   CM   &  Ca XI     &   5686   &   CH   &  Mn XVII   &   3      &   CH   &  Cu XXVII  &  -       &   -    \\   
Al VI     &   6964   &   CM   &  Ca XII    &   4464   &   CH   &  Mn XVIII  &   24     &   CH   &  Cu XXVIII &  -       &   -    \\   
Al VII    &   12874  &   CM   &  Ca XIII   &   811    &   CH   &  Mn XIX    &   59     &   CH   &  Cu XXIX   &  9       &   N    \\   
Al VIII   &   11469  &   CM   &  Ca XIV    &   1340   &   CH   &  Mn XX     &   25845  &   CH   &  Zn I      &  16      &   N    \\   
Al IX     &   5494   &   CM   &  Ca XV     &   28156  &   CH   &  Mn XXI    &   5334   &   CH   &  Zn II     &  22      &   N    \\   
Al X      &   4501   &   CM   &  Ca XVI    &   5258   &   CH   &  Mn XXII   &   3405   &   CH   &  Zn III    &  -       &   -    \\   
Al XI     &   22606  &   CH   &  Ca XVII   &   3487   &   CH   &  Mn XXIII  &   4061   &   CH   &  Zn IV     &  -       &   -    \\   
Al XII    &   5543   &   CH   &  Ca XVIII  &   18577  &   CH   &  Mn XXIV   &   8      &   N    &  Zn V      &  -       &   -    \\   
Al XIII   &   67     &   CH   &  Ca XIX    &   5045   &   CH   &  Mn XXV    &   9      &   N    &  Zn VI     &  -       &   -    \\   
Si I      &   16479  &   CM   &  Ca XX     &   68     &   CH   &  Fe I      &   126979 &   CM   &  Zn VII    &  -       &   -    \\   
Si II     &   3624   &   CM   &  Sc I      &   258    &   N    &  Fe II     &   489674 &   CM   &  Zn VIII   &  -       &   -    \\   
Si III    &   664    &   CM   &  Sc II     &   70559  &   CM   &  Fe III    &   133271 &   CM   &  Zn IX     &  -       &   -    \\   
Si IV     &   2626   &   CH   &  Sc III    &   507    &   CM   &  Fe IV     &   71318  &   CM   &  Zn X      &  -       &   -    \\  
Si V      &   5237   &   CH   &  Sc IV     &   4      &   N    &  Fe V      &   70874  &   CM   &  Zn XI     &  -       &   -    \\  
Si VI     &   6227   &   CM   &  Sc V      &   4      &   N    &  Fe VI     &   178638 &   CM   &  Zn XII    &  -       &   -    \\  
Si VII    &   8630   &   CM   &  Sc VI     &   -      &   -    &  Fe VII    &   83506  &   CM   &  Zn XIII   &  -       &   -    \\  
Si VIII   &   686    &   CH   &  Sc VII    &   15     &   N    &  Fe VIII   &   20508  &   CH   &  Zn XIV    &  -       &   -    \\  
Si IX     &   402    &   CH   &  Sc VIII   &   16     &   N    &  Fe IX     &   42467  &   CH   &  Zn XV     &  -       &   -    \\  
Si X      &   4934   &   CH   &  Sc IX     &   15     &   N    &  Fe X      &   41098  &   CM   &  Zn XVI    &  -       &   -    \\  
Si XI     &   3346   &   CH   &  Sc X      &   -      &   -    &  Fe XI     &   69128  &   CH   &  Zn XVII   &  -       &   -    \\  
Si XII    &   22468  &   CH   &  Sc XI     &   53     &   N    &  Fe XII    &   64911  &   CH   &  Zn XVIII  &  -       &   -    \\  
Si XIII   &   5408   &   CH   &  Sc XII    &   5      &   N    &  Fe XIII   &   46117  &   CH   &  Zn XIX    &  9458    &   CH   \\  
Si XIV    &   66     &   CH   &  Sc XIII   &   17     &   N    &  Fe XIV    &   38951  &   CH   &  Zn XX     &  2456    &   CH   \\  
P I       &   45     &   N    &  Sc XIV    &   52     &   N    &  Fe XV     &   9368   &   CH   &  Zn XXI    &  5758    &   CH   \\  
P II      &   195275 &   CM   &  Sc XV     &   42     &   N    &  Fe XVI    &   2571   &   CH   &  Zn XXII   &  -       &   -    \\  
P III     &   5128   &   CM   &  Sc XVI    &   63     &   N    &  Fe XVII   &   4504   &   CH   &  Zn XXIII  &  33      &   CH   \\  
P IV      &   1980   &   CM   &  Sc XVII   &   41     &   N    &  Fe XVIII  &   10530  &   CH   &  Zn XXIV   &  680     &   CH   \\  
P V       &   2644   &   CH   &  Sc XVIII  &   24     &   N    &  Fe XIX    &   24702  &   CH   &  Zn XXV    &  23082   &   CH   \\  
P VI      &   5468   &   CH   &  Sc XIX    &   57     &   N    &  Fe XX     &   18375  &   CH   &  Zn XXVI   &  -       &   -    \\  
P VII     &   3      &   CH   &  Sc XX     &   8      &   N    &  Fe XXI    &   25892  &   CH   &  Zn XXVII  &  18      &   CH   \\  
P VIII    &   25     &   CH   &  Sc XXI    &   9      &   N    &  Fe XXII   &   13193  &   CH   &  Zn XXVIII &  14197   &   CH   \\  
P IX      &   59     &   CH   &  Ti I      &   490    &   N    &  Fe XXIII  &   3565   &   CH   &  Zn XXIX   &  4390    &   CH   \\  
P X       &   78     &   CH   &  Ti II     &   87826  &   CM   &  Fe XXIV   &   15672  &   CH   &  Zn XXX    &  9       &   N    \\  
P XI      &   5038   &   CH   &  Ti III    &   20656  &   CM   &  Fe XXV    &   4613   &   CH   &  -         &  -       &   -    \\
\enddata

\end{deluxetable*}

\section{Quantitative Comparison to Other Photoionization Codes}\label{quantitative comparisons appendix}
\setcounter{table}{0}

Here we provide a comparison of some key plasma properties (e.g.,temperature and ionization state) to other
commonly used radiation transfer photoionization codes, namely CLOUDY\footnote{\url{http://www.nublado.org/}} \citep{Hazy1983,CLOUDY1998,CLOUDY2017,CLOUDY2023}, SPEX\footnote{\url{http://www.sron.nl/spex}} \citep{SPEX1}, and XSTAR\footnote{\url{https://heasarc.gsfc.nasa.gov/xstar/xstar.html}} \citep{XSTAR1,XSTAR2,XSTAR3}. The data for these comparisons were taken from \cite{Mehdipour+2016}, which also provides a detailed side-by-side comparison of the performance of these three codes.

\begin{figure}
    \centering
    \includegraphics[width=\linewidth]{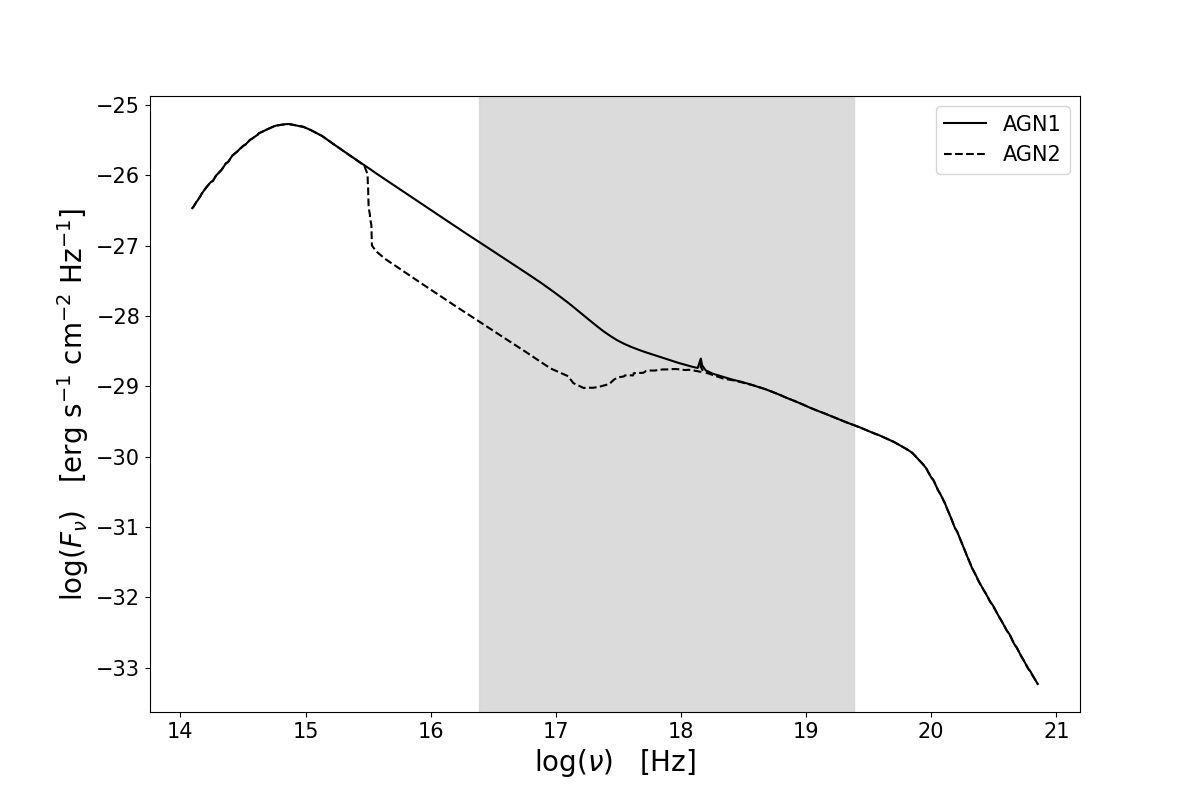}
    \caption{The source SEDs of NGC 5548, adapted from \cite{Mehdipour+2015,Mehdipour+2016}. The solid line (AGN1) depicts the unobscured AGN, and the dashed line (AGN2) represents the obscured SED of the same source. The range of frequency used in Section \ref{subsec:Therm. Eq.} to calculate $T_X$, corresponding to 0.1 keV--100 keV, is shaded in gray.}
    \label{fig: AGN1+AGN2 SED}
\end{figure}

Figure \ref{fig: AGN1+AGN2 SED} plots the two source SEDs used in this section. As in \cite{Mehdipour+2016}, we use NGC 5548 to represent a typical Seyfert galaxy, and we adapt the SEDs of this source that are detailed in \cite{Mehdipour+2015}. 
To maintain comparability, for the calculations described below we use the unobscured SED (labeled AGN1 in Figure \ref{fig: AGN1+AGN2 SED}), unless noted otherwise.

\subsection{Temperatures} \label{AppB: Temps}
Here, we have calculated the temperature $T$ at each radial step for the two SEDs provided in \cite{Dannen+2019ApJ}, labeled as AGN1 and AGN2 in Figure \ref{fig: AGN1+AGN2 SED}. For these temperature comparisons, we neglect adiabatic cooling, as in panel (b) of Figure \ref{fig: temp}.
\begin{deluxetable}{cccc}[H] \label{Table: freq_combos}
    \tablecolumns{4}
    \tablewidth{\linewidth}
    \tablecaption{Energy limits and the resulting values of $T_X$ used in calculating the temperature curves shown in Figure \ref{fig: temp_comparison}.}
    \tablehead{
    \colhead{$E_{\rm X, min}$} & \colhead{$E_{\rm X, max}$} & \colhead{$\log(T_X)$} & \colhead{$\log(T_X)$} \\
    \colhead{(keV)}            & \colhead{(keV)}            & \colhead{AGN1 (K)}    & \colhead{AGN2 (K)}}
    \startdata
    0.1  & 100  & 8.38   &  8.49  \\
    1.0  & 100  & 8.5    &  8.51  \\
    0.01 & 100  & 8.23   &  7.96  \\
    \hline
    0.1  & 1000 & 9.12   &  9.25  \\
    1.0  & 1000 & 9.19   &  9.26  \\
    0.01 & 1000 & 9.03   &  9.22  \\
    \hline
    0.1  & 10   & 7.35   &  7.63  \\
    1.0  & 10   & 7.65   &  7.69  \\
    0.01 & 10   & 7.07   &  7.35  \\     
    \enddata
\end{deluxetable}

\begin{figure}
    \centering
    \includegraphics[width=\linewidth]{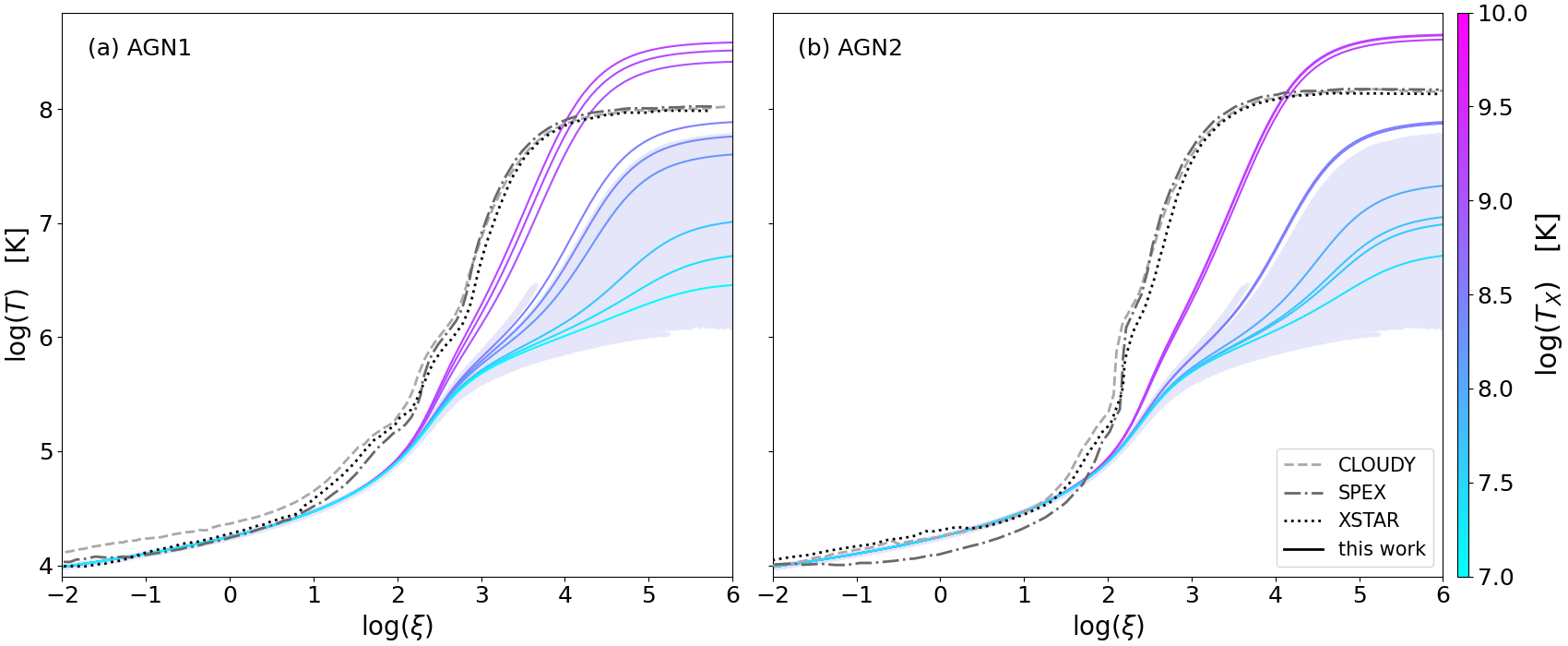}
    \caption{Temperature as a function of ionization parameter $\xi$, calculated for the source SEDs given in Figure \ref{fig: AGN1+AGN2 SED}. Panel (a) shows the values calculated when using the AGN1 initial SED, and panel (b) shows those found using AGN2. Dashed, dot-dashed, and dotted lines indicate the temperature equilibrium curves produced by CLOUDY, SPEX, and XSTAR respectively \citep{Mehdipour+2016}. Solid colored curves show the thermal equilibrium temperatures calculated via the method described above and in Section \ref{subsec:Therm. Eq.}. The blue shaded region indicates temperature values for the full set of models discussed in the body of this work.}
    \label{fig: temp_comparison}
\end{figure}

For the initial source SED, we calculated $T$ along the primary LOS. For this calculation, the temperature depends only on the ionization parameter and the shape of the input SED.
Thus, we represent the LOS using a grid of 500 values of $\xi$, evenly distributed in logspace. We set the limits of $\xi$, corresponding to the first and last radial zones, as $\log(\xi_{\rm in}) = 6$ and $\log(\xi_{\rm out}) = -2$, in accordance with the ionization parameter range used in \cite{Mehdipour+2016}.
Using this even spacing in $\xi$, we then computed the temperature using the method described in Section \ref{subsec:Therm. Eq.}. We performed this calculation using both AGN1 and AGN2 as the initial input SED. Figure \ref{fig: temp_comparison} shows the temperature curves corresponding to those produced by CLOUDY, SPEX, and XSTAR for AGN1 and AGN2. Comparative curves produced by our code are also plotted for several values of the X-ray temperature $T_X$. The various energy limits used in determining $T_X$ are shown in Table \ref{Table: freq_combos}.

We see from Figure \ref{fig: temp_comparison} that our curves produce a closer fit to the CLOUDY, SPEX, and XSTAR curves at higher values of $T_X$. When neglecting adiabatic cooling, our thermal equilibrium method depends solely on $T_X$, which in turn depends only on the chosen X-ray frequency limits and the shape of the initial SED. When using our default frequency limits $\nu_{\rm min}$ and $\nu_{\rm max}$,
corresponding to energies $E_{\rm min} = 100$ eV and $E_{\rm max} = 100$ keV, we find $\log(T_X) \approx 8.38$ for AGN1 and $\log(T_X) \approx 8.49$ for AGN2. These fall near the middle of the curves shown in Figure \ref{fig: temp_comparison}. As the value of $T_X$ increases, the agreement between the calculated temperatures and those found by the three comparison codes improves. Generally, these higher values of $T_X$ correspond to the use of wider frequency ranges in calculating the X-ray temperature when using Equation (\ref{eq: TX}).

The discrepancies between our calculated temperature values and those computed by CLOUDY, SPEX, and XSTAR are likely due to differences in the treatment of the considered heating and cooling processes, as we see similar differences to the values calculated by \cite{Dannen+2019ApJ} (see Figure \ref{fig: temp}), who used XSTAR to calculate their temperature values. Additionally, we note that each heating and cooling process incorporated into Equation (\ref{eq: Lambda-Gamma}) has its own unique dependence on the shape of the SED. Although Equation (\ref{eq: Lambda-Gamma}) generalizes these dependencies to a single value of $T_X$ for the sake of simplicity, we consider any inaccuracies that may arise as a result to be negligible once the effects of adiabatic cooling are included (as in Figure \ref{fig: temp}(a)), since this has a much greater effect on the thermal equilibrium. 

\subsection{Ionic Fractions}
We also present the ionic fractions, defined as $n_{\rm ion}/n_{\rm el}$, for select important elements. We note that this is a related but separate quantity from the mean ionization number $\mathcal{I}$, shown in Figure \ref{fig: mean_ion_frac2}. The ionic fractions detailed here represent the fraction of ions in a given ionization state when compared to the total number density of the given element. We compare these to the values taken from \cite{Mehdipour+2016}.
When calculating the ionization balance from the unobscured AGN1 SED for comparison to the results from CLOUDY, SPEX, and XSTAR we neglect the attenuation of the SED due to the intervening medium. While this attenuation is included in our main set of models, we make this modification in order to maintain similar conditions within the wind material as those in \cite{Mehdipour+2016}.

The inner and outer edges of the radial grid along the line of sight are given by
\begin{equation}
    r_{\rm in} = \sqrt{\frac{L_X}{n_H\xi_{\rm in}}}     
    \textrm{   and }
    r_{\rm out} = \sqrt{\frac{L_X}{n_H\xi_{\rm out}}}  
\end{equation}
respectively, where $\log(\xi_{\rm in}) = 6$ and $\log(\xi_{\rm out}) = -2$, as defined in Section \ref{AppB: Temps} from \cite{Mehdipour+2016}. 
We also adopt their convention of using a constant value of the hydrogen density, with $n_H = 10^8$ cm$^{-3}$, and set the ionizing luminosity $L_X = 1.17\times 10^{44}$ erg s$^{-1}$ for the unobscured AGN1 \citep{Mehdipour+2015,Mehdipour+2016}. 

\begin{figure}
    \centering
    \includegraphics[width=\linewidth]{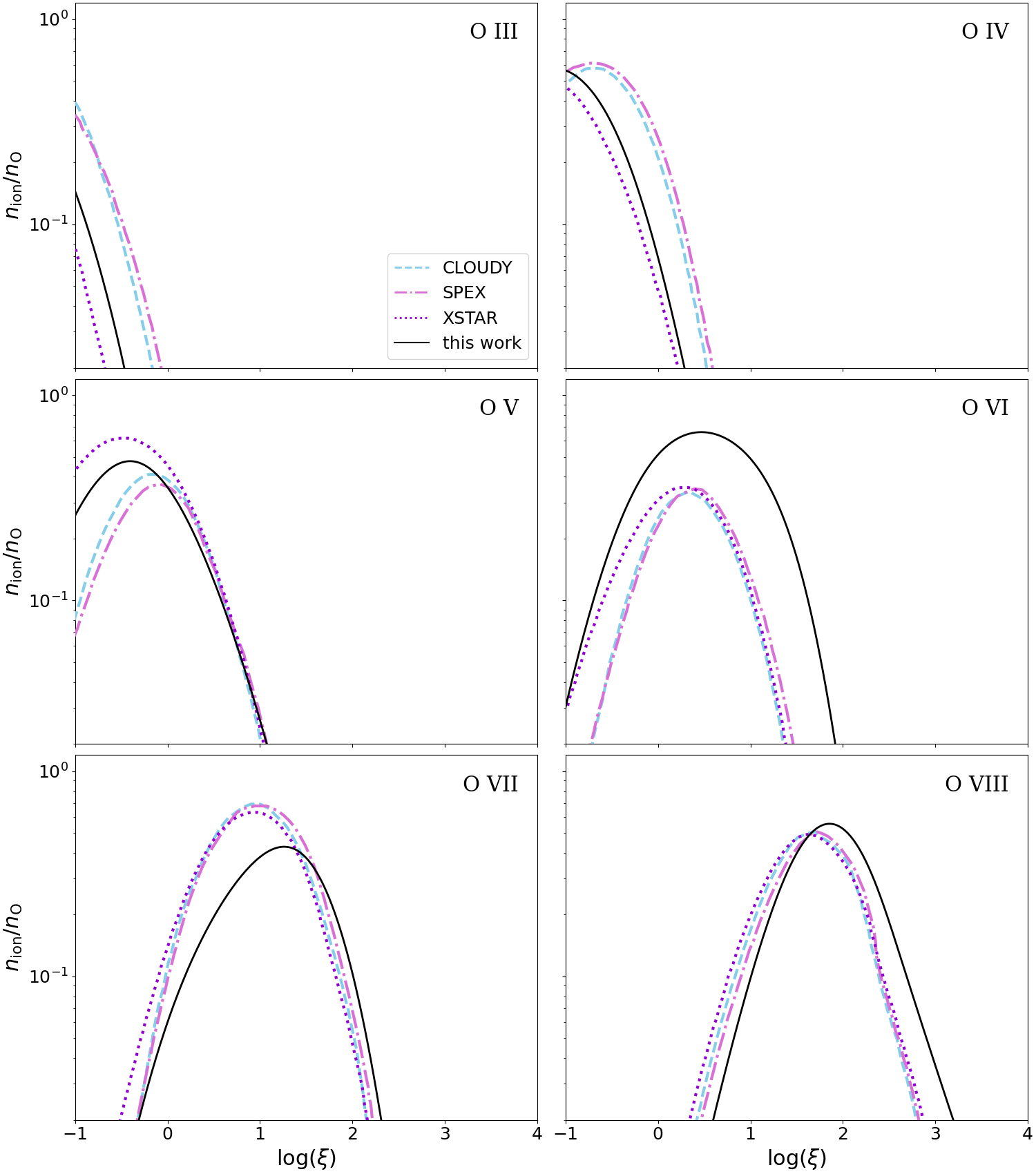}
    \caption{Ionic fractions $n_{\rm ion}/n_{\rm O}$ for oxygen ($Z=8$) ions. Colored dashed, dot-dashed, and dotted lines represent values calculated by CLOUDY, SPEX, and XSTAR respectively. Solid black lines indicated values found from the ionization balance described in Section \ref{subsec:Ion. Bal.}.}
    \label{fig: O_ionic_frac}
\end{figure}

\begin{figure}
    \centering
    \includegraphics[width=\linewidth]{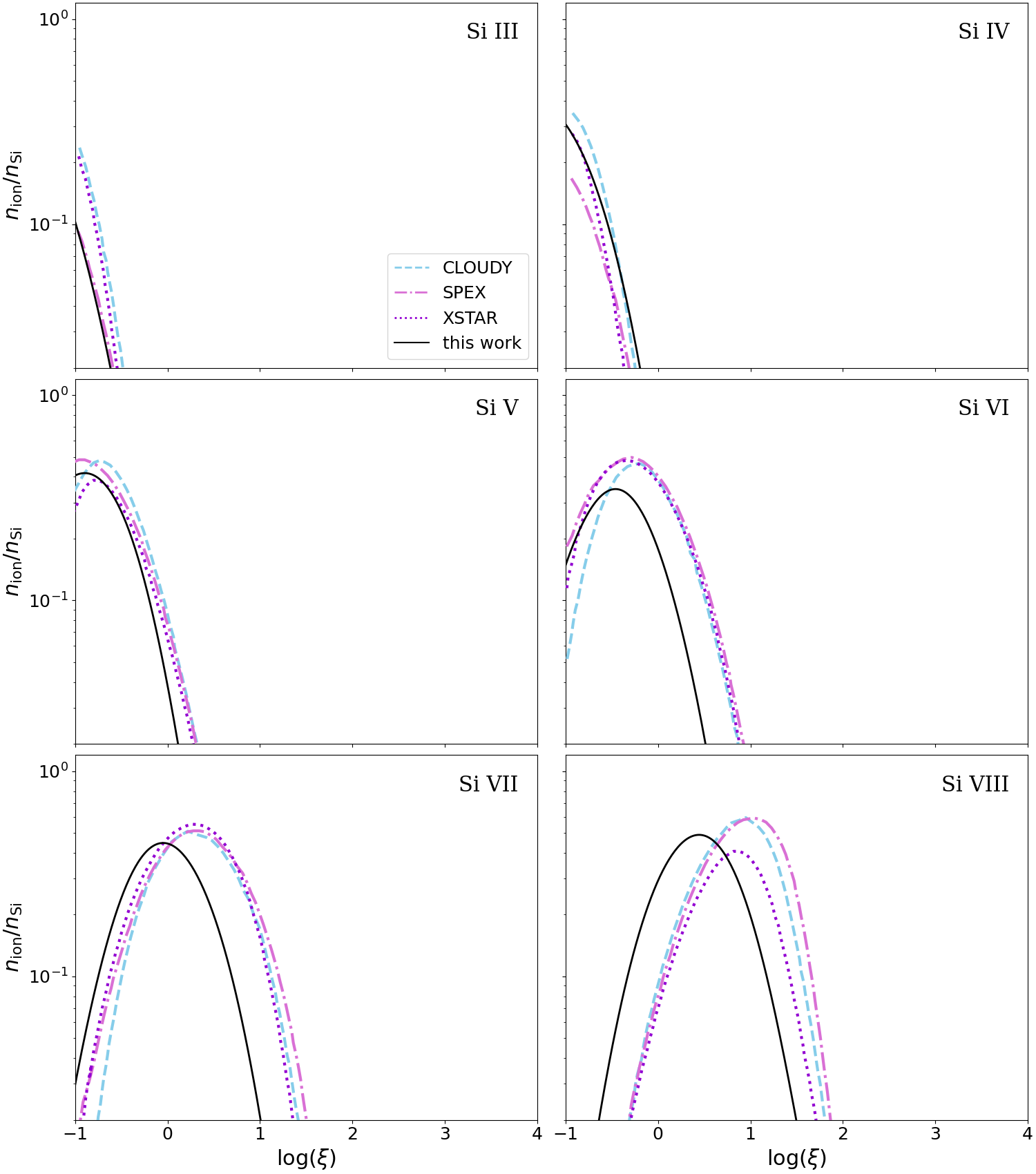}
    \caption{Ionic fractions $n_{\rm ion}/n_{\rm Si}$ for silicon ($Z=14$) ions. Colored dashed, dot-dashed, and dotted lines represent values calculated by CLOUDY, SPEX, and XSTAR respectively. Solid black lines indicated values found from the ionization balance described in Section \ref{subsec:Ion. Bal.}.}
    \label{fig: Si_ionic_frac1}
\end{figure}

\begin{figure}
    \centering
    \includegraphics[width=\linewidth]{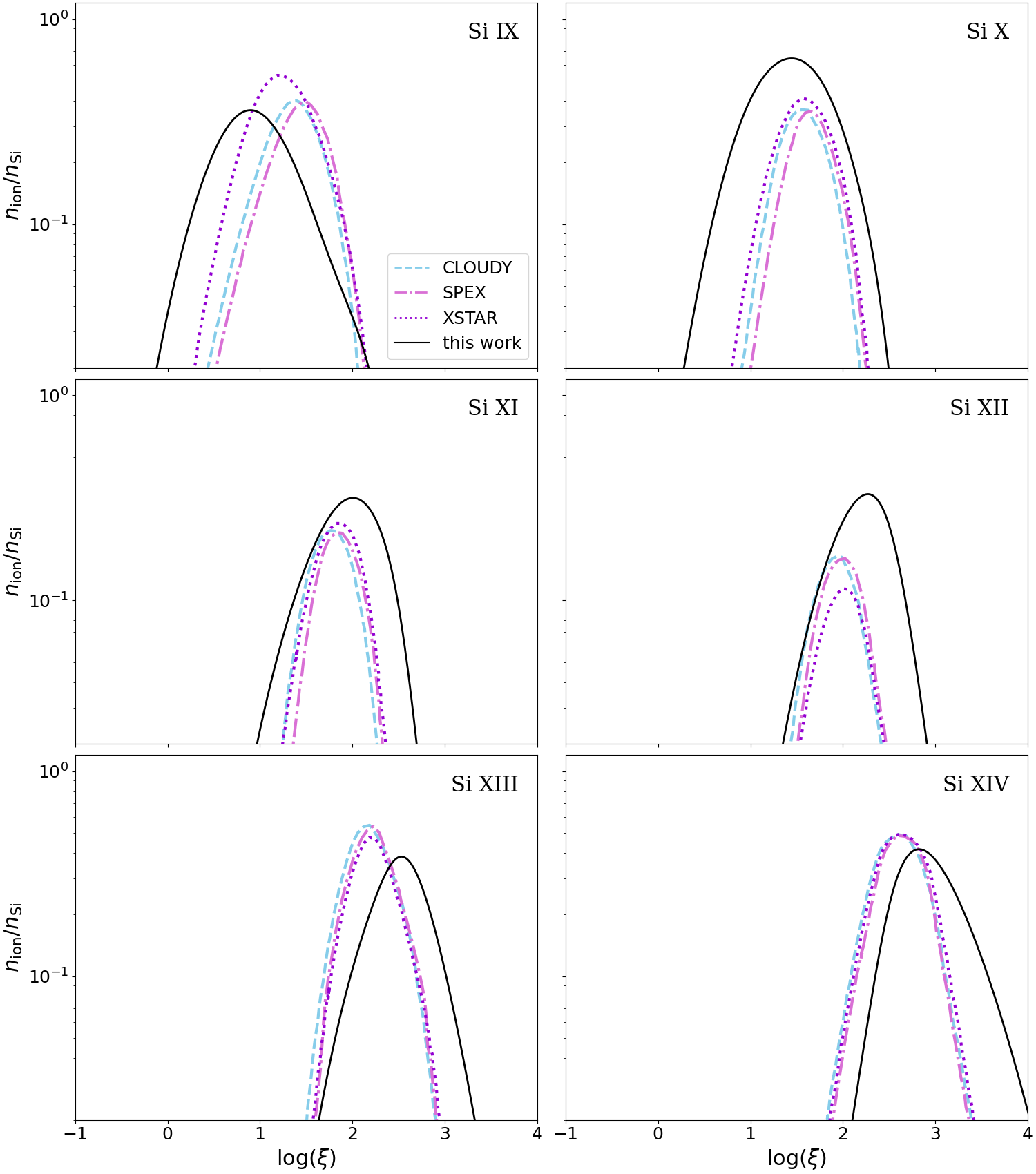}
    \caption{Ionic fractions for silicon ($Z=14$), continued.}
    \label{fig: Si_ionic_frac2}
    \end{figure}

\begin{figure}
    \centering
    \includegraphics[width=\linewidth]{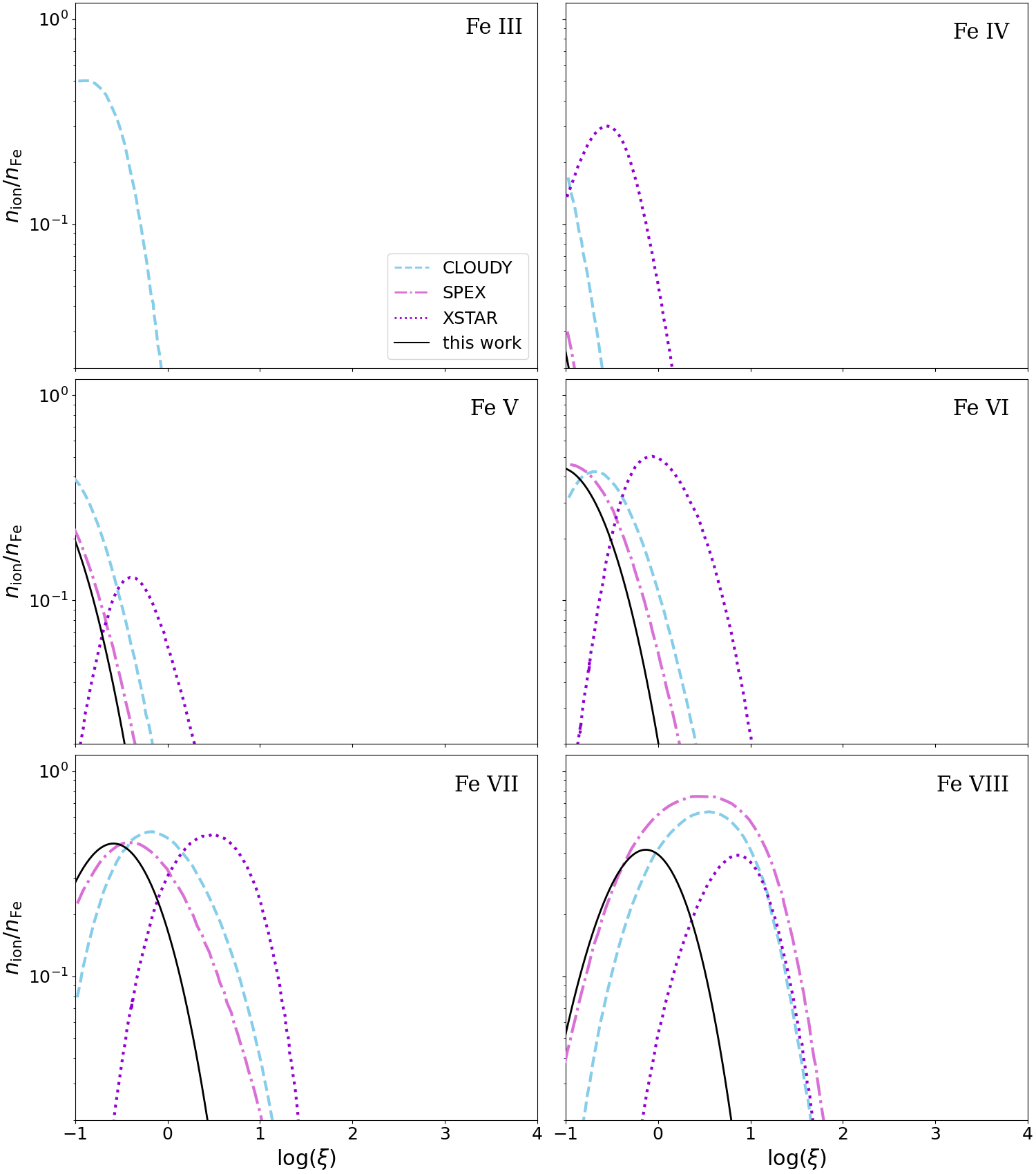}
    \caption{Ionic fractions $n_{\rm ion}/n_{\rm Fe}$ for iron ($Z=26$) ions. Colored dashed, dot-dashed, and dotted lines represent values calculated by CLOUDY, SPEX, and XSTAR respectively. Solid black lines indicated values found from the ionization balance described in Section \ref{subsec:Ion. Bal.}.}
    \label{fig: Fe_ionic_frac1}
\end{figure}

\begin{figure}
    \centering
    \includegraphics[width=\linewidth]{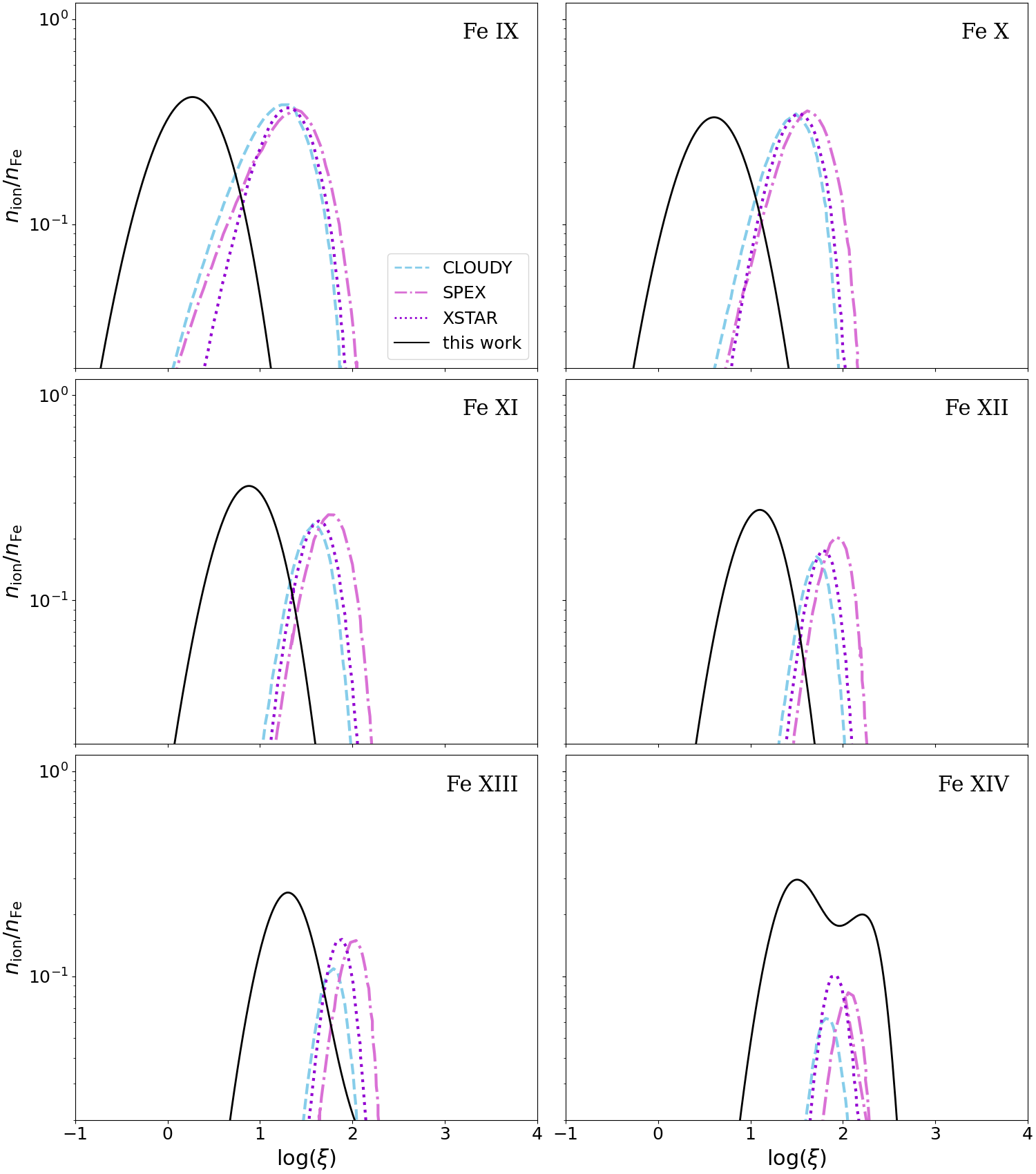}
    \caption{Ionic fractions for iron ($Z=26$), continued.}
    \label{fig: Fe_ionic_frac2}
\end{figure}

\begin{figure}
    \centering
    \includegraphics[width=\linewidth]{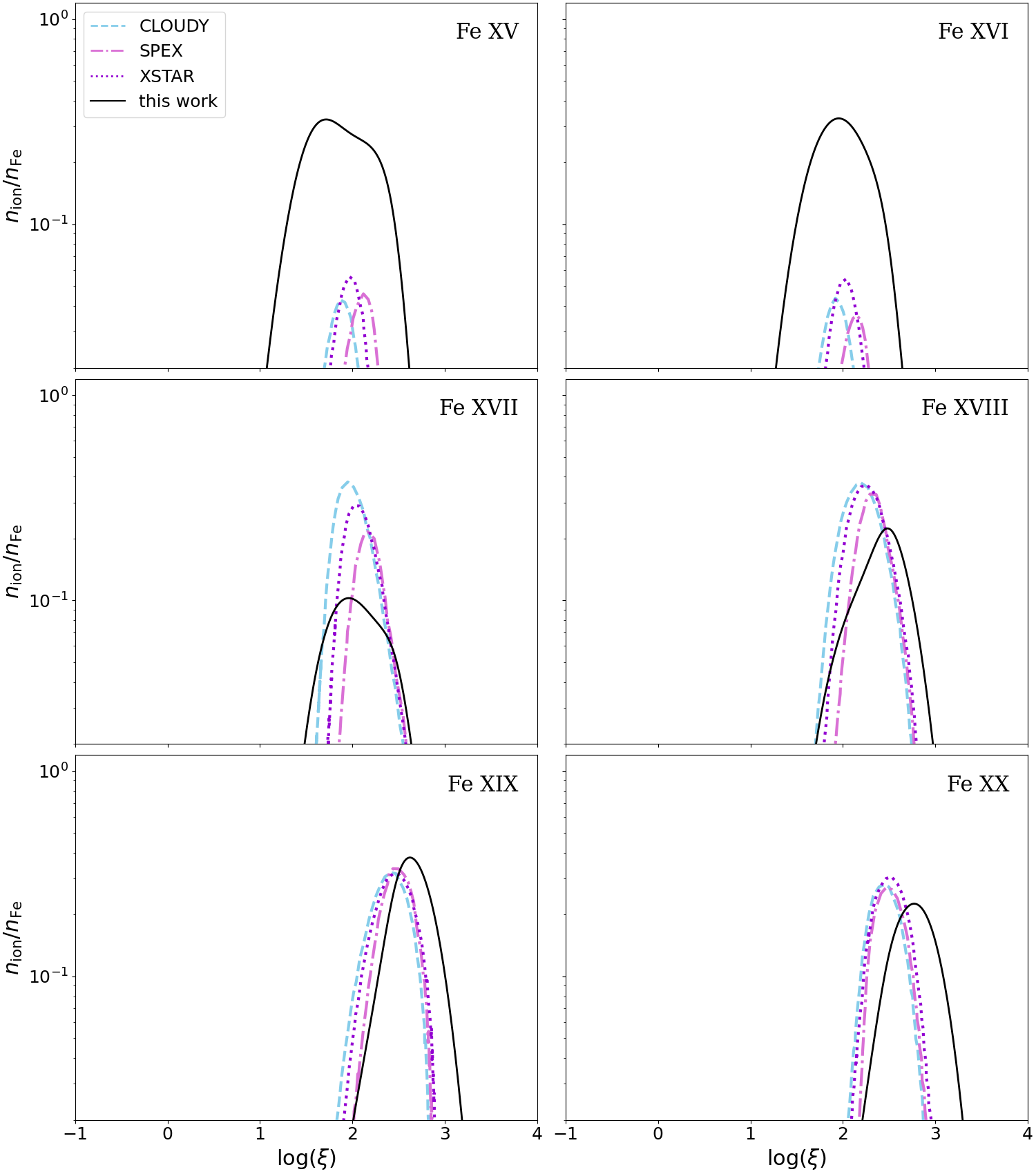}
    \caption{Ionic fractions for iron ($Z=26$), continued.}
    \label{fig: Fe_ionic_frac3}
\end{figure}

\begin{figure}
    \centering
    \includegraphics[width=\linewidth]{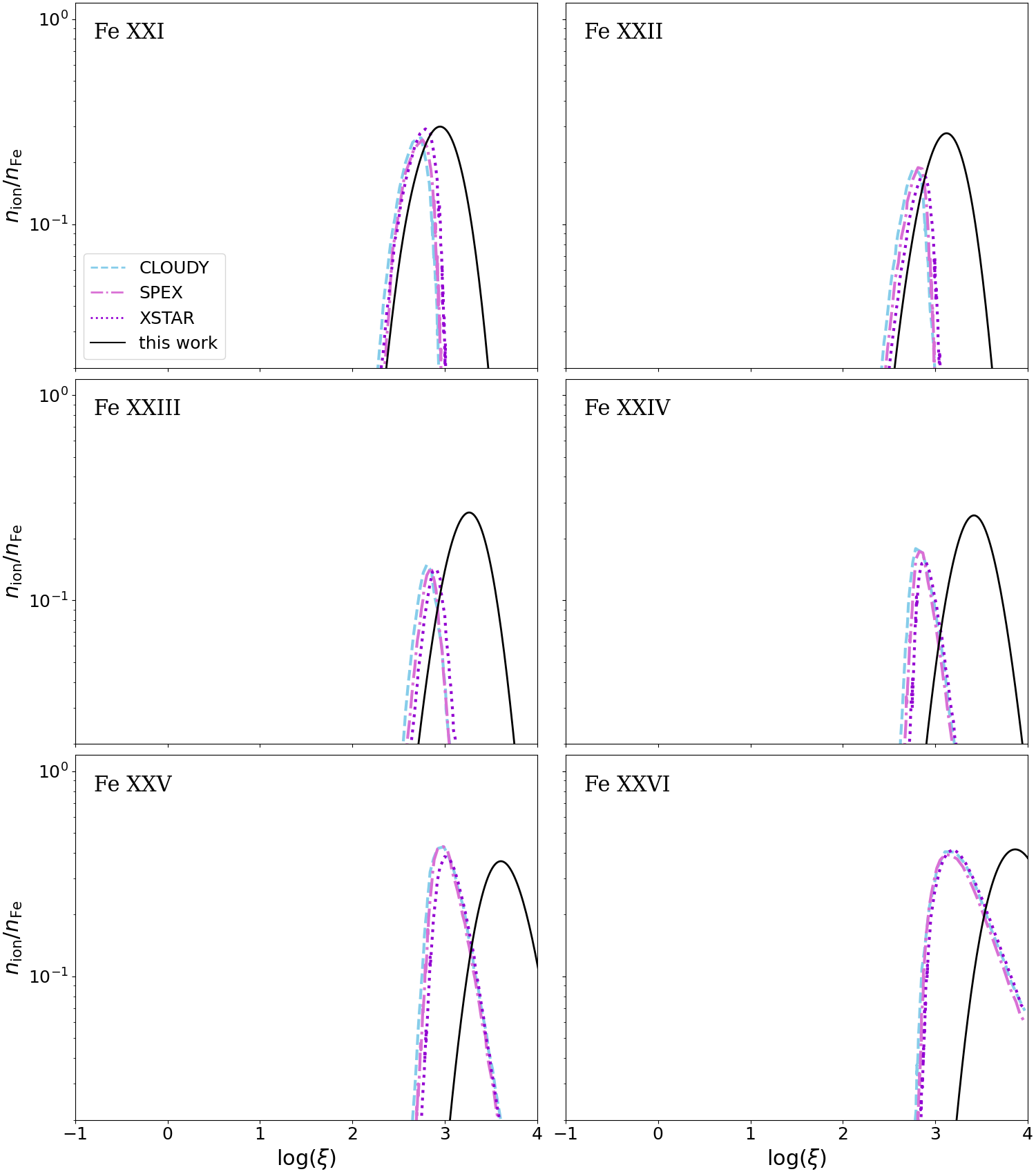}
    \caption{Ionic fractions for iron ($Z=26$), continued.}
    \label{fig: Fe_ionic_frac4}
\end{figure}

Figures \ref{fig: O_ionic_frac}-\ref{fig: Fe_ionic_frac4} show the calculated ion fractions, alongside those taken from \cite{Mehdipour+2016} for elements O, Si, and Fe. These elements were chosen to be consistent with Figure \ref{fig: mean_ion_frac2} wherever possible. 

We see from Figures \ref{fig: O_ionic_frac}-\ref{fig: Fe_ionic_frac4} that our ionization balance calculations tend to produce results in good agreement with the three comparison codes for lower-Z elements. This trend is also shown for low ionization states. While the agreement between the four plotted curves begins to decrease for high-Z and high ionization states, such as high Fe ionization states, this is not unexpected given the deviations found between the different thermal equilibrium curves in Section \ref{AppB: Temps}. We note also that the three comparison codes tend to show higher degrees of mutual disagreement for these same ions (see, for example, Fe~VII). 

The discrepancies between our ionic fractions and those calculated by CLOUDY, SPEX, and XSTAR are likely due to a combination of propagating effects from the thermal equilibrium calculation as well as differences in the treatment of the cloud of wind material. For example, \cite{Mehdipour+2016} assumes a geometry of a central AGN with a single SED (AGN1) obscured by a surrounding cloud, which in turn produces a separate SED (AGN2). However, our models treat these two components together, making a direct comparison between the two difficult. Additionally, \cite{Mehdipour+2016} used different set of elemental abundances (those from \cite{Lodders2009}) in their calculation of the photoionization equilibria resulting from the AGN1 and AGN2 SEDs. It is worth noting that these abundances, given in Table 1 of \cite{Mehdipour+2015}, do not include a number of elements that are included in \cite{Asplund2021}, which we use in our calculations. This may also result in inconsistencies in the resulting ionization balances. Given these factors, we consider the agreement between the calculated ionization balance and those found by the three comparison codes to be within acceptable limits for the purposes of this work.

\bibliography{BIB.bib}
\bibliographystyle{aasjournal}

\end{document}